\documentclass[onecolumn]{aastex631}
\usepackage{url}
\usepackage{amsmath}
\usepackage{natbib}
\usepackage{graphicx}
\usepackage{psfig}
\usepackage{ifthen}
\usepackage{hyperref}
\usepackage[all]{hypcap}
\newcommand{\stimes}{{\kern-0.08em\times\kern-0.1em}}

\newcommand{\eb}[1]{\begin{equation}\label{eq:#1}}
\newcommand{\en}{\end{equation}}
\newcommand{\eab}[1]{\begin{eqnarray}\label{eq:#1}}
\newcommand{\ean}{\end{eqnarray}}

\newcommand{\grad}{\nabla}
\newcommand{\curl}{\nabla\times}
\newcommand{\ddt}[1]{\frac{\partial #1}{\partial t}}
\newcommand{\ddx}[1]{\frac{\partial #1}{\partial x}}
\newcommand{\ddy}[1]{\frac{\partial #1}{\partial y}}

\newcommand{\ddi}[1]{\frac{\partial #1}{\partial x_i}}
\newcommand{\ddj}[1]{\frac{\partial #1}{\partial x_j}}

\DeclareMathAlphabet{\mathbfit}{OML}{cmm}{b}{it}
\renewcommand{\vec}[1]{\mathbfit{#1}}

\renewcommand{\div}{\nabla\cdot}
\renewcommand{\d}{{\rm d}}
\newcommand{\dd} [2]{{\frac{{\rm d}{#1}}{{\rm d}{#2}}}}
\newcommand{\bin}{{\{\lambda\}_i}}

\newcommand{\uu}{\vec{u}}

\newcommand{\Om}{\vec{\Omega}}

\newcommand{\Qvisc}{Q_{\rm visc}}
\newcommand{\Tvisc}{\vec{\tau}_{\rm visc}}
\newcommand{\Tij}{\tau_{ij}}

\newcommand{\Bl}{B_{\lambda}}

\newcommand{\Il}{I_{\lambda}}
\newcommand{\Jl}{J_{\lambda}}

\newcommand{\kv}{\kappa_{\nu}}
\newcommand{\kl}{\kappa_{\lambda}}
\newcommand{\sigl}{\sigma_{\lambda}}
\newcommand{\al}{a_{\lambda}}
\newcommand{\half}{\frac{1}{2}}

\newcommand{\BB}{\vec{B}}
\newcommand{\EE}{\vec{E}}
\newcommand{\JJ}{\vec{J}}
\newcommand{\grav}{\vec{g}}

\newcommand{\dxdy}[2]{\frac{ \partial #1 }{ \partial #2 }}

\newcounter{saveeqn}
\newcounter{temp}
\newcommand{\alpheqn}{\setcounter{saveeqn}{\value{equation}}%
  \stepcounter{saveeqn}\setcounter{equation}{0}\let\theorigeqn=\theequation%
  \renewcommand{\theequation}{\setcounter{temp}{\value{equation}}%
    \setcounter{equation}{\value{saveeqn}}\theorigeqn%
    \setcounter{equation}{\value{temp}}\alph{equation}}}
\newcommand{\reseteqn}{\setcounter{equation}{\value{saveeqn}}
  \let\theequation=\theorigeqn}
\newcommand{\be}[1]{\begin{equation}\label{#1}}
\newcommand{\ee}{\end{equation}}
\newcommand{\bea}[1]{\begin{eqnarray}\label{#1}}
\newcommand{\eea}{\end{eqnarray}}
\newcommand{\beaa}[1]{\alpheqn\begin{eqnarray}\label{#1}}
\newcommand{\eeaa}{\end{eqnarray}\reseteqn}
\newcommand{\mideqn}{\nonumber\\[-2\jot] &{\vbox{\hsize=0pt}}&
	\nonumber\\[-1.25\baselineskip] &{\vbox{\hsize=0em}}&\\[-0.75\baselineskip]}
\newcommand{\gwig}{{\leavevmode\kern0.3em\raise.3ex\hbox{$>$}
                    \kern-0.8em\lower.7ex \hbox{$\sim$}\kern0.3em}}
\newcommand{\lwig}{{\leavevmode\kern0.3em\raise.3ex\hbox{$<$}
                    \kern-0.8em\lower.7ex \hbox{$\sim$}\kern0.3em}}
\newcommand{\mfigur}[3]
    {\begin{figure}[htb]
         \centerline{\includegraphics[height=#2,clip=true]{#1.jpg}}
        \caption{\label{fig:#1}#3}
    \end{figure}}

\def\aatab#1#2#3#4#5#6{\ifthenelse{\equal{*}{#1}}
{\begin{table*}[htbp]\caption{\label{#2} #3}
    \centering
    \begin{tabular}{#4}
      \hline\noalign{\smallskip} #5
          \noalign{\smallskip} \hline \noalign{\smallskip} #6
          \noalign{\smallskip} \hline
        \end{tabular}
\end{table*}}
{\begin{table}[htbp]\caption{\label{#1} #2} 
    \centering
    \begin{tabular}{#3}
      \hline\noalign{\smallskip} #4
      \noalign{\smallskip} \hline \noalign{\smallskip} #5
      \noalign{\smallskip} \hline
    \end{tabular}
\end{table}}}

\makeatletter
\def\cases#1{\left\{\,\vcenter{\normalbaselines\m@th
    \ialign{$##\hfil$&\quad{##}\hfil\crcr#1\crcr}}\right.}
\makeatother

\makeatletter
\def\cases#1{\left\{\,\vcenter{\normalbaselines\m@th
    \ialign{$##\hfil$&\quad{##}\hfil\crcr#1\crcr}}\right.}
\makeatother

\begin{document}
	\title{The Stagger Code for Accurate and Efficient, Radiation-Coupled
           MHD Simulations}
    
    \author{Robert F. Stein}
	\affil{Department of Physics and Astronomy, Michigan State University,
                     East Lansing, MI 48824, USA}
	\email{steinr@msu.edu}

	\author{{\AA}ke Nordlund}
	\affil{University of Copenhagen, Niels Bohr Institute, 
                     Copenhagen, DK-2100, Denmark}
	\email{aake@ nbi.ku.dk}
	
    \author{Remo Collet}
	\affil{Stellare Astrophysics Centre, Department of Physics and
Astronomy, Aarhus University, Denmark}
	\email{remo.collet@gmail.com}
	
    \author{Regner Trampedach}
	\affil{Space Science Institute, 4765 Walnut Street,
           Boulder, CO 80301 USA}
	\email{rtrampedach@SpaceScience.org}

%	\offprints{R.\,Trampedach}

\begin{abstract}
We describe the Stagger Code for simulations of magneto-hydrodynamic
(MHD) systems. This is a modular code with a variety of physics
modules that will let the user run simulations of deep stellar
atmospheres, sunspot formation, stellar chromospheres and coronae,
proto-stellar disks, star formation from giant molecular clouds and
even galaxy formation. The Stagger Code is efficiently and highly
parallelizable, enabling such simulations with large ranges of both
spatial and temporal scales. We, describe the methodology of the code, 
and present the most important of the physics modules,  as well as
its input and output variables. We show results of a number of
standard MHD tests to enable comparison with other, similar
codes. In addition, we provide an overview of tests that have been
carried out against solar observations, ranging from spectral line
shapes, spectral flux distribution, limb darkening, intensity and
velocity distributions of granulation, to seismic power-spectra and
the excitation of p modes.  The Stagger Code has proven to be a
high fidelity code with a large range of uses.  \end{abstract}

\keywords{Magneto-hydrodynamics (MHD) -- Atomic processes -- Equation of
state -- Methods: numerical}

%%%%%%%%%%%%%%%%%%%%%%%%%%%%%%%%%%%%%%%%%%%%%%%%%%%%%%%%%%%%%%%%%%%%%%%%%%%%%%%
\section{Introduction}
%%%%%%%%%%%%%%%%%%%%%%%%%%%%%%%%%%%%%%%%%%%%%%%%%%%%%%%%%%%%%%%%%%%%%%%%%%%%%%%
\label{sect:intro}

Many astrophysical objects are hydrodynamical in nature, and since
hydrodynamics eludes analytical analysis in all but the simplest cases,
simulations are indispensable for the analysis and interpretation of
observations of these. We present here a code for such simulations, the Stagger
Code, named after the staggering between dynamical and thermodynamical
variables in the simulation domain. This code is modular, highly parallelizable, has accurate MHD solvers and is adaptable to many kinds of astrophysical systems.

Simulations carried out with the Stagger Code have been applied to a variety
of astrophysical problems, including
super granulation in the Sun \citep{bob:SuperGran};
solar abundance analysis
\citep{AGSS2009,scott:SolarNa-Ca,scott:SolarFeGroup,grevesse:AGSS09-HeavyElems,amarsi:3DnLTE-O,asplund:AAG2021};
scattering in the atmospheres of metal-poor stars \citep{remo:3Dscatter};
abundance analysis of metal-poor giants \citep{collet:3DabundsHaloGiant2};
analysis of convective variability in disk-integrated light \citep{chiavassa:3D-RGinterfero}; a grid of 3D convective stellar atmospheres for general use \citep{magic:stagger-grid,remo:StaggerGrid,rodriguezdiaz:StaggerGrid};
local helioseismology \citep{braun:3DsimHolography};
temporal granulation spectra as asteroseismic noise \citep{rodriguezdiaz:3DgranNoise};
non-adiabatic helio- and asteroseismology \citep{zhou:SolarModeDampExci,zhou:PhotVsSpctrModeAmpl};
magnetoconvective helioseismology \citep{piau:3D-SunSeism};
solar magnetic flux emergence \citep{stein:BEmergeSim};
high-resolution solar simulations for interpretation of DKIST observations and
for understanding physics at those small scales \citep{collet:CaseForUHDsunSim};
the dynamics and magnetic fields of molecular clouds \citep{lunttila:SuperAlfvenicMolClouds}; star formation from self-gravitating knots in molecular clouds \citep{padoan:StarFormationSim,padoan:3DprotostarLsolved}.

The Stagger Code has also taken part in code comparison and
verification efforts, most notably by \citet{beeck:3DsimComp} and \citet{kritsuk:ComparingMHDcodes}.

The Stagger Code has previously been described briefly by \cite{nordlund:StaggerCode},
with the present document being a comprehensive update, and
aiming for a more complete presentation. The structure and organization
of the code is explained in Sect.\ \ref{sect:Structure}, the governing
equations are laid out in Sect.\ \ref{sect:eqns}, how time is advanced
is described in Sect.\ \ref{sect:Time}, the options for artificial
viscosity are presented in Sect.\ \ref{sect:Diffus}, the physical
heating and cooling mechanism are described in Sect.\ \ref{sect:transf},
the available micro physics is outlined in Sect.\ \ref{sect:Opac+EOS},
external and internal forcing options are listed in
Sect.\ \ref{sect:Forcing}, boundary conditions are discussed in
Sect.\ \ref{sect:Bdry}, file-handling is explained in
Sect.\ \ref{sect:io}, a large range of numerical and observational
tests are analyzed in Sect.\ \ref{sect:tests}, and we provide some
perspectives in Sect.\ \ref{sect:conclusion}.

%%%%%%%%%%%%%%%%%%%%%%%%%%%%%%%%%%%%%%%%%%%%%%%%%%%%%%%%%%%%%%%%%%%%%%%%%%%%%%%
\section{Structure and philosophy of the Code}
%%%%%%%%%%%%%%%%%%%%%%%%%%%%%%%%%%%%%%%%%%%%%%%%%%%%%%%%%%%%%%%%%%%%%%%%%%%%%%%
\label{sect:Structure}

The Stagger Code solves the magneto-hydrodynamic (MHD) equations.
Its basic philosophy is efficient parallelization and modularity.  
This involves using MPI and
segregating different functionality and environment dependent aspects.
The goal was to enable its easy use to simulate a wide variety of 
scenarios on computers ranging from laptops to massively parallel machines.

All operating system (OS) dependent aspects are segregated into files
containing any system dependent Fortran functions, appropriate compiler
commands and flags, and system commands. 
MPI routines are called only from the differentiation, extrapolation and
I/O routines, which are contained in separate subroutines and directories.
The MPI routines themselves are in a separate directory.  The code to 
calculate the time derivatives of the physical variables is in two subroutines,
one for the fluid and the other for the magnetic field.  Time stepping 
procedure, equation of state application, heating and cooling, external 
forcing are all done in separate subroutines.  Each experiment has a subdirectory
with its own specialized routines, including any specialized boundary
conditions.

The Stagger Code routines are organized in a tree.  At the top are the
basic routines common to all calculations -- those that calculate
the time derivatives, do the time stepping, do the I/O and the MPI
operations.  At the next level are separate directories for: 
the spatial derivatives and interpolation, which call MPI
operations; the equation of state; heating/cooling; external forcing,
e.g. gravity; initial setup; utilities such as FFTs; and finally a
tree of setup instructions and specific routines for different
astrophysical experiments, e.g.  magneto-convection, star formation
and galactic structure.

This allows the Stagger Code to be applied to many different situations
with different physics, different geometry or computational mesh, and
for it to be run on a wide variety of computational platforms, with a
need to change only a few routines.  It can be run on most computers
from PCs to massively parallel supercomputers.  The modularity also
makes it easier to maintain and debug the code, as problems can be
localized and addressed quickly, considering the {$\sim$}280k lines of
code involved.  The Stagger Code is maintained in a {\tt git}
repository {\tt git@bitbucket.org:aanordlund/stagger.git}.  The version 
used for testing is tagged {\tt 2024-ApJ}.

%%%%%%%%%%%%%%%%%%%%%%%%%%%%%%%%%%%%%%%%%%%%%%%%%%%%%%%%%%%%%%%%%%%%%%%%%%%%%%%
\section{Equations}
%%%%%%%%%%%%%%%%%%%%%%%%%%%%%%%%%%%%%%%%%%%%%%%%%%%%%%%%%%%%%%%%%%%%%%%%%%%%%%%
\label{sect:eqns}

\label{section:equations}

The Stagger Code calculates the non-split time derivatives from the conservation equations 
for mass, momentum, internal energy and the induction equation for the 
magnetic field, together with Ohm's law for the electric field.

Mass conservation controls the topology of stratified convection,
\eb{mass}
    \ddt{\rho} = -\div(\rho\uu) ,
\en
where $\rho$ is the density and $\uu$ the velocity.

Momentum conservation controls the plasma motions,

\eb{mom}
\ddt{(\rho\uu)} = -\div(\rho\uu\uu)
    - \nabla P - \rho\grav + \JJ\times\BB -2\rho\Om\times\uu
    - \div\Tvisc.
\en

Here $P$ is the pressure, $\grav$ is the gravitational acceleration,
$\BB$ is the magnetic field, $\JJ=\nabla\times\BB/\mu$ is the
current and $\mu$ is the permeability (magnetic constant),
$2\rho\Om\times\uu$ is the Coriolis force and 

\eab{itv}
\Tvisc = {\mathbf \Tij} & = & \rho\nu_{ij} \left[S_{ij}+ c_1 \div\uu\delta_{ij}\right]  \\ \nonumber
      & = & \rho\nu_{ij} \left[\half \left(\ddj{u_i}+\ddi{u_j}\right) + c_1 \div\uu\delta_{ij}\right]
\ean
is the viscous stress tensor, $S_{ij}=\half \left(\ddj{u_i}+\ddi{u_j}\right)$  is the strain tensor, 
$\nu_{ij}$ the coefficient of viscosity, and $c_1$ is an input parameter.  
Typically we take $c_1=0$.  If no viscous pressure resistance to overall contraction is desired then 
$c_1=-1/3$.
In most of the computational domain we use sixth order derivatives in the viscous stress tensor.
However, where the magnitude of the second order differences $\left|\Delta^2 \ln \rho\right|$ or 
$\left|\Delta^2 \ln e\right|$ are large, we switch to second order derivatives.

At large depths where fluctuations are small there can be issues with roundoff error in the
derivatives of the pressure.  To address this, Stagger has a switch to do all the
derivative and centering operations on $\ln(P/P_{\rm aver})$, where $P_{\rm aver}$ is the
horizontally averaged gas pressure at the given vertical position.  For instance, the
vertical derivative of the pressure becomes
\eb{pressure}
\left(\ddy{P}\right)_{j-1/2} \rightarrow
   \exp{\ln(P_{\rm aver})_{j-1/2}}\exp\ln(P/P_{\rm aver})_{j-1/2}
   \left[(\ddy{\ln{P/P_{\rm aver}}})_{j-1/2}+(\ddy{\ln P_{\rm aver}})_{j-1/2}\right] \ .
\en
The horizontal derivatives are similar except that the horizontal derivative of $P_{\rm aver}$ vanishes.

Internal energy is changed by transport, by $P\d V$ work, by Joule heating,
by viscous dissipation and by heating and cooling.  It is the
fluid version of the second law of thermodynamics and (together with
the density) determines the plasma temperature, pressure and entropy. 
\eb{ien}
    \ddt{\rho e} = -\div (\rho e\uu) - P (\div\uu) + Q + \Qvisc + \eta \JJ^2 ,
\en
where $e$ is the internal energy per unit mass.
$Q$ is the heating or cooling.  It can be Newtonian cooling 
or by thermal conduction, optically thin radiation, 
or by radiative transfer.

The viscous dissipation is
\eb{vdiss}
    \Qvisc  =  \sum_{ij} \Tij \ddj{u_i} = \sum_{ij} \Tij S_{ij}
\en

Convection influences the magnetic field via the curl $\EE$ term in the induction equation,
\eb{indu}
    \ddt\BB = - \nabla\times\EE\ ,
\en
where the electric field is given by Ohm's Law.  In a one-fluid MHD system, it is
\eb{ohm}
    \EE = -\uu\times\BB + \eta \JJ + \frac{1}{e n_{\rm e}}\left(\JJ\times\BB-\grad P_{\rm e}\right) ,
\en
and $\eta$ is the resistivity, $n_{\rm e}$ is the electron number density,
$P_{\rm e}$ is the electron pressure and $e$ is the electron charge.
The last two (Hall and pressure) terms are usually neglected, but the Hall
term may be important in the weakly ionized photosphere.  
$\BB$ is periodically cleaned to keep $\div\BB=0$ \citep{Brackbill&Barnes80,Toth00}.  
Since $\BB=\curl{\bf A} + \nabla \phi$\ (where $\bf A$ is the vector potential and $\phi$
is the scalar potential for the magnetic field), 
\eb{clean1}
    \div\nabla\phi=\div \BB \ .
\en
This is solved for $\phi$ in Fourier space and the corrected $\BB$ is 
\eb{clean2}
    \BB=\BB-\nabla\phi\ .
\en
In addition, a term can be added to the magnetic field time derivative,
$C_{\rm divb}{\min({\rm d}x,{\rm d}y,{\rm d}z)^2}/{\rm d}t \div\BB$, 
(where $C_{\rm divb}$ is an input constant $< 0.11$ for stability,
d$x$,d$y$,d$z$ are the local grid spacings in their respective directions and d$t$ the time step) 
to diffuse away the divergence of $\BB$ and ensure it is  zero to machine accuracy at all times.

The time derivatives of density, momenta and internal energy are 
calculated in one routine and the time derivative of the magnetic field in
another to enable both fluid and MHD calculations to be easily performed.

%%%%%%%%%%%%%%%%%%%%%%%%%%%%%%%%%%%%%%%%%%%%%%%%%%%%%%%%%%%%%%%%%%%%%%%%%%%%%%%
\section{Space Operations}
%%%%%%%%%%%%%%%%%%%%%%%%%%%%%%%%%%%%%%%%%%%%%%%%%%%%%%%%%%%%%%%%%%%%%%%%%%%%%%%
\label{sect:Space}

%{
%(
%++++++++++++++++++++++++++++++++++++++++++++++++++++++++++++++++++++++++++++++
\subsection{Centering}
%++++++++++++++++++++++++++++++++++++++++++++++++++++++++++++++++++++++++++++++
\label{sect:centerin}

\begin{figure}[!htb]
  \includegraphics[width=0.6\textwidth]{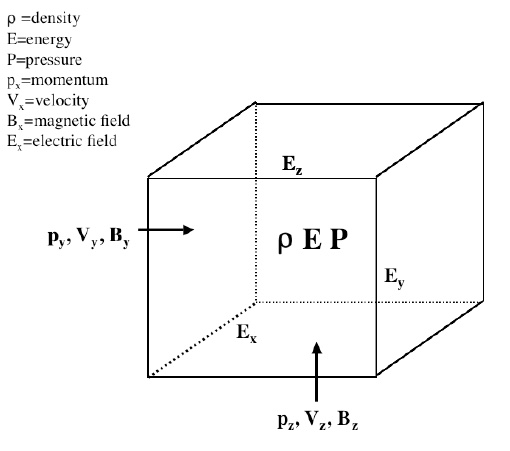}
  \caption
  {Computational cell showing the location of the physical variables.
  \label{fig:centering}}
\end{figure}

The physical variables are staggered.  Density, energy, pressure are located at cell centers.  Momenta,
velocities, magnetic fields are centered on cell faces at the lower index side.  Electric fields are
centered on the cell edges (figure \ref{fig:centering}).

%++++++++++++++++++++++++++++++++++++++++++++++++++++++++++++++++++++++++++++++
\subsection{Derivatives}
%++++++++++++++++++++++++++++++++++++++++++++++++++++++++++++++++++++++++++++++
\label{sect:SpaceDerivs}

Derivatives, except possibly in diffusion terms, are evaluated by sixth order, centered differences, 
in index space with the Jacobian of the transformation back to physical space,
\eab{deriv}
 \ddx{f} (i+1/2  ,j,k) & = & \ddx{i}\bigg[
                       c\left\{f(i+3 ,j,k)-f(i-2 ,j,k)\right\} \nonumber \\
                 & + & b\left\{f(i+2 ,j,k)-f(i-1 ,j,k)\right\} \\ \nonumber
                 & + & a\left\{f(i+1 ,j,k)-f(i   ,j,k)\right\}\bigg]\ .
\ean
The weighting coefficients are, 
\eab{derivcoef}
  c & = & \frac{-1+(3^5-3)/(3^3-3)}{5^5-5-5(3^5-3)} = \frac{3}{640} \nonumber \\
  b & = & \frac{-1-120c}{24} = -\frac{25}{384}\\ \nonumber
  a & = &  1-3b-5c = \frac{225}{192}\ .
\ean
Similar formulas are used for each dimension and for downward derivatives..
The Jacobian for equally spaced directions is simply $1/\delta x_{i}$.  For stretched grid directions
the Jacobian is calculated by first making a spline fit to the grid spacing and taking the Jacobian as
the inverse of the fitted grid interval at the center of the derivative.
All the derivative routines assume periodicity and take variable values falling outside the boundary
from the corresponding locations in the opposite boundary.  
MPI routines are used get these values.
In non-periodic directions boundary conditions load values into ghost zones
at each boundary.

%++++++++++++++++++++++++++++++++++++++++++++++++++++++++++++++++++++++++++++++
\subsection{Interpolations}
%++++++++++++++++++++++++++++++++++++++++++++++++++++++++++++++++++++++++++++++
\label{sect:interpolations}

Interpolations to properly center variables are calculated by a fifth order, centered template in index
space,

\eab{interp}
f(i+1/2,j,k) & = & c[f(i-2,j,k)+f(i+3 ,j,k)] \nonumber \\
             & + & b[f(i-1,j,k)+f(i+2 ,j,k)] \\
             & + & a[f(i  ,j,k)+f(i+1 ,j,k)]\ , \nonumber
\ean
where
\eab{interpcoef}
  c & = & 3/256 \nonumber   \\
  b & = & -25/256 \\
  a & = & \half-b-c = \frac{150}{256}\nonumber\ .
\ean
Similar equations are used in each dimension and for downward interpolations.
As for the derivatives, MPI routines are used to get values at the opposing periodic boundaries and
ghost zones are used for boundary values in non-periodic directions.

%++++++++++++++++++++++++++++++++++++++++++++++++++++++++++++++++++++++++++++++
\subsection{Coordinate System}
%++++++++++++++++++++++++++++++++++++++++++++++++++++++++++++++++++++++++++++++
\label{sect:coord}

In stratified calculations (e.g. stellar convection) the stratified direction is taken as the 
$y$-direction.  The $x-$ and $z$-directions are periodic.  In stellar calculations the indices 
increase downward into the star.  Hence the lower boundary in index space is in the photosphere 
and the upper boundary deep in the convection zone.
Together with the horizontal $x$- and $z$-directions, this forms a right-handed coordinate system

%)
%}

%%%%%%%%%%%%%%%%%%%%%%%%%%%%%%%%%%%%%%%%%%%%%%%%%%%%%%%%%%%%%%%%%%%%%%%%%%%%%%%
\section{Time Advance}
%%%%%%%%%%%%%%%%%%%%%%%%%%%%%%%%%%%%%%%%%%%%%%%%%%%%%%%%%%%%%%%%%%%%%%%%%%%%%%%
\label{sect:Time}

%{
%(

Time advance is by a low-memory usage, third order Runge-Kutta scheme \cite{kennedy:LowMemRungeKutta}.
The time derivatives of the variables are evaluated in a non-split manner, three times for each full
time step, using equations \ref{eq:mass}--\ref{eq:ohm}. The time advance scheme is the following:
Suppose the time derivatives are given by a function F, i.e.
\eb{dfdt}
\left(\ddt{f}\right)^i = F(f^i)
\en
In a cycle of three iterations, $i=1,3$, updated derivatives are calculated and added to a 
fraction $\alpha_i$ of the previous derivatives and stored in the same location.  
\eb{ddt}
\left(\ddt{f}\right)^i  = \alpha_i\left(\ddt{f}\right)^{i-1}+F(f^i)  
\en
Note that $\left(\partial f/\partial t\right)^{i-1}$ and 
$\left(\partial f/\partial t\right)^{i}$ are actually the same location in memory..  
First the existing time derivative is multiplied by $\alpha_i$ and restored 
in the same location.  Then each term in the new time derivative is 
added to $\left(\partial f/\partial t\right)^{i}$ as they are calculated.
In the first iteration $\alpha_1=0$, so for each time step all the derivatives 
are first initialized to 0.  After the update of the time derivatives is
completed, the variables $f$ are updated
\eb{tupdate}
f^{i+1}  =  f^i + \beta_i \Delta t \left(\ddt{f}\right)^i, 
\en
and the next iteration begun until after three cycles the variables for the next time step are
obtained.
The constants are
\eab{tconst}
\alpha_i & = & [0      , -0.641874,-1.31021]    \mideqn 
\beta_i  & = & [0.46173,  0.924087, 0.390614]   \nonumber
\ean
Only one set of variables and one set of their time derivatives needs to be stored.

%-----------------------------------------------------------------------------
\subsection{Courant Condition}
%-----------------------------------------------------------------------------
\label{sect:courant}

The time step is set by a combination of the usual Courant condition with the wave speed
plus fluid velocity (in the case of magnetic field the fluid velocity perpendicular to
the field), 
together with a limit depending on the rate of change of the density and energy 
and a limit depending on the rate of dissipation (Eqn. \ref{eq:courantcond}).

\eb{courantcond}
dt_{\rm new}  =  dt \min_{ijk}\left[{C_{\rm dt} \over
{\max_{ijk}[(C_s+C_A+|\uu_{\perp\BB}|)(dt/d\ell_{\rm min})]}},
{C_{\rm dtr} \over {\max_{ijk}[{dt\over\rho}\ddt\rho,{dt\over e}\ddt e]}},
{C_{\rm dtd} \over {\max_{ijk}[(\nu, \eta d\ell_{\rm max}/d\ell_{\rm min})(dt/d\ell_{\rm
min})] }}\right] \ .
\en
Here $\min_{ijk}$ and $\max_{ijk}$ are the minima and maxima over the entire
computational valume (excluding ghost zones).
$C_{\rm dt}$, $C_{\rm dtr}$ and $C_{\rm dtd}$ are input parameters (see table \ref{sect:RUNparms}). 
$C_{\rm s}$ is the sound speed and $C_{\rm A}$ is the Alfv{\' e}n speed.
$\nu$ and $\eta$ are the viscosity and resistivity.
${\rm d}\ell_{\rm min} = \min({\rm d}x,{\rm d}y,{\rm d}z)$ 
and ${\rm d}\ell_{\max}$ is the corresponding maximum.

%)
%}

%%%%%%%%%%%%%%%%%%%%%%%%%%%%%%%%%%%%%%%%%%%%%%%%%%%%%%%%%%%%%%%%%%%%%%%%%%%%%%%
\section{Viscosity and Diffusion}
%%%%%%%%%%%%%%%%%%%%%%%%%%%%%%%%%%%%%%%%%%%%%%%%%%%%%%%%%%%%%%%%%%%%%%%%%%%%%%%
\label{sect:Diffus}

%++++++++++++++++++++++++++++++++++++++++++++++++++++++++++++++++++++++++++++++
\subsection{Viscosity}
%++++++++++++++++++++++++++++++++++++++++++++++++++++++++++++++++++++++++++++++
\label{sect:viscosity}

Higher order schemes (higher than linear) are inherently unstable, and
need a mechanism to suppress the growth of unphysical features. One such
mechanism is an artificial viscosity, to dampen unphysical oscillations between
neighboring grid points, ringing at sharp changes and to stabilize sound and
Alfv{\`e}n waves, as well as shocks.

There are two philosophically different approaches to artificial viscosity: One
is for the viscosity to describe the effects at the grid-scale, of the actual,
molecular viscosity acting on the dissipation-scale of the plasma (see Sect.\
\ref{sect:SmagDiffus}); Another is the entirely utilitarian approach of
stabilizing the simulation with a minimal amount of viscosity, of which Sect.\
\ref{sect:RichtMortDiffus} gives an example.

There are input variable switches to choose different types of viscosity (see table \ref{sect:PDFparms}).  The simplest is
a constant viscosity $\nu_{ij}=const$.  This is useful for testing purposes, or
for making direct numerical simulations (DNS), but is otherwise an inefficient
use of a given number of grid-points in a simulation.  In general,
we use a Richtmyer \& Morton type of viscosity proportional to the wave and fluid speeds.

%-----------------------------------------------------------------------------
\subsubsection{Richtmyer \& Morton Type Viscosity}
%-----------------------------------------------------------------------------
\label{sect:RichtMortDiffus}

The viscosity model by \citet{richtmyer:ViscosityModel} is
\eb{RMnu}
    \nu=c_{\rm v} |\uu| + c_{\rm w} (s + a) - {\mathbf c_{\rm sh} \div\uu} \ ,
\en
where $s$ is the sound speed and $a$ is the Alfv{\`e}n speed and $c_{\rm v}$, 
$c_{\rm sh}$ and $c_{\rm w}$ are input parameters (nu1,nu2 and nu3).
The viscosity is scaled by the grid-spacing, $\Delta x_i$,
which accommodates non-isotropic grids, and also makes the choice of diffusion
constants, e.g., nu{1,2,3} independent of resolution to first order.
The influence of the local viscosity is spread by taking its 
maximum over three or five adjacent cells and then smoothing it by taking the average
over three adjacent cells in each direction, denoted by $\left<\nu\right>_{\rm max\ smooth}$.
\eab{nuij}
    \nu_{ii} & = & (\Delta x_i) \left<\nu\right>_{\rm max\ smooth} \\ \nonumber
    \nu_{i\ne j} & = & \max(\Delta x_i, \Delta x_j) \left<\nu\right>_{\rm max\ smooth} \ .
\ean
For the off-diagonal viscosity the averaging is carried out on $\ln \nu$ and then
the exponential is taken, in order to avoid edge effects of large gradients.

%-----------------------------------------------------------------------------
\subsubsection{Smagorinsky Type Viscosity}
%-----------------------------------------------------------------------------
\label{sect:SmagDiffus}

Another possible form for the viscosity is the
\citet{smagorinsky:GenCircExpsPrimitiveEqs} type. We take 
\eb{nusmag}
    \nu = c_{\rm Smag} \left[2 \left( \sum_i S_{ii}^2 - \frac{1}{3}(\nabla \cdot \uu)^2 +
    \sum_{ij}S_{ij}^2\right)\right]^{1/2}\ .
\en
This viscosity is sometimes reduced with depth in stratified atmospheres by a factor
$1/[1+(\langle\rho\rangle/\rho_0)^{0.3}]$, where $\rho_0$ is an input reference density.  In this case also
the viscosities influence is spread by taking the maximum over three or five adjacent cells and 
then smoothed by averaging over three adjacent cells.  
\eb{nuij2}
    \nu_{{\mathbf ii}} = (\Delta x_i)^2 \left<\nu\right>_{\rm max\ smooth} \ .
\en

Application of the Smagorinsky approach is often justified by it describing the
viscosity in the case of a self-similar turbulent cascade of isotropic
turbulence. For a numerical simulation to exhibit isotropy at the grid-scale,
it would need a large inertial range, though, making this argument spurious
in many cases. This does not, however, affect its utility in stabilizing a 
numerical scheme with little effect in smooth regions.

We often include both the Richtmyer \& Morton and the Smagorinsky viscosities
in the same simulation in varying proportions.

%++++++++++++++++++++++++++++++++++++++++++++++++++++++++++++++++++++++++++++++
\subsection{Quenching}
%++++++++++++++++++++++++++++++++++++++++++++++++++++++++++++++++++++++++++++++
\label{sect:quenching}

We reduce the viscous stresses in smooth regions (and enhance it for two zone wiggles) by the ratio of 
the second differences to the absolute
value of each viscous stress tensor component, $T$.  
\eab{quench}
q_i & = & \Delta^2 T / \left[|T| + \Delta^2 T/q_{\rm max}\right] \\ \nonumber
T & = & q_i \min\left[\max (q_{i+1},q_i,q_{i-1}),q_{\rm lim}\right] \ .
\ean
$q_{\rm max}$ and $q_{\rm lim}$ are input parameters (see table \ref{sect:QUENCHparms}).

%++++++++++++++++++++++++++++++++++++++++++++++++++++++++++++++++++++++++++++++
\subsection{Diffusion in the Energy Equation}
%++++++++++++++++++++++++++++++++++++++++++++++++++++++++++++++++++++++++++++++
\label{sect:energydiffuse}

The momentum equation (\ref{eq:mom}) is stabilized by dissipation (Eq. \ref{eq:itv}), a diffusive process..  This 
dissipation acts as a heat source in the energy equation (\ref{eq:ien}).  To keep the energy equation (\ref{eq:ien}) 
stable requires the addition of some diffusion,
\eb{ediffuse}
\ddt{\rho e} {\mathbf \rightarrow} \ddt{\rho e} + c_E\sum_i{\frac{\partial}{\partial x_i}} 
\left[ \rho \nu_{ii} \Delta x_i e \frac{\partial \ln e}{\partial x_i}\right] \ ,
\en
where $c_E$ is an input parameter (nuE in table \ref{sect:PDFparms}).
When there is a mean gradient in the energy, we want to only diffuse fluctuations about the mean.  In
that case the energy diffusion is 
\eb{ediffav}
\ddt{\rho e} {\mathbf \rightarrow} \ddt{\rho e} +  c_E\sum_i\frac{\partial}{\partial x_i}
     \left[ \rho \nu_{ii} \Delta x_i e 
     \frac{\partial (\ln e - \ln\langle e\rangle_{\rm horiz})}{\partial x_i}\right] \ ,
\en
where $\ln\langle e\rangle_{\rm horiz}$ is the horizontal average of the internal energy per unit mass.

%%%%%%%%%%%%%%%%%%%%%%%%%%%%%%%%%%%%%%%%%%%%%%%%%%%%%%%%%%%%%%%%%%%%%%%%%%%%%%%
\section{Heating and Cooling}
%%%%%%%%%%%%%%%%%%%%%%%%%%%%%%%%%%%%%%%%%%%%%%%%%%%%%%%%%%%%%%%%%%%%%%%%%%%%%%%
\label{sect:transf}

%++++++++++++++++++++++++++++++++++++++++++++++++++++++++++++++++++++++++++++++
\subsection{Newtonian Cooling}
%++++++++++++++++++++++++++++++++++++++++++++++++++++++++++++++++++++++++++++++
\label{sect:NewtonCool}

Newtonian heating/cooling makes the energy tend toward a fixed value $e_{\rm cool}$ on 
a time-scale $t_{\rm cool}$,
\eb{NewtonCool}
  Q_{\rm Newton} = -(e-e_{\rm cool})/t_{\rm cool}.
\en
This can be multiplied by a function of density to smoothly reduce the cooling at either
low or high densities $\rho_{\rm cool}$, e.g.,
\eb{rhocool}
  \frac{(\rho/\rho_{\rm cool})^2}{1+(\rho/\rho_{\rm cool})^2} \quad (\rm low\  cutoff) \\  \nonumber
  \quad {\rm or} \quad \quad\\
  \\ \frac{1}{1+(\rho/\rho_{\rm cool})^2} \quad (\rm high\ cutoff) \nonumber \ .
\en
This is useful in, e.g.,
accretion disk simulations, to allow for a hot corona around the disk.

It is also sometimes used if only a crude chromospheric model is desired. Then
\be{eq:ChromoCool}
    Q_{\rm Chromosphere} = -\rho\left(\frac{\rho}{\rho_{\rm bot}}\right)^{5/3}
	    \frac{e-\langle e\rangle}{t_{\rm cool}}\ ,
\ee
\citet{schrijver:SunStarActivity}, where the cooling time, $t_{\rm cool}$, 
defaults to 0.1\,s.

%++++++++++++++++++++++++++++++++++++++++++++++++++++++++++++++++++++++++++++++
\subsection{Thermal Conduction}
%++++++++++++++++++++++++++++++++++++++++++++++++++++++++++++++++++++++++++++++
\label{sect:SpitzerConduct}

Heating/cooling by thermal conduction is give by
\eb{conduction}
    Q_{\rm cond} = \div\vec{F}_{\rm cond} \ . 
\en
The conductive flux, $\vec{F}_{\rm cond}$, in the presence of a magnetic field,
$\BB$, is
\eb{condflux}
    \vec{F}_{\rm cond} = - \kappa \nabla \ln \left(\rho e\right)
            \cdot\vec{\widehat{B}}\ ,
%           \cdot\BB[\BB/\BB^2]
\en
where $\vec{\widehat{B}}$ is the unit vector in the direction of the magnetic
field.
Spitzer conduction \citep[][p.\ 222]{schrijver:SunStarActivity} for a dilute
plasma is implemented as
\be{eq:SpitzerConduct}
    Q_{\rm Spitzer}  =  \nabla\cdot \vec{F}_{\rm Spitzer} 
       =  -\nabla\cdot\left[\frac{(kT)^{5/2}}{m_{\rm e}^{1/2}e^4}\nabla(kT)\right]
      \simeq -8\times 10^{-7} \nabla\cdot\left[e^{5/2}\nabla e\right]\ ,
\ee
where $k$ is Boltzmann's constant, $m_{\rm e}$ the electron mass, $e$ the
charge of an electron, and the temperature is approximated as
\be{eq:SpitzerTemp}
    T \simeq (\gamma-1)e\ ,
\ee
and, for this particular case, the adiabatic exponent is approximated as a constant, $\gamma=5/3$.

%++++++++++++++++++++++++++++++++++++++++++++++++++++++++++++++++++++++++++++++
\subsection{Optically Thin Cooling}
%++++++++++++++++++++++++++++++++++++++++++++++++++++++++++++++++++++++++++++++
\label{sect:ChromoCool}

If radiation can easily escape from the entire volume, then its only effects is
to cool the plasma, in a way that depends exclusively on local conditions,
\eb{optthin}
    Q=\Lambda(\rho,e)
\en
(where $\Lambda$ is a function of the local density and energy).
This is appropriate in, e..g, the solar corona and the interstellar medium.
For the corona, the formula used is
\be{eq:CoronaCool}
    Q_{\rm Corona} = 1.33\times 10^{-19} T^{-1/2}
                     \frac{\rho^2}{m_{\rm a}^2}
                     e^{-(T/10^4\,{\rm K})^2}
                                    \,{\rm erg}\,{\rm cm}^{-3}\,{\rm s}^{-1}\ ,
\ee
\citet{kahn:CoolingOfSNR}, where the number density of all atoms and ions for
a representative composition is
approximated with $\sum_i n_i = \rho/(2\times 10^{-24}\,{\rm g})$.
The exponential quenching factor, suppresses the cooling for $T<10^4$\,K,
to better approximate the basis for \citet{kahn:CoolingOfSNR}'s original
expression without it.

The interstellar medium is another milieu in which optically thin cooling 
by various metal atoms and ions is appropriate. In this case the cooling
function is more complicated with different physical processes important in
different temperature regions, see \citep{dalgarno:HIregionCool}.

%++++++++++++++++++++++++++++++++++++++++++++++++++++++++++++++++++++++++++++++
\subsection{Radiative Transfer}
%++++++++++++++++++++++++++++++++++++++++++++++++++++++++++++++++++++++++++++++

When neither the optically thin nor the optically thick diffusion limit is appropriate,
the radiative heating/cooling is 
\eb{qrad}
    Q_{\rm rad} = {\rm absorption - emission}
    = \int_{\lambda}\d\lambda \int_{\Om} \d\Omega (\rho\kl\Il - \al\Bl - \sigl\Jl) \,
\en

where $\Il$ is the monochromatic, radiation intensity in a beam $\d\Om$ in a given direction, 
and $\Jl$ is the mean intensity averaged over all directions and $B_\lambda$ is the Planck function.
The total monochromatic opacity, $\rho\kl = \al + \sigl$
is composed of true, thermalizing absorption, $\al$,
and isotropic and coherent (no change of photon wavelength) scattering, $\sigl$
Both absorption- and scattering coefficients are per volume,
but the total extinction, $\kappa$, is per mass, hence the factor of density.

The radiative transfer equation, describing $I_\lambda$, is
\be{eq:RTlmbd}
    \dxdy{I_\lambda}{\ell}  = \sigma_\lambda J_\lambda + a_\lambda B_\lambda
                        - \rho\kappa_\lambda I_\lambda \ ,
\ee
where $\ell$ is the path length along the ray. Divide by the opacity $\rho\kappa_\lambda$ to
get the transfer equation in terms of optical depth, which is the form used to solve it,
\eb{RTtau}
    \dxdy{I_\lambda(\mu,\phi)}{\tau_\lambda(\mu,\phi)}=(1-\epsilon_\lambda)J_\lambda 
       + \epsilon_\lambda B_\lambda - I_\lambda (\mu,\phi) \ ,
\en
where ${\rm d}\tau_\lambda = \rho\kappa_\lambda {\rm d}\ell$ is the optical depth (the distance
in units of the photon mean free path). 
$\mu$ and $\phi$ specify the direction of the ray along which the transfer occurs.
Here,
\be{eq:eps-lmbd}
    \epsilon_\lambda = a_\lambda / \rho\kv\ ,
\ee
is the probability a photon will be destroyed (i.e., thermalized).

So far, the Stagger Code assumes local thermodynamic equilibrium (LTE),
which means $\epsilon_\lambda=1$, but the option of $\epsilon_\lambda\ne 1$
is left open in the radiative transfer tables of Sect.\ \ref{sect:NewAtmTabs}.
A non-LTE treatment was devised by \cite{hayek:Parallel3Dscatter} for the
Stagger Code derived Bifrost code \citep{gudiksen:Bifrost}.

The radiative transfer employed by the Stagger Code has been described in
detail by, e..g, \citet{aake:numsim1}
and \citet{stell-gran3} for earlier simulations. The forward solution is performed on long
characteristics, using a {\em Feautrier technique}
\citep{mihalas:stel-atm}, modified to give more accurate solutions in the
optically deep layers \citep{aake:numsim1}.
The monochromatic solution is prohibitively expensive for MHD
calculations involving thousands of time steps. The Stagger Code
speeds up the solution enormously by reducing the many wavelengths
used in the solution by grouping them into bins of similar opacity
(opacity binning or multi-group method) with only a few bins and
by approximating the 3D angular distribution with a vertical ray
and only a few azimuthal rays.

The angular integration in the azimuthal direction, $\varphi$, is carried out
with a rotating set of equidistant angles in order to minimize directional biases. 
The integration over inclination angle $\mu=\cos\theta$
($\mu=1$ is the vertical ray) is performed with Radau quadrature
\citep{radau:quadrature,abramowitz},
which includes the vertical ray, and $N_\mu-1$ slanted rays. It is implemented for 
$N_\mu=1$--10. A common choice is $N_\mu=3$, $N_\varphi=4$.
%which needs to be calculated anyway, thereby increasing the order of the
%solution by one, at no extra cost, compared to the $(N_\theta-1)$-point
%Gaussian integration that excludes the vertical ray.
%%The Radau method is implemented for $N_\varphi=1$--10.
The quadrature scheme by \citet{carlson:quadrature,bruls:QradOnUnstructGrid} is
also available, for $N_\mu=2$--4.

Radiative transfer is solved in detail for layers where
%%
%% CHECK: is this a bin-wise \min(\tau_i) ???
%% This is the minimum over a horizontal layer, and not the minimum of a
%% horizontal average. And \tau_\max = 200.
$\min(\tau)_{\rm horiz} \le \tau_{\max}=500$, and ignored in layers deeper than that.
In the deeper layers the diffusion approximation is valid, but radiative heating and cooling is insignificant there for cooler stars.
For stars hotter than $T_{\rm eff}\simeq 7\,000\,$K the
radiative heating/cooling in the diffusion approximation
can be included as the divergence of the radiative flux
\eb{raddif}
  Q_{\rm rad}^{\rm diff} = \nabla \cdot \vec{F}_{\rm rad}^{\rm diff} = 
     \frac{16\sigma}{3}\nabla \cdot 
     \left[\frac{T^3}{\rho\kappa_{\rm Ross}}\nabla T\right] \ ,
\en
%{\tiny check signs}
where here $\sigma$ is the Stefan-Boltzmann constant and $\kappa_{\rm Ross}$ is the Rosseland mean opacity.
\be{eq:kapR}
    \frac{1}{\kappa_{\rm Ross}} =
        \frac{\int \kappa_\lambda^{-1}({\rm d}B_\lambda/{\rm d}T) {\rm d}\lambda}
             {\int{\rm d}B_\lambda/{\rm d}T){\rm d}\lambda}\ .
\ee
Being a harmonic mean, the Rosseland opacity is heavily
weighted towards the lowest opacities, typically located in the
optical part of the spectrum. The Planck mean, on the other hand,
is a straight average
\be{eq:kapB}
    \kappa_B = \int_0^\infty\kappa_\lambda B_\lambda{\rm d}\lambda\ .
\ee
%The diffusion approximation will be implemented for deep layers in a future
%update to the Stagger Code.

%There is also an option to completely ignore radiative heating and cooling,
%for atmospheric layers high enough that
%$\max(\tau)_{\rm horiz} \le \tau_{\min}=10^{-6}$.
%

%++++++++++++++++++++++++++++++++++++++++++++++++++++++++++++++++++++++++++++++
\subsection{General Considerations of Opacity Binning}
%++++++++++++++++++++++++++++++++++++++++++++++++++++++++++++++++++++++++++++++

Opacity binning is a method for dramatically reducing the computational cost of
realistic radiative transfer from that of the monochromatic case with of
order $10^5$ wavelength points to a binned case with a mere dozen bins,
$N_{\rm bin}$.
Each bin, labeled $i$, has a weight, $w_i$ of dimension wavelength. 

The bin-wise Planck functions, $B_i$, and bin-wise weights, $w_i$,
are computed as
\be{eq:Bi}
    B_i = \sum_\bin w_\lambda B_\lambda\ ,\qquad w_i=\sum_\bin w_\lambda
\ee
where $\bin$ denotes the subset of wavelength points that belong to bin $i$,
and $w_\lambda$ is the weight of wavelength $\lambda$.
The total Planck function is a simple sum of the bin-wise contributions
\be{eq:Btot}
    B = \sum_i B_i\ ,
\ee
as is also the case for 
the total specific intensity, $I$, angular-averaged intensity, $J$, flux,
$\vec{F}$ and heating, $Q$.
These latter four quantities are all the result of radiative
transfer calculations on the bin-wise $B_i$ and bin-wise $\tau_i$-scales, and
the $\tau_i$ in turn results from the
integrations of the bin-wise opacities, $\kappa_i$.

Different binning schemes are defined by how bin member-ship is decided for each
wavelength, and how bin-wise opacities and source functions are computed.
These are based on a monochromatic reference calculation
(see Sect.\ \ref{sect:meanrhoT}), and the binning schemes aim to reproduce
the cooling and heating of this reference calculation.
The Stagger Code currently offers the choice of two schemes, as detailed in
Sects.\ \ref{sect:OpacBinScaled} and \ref{sect:OpacBinIndiv}.

%++++++++++++++++++++++++++++++++++++++++++++++++++++++++++++++++++++++++++++++
\subsection{Monochromatic Reference Calculation}
%++++++++++++++++++++++++++++++++++++++++++++++++++++++++++++++++++++++++++++++
\label{sect:meanrhoT}

The wavelengths are sorted into bins based on a monochromatic, 1D, reference
radiative transfer calculation, performed on the averaged atmosphere structure of a
simulation. 
The average structure is a horizontal mean of $\langle\rho\rangle$ and
$\langle T\rangle$, averaged in time to reduce p-mode effects.
%over an integer number of periods of the dominant p mode excited in the
%simulation, to minimize the effects of those p modes. To further stabilize the
%averages, we evaluate the temporal average on a reference, horizontally
The temporal averages are performed on a horizontally
averaged column mass scale and interpolated back to
the geometric scale. This is a \emph{pseudo Lagrangian} temporal average, that follows
the mean vertical motion, effectively filtering out first order effects of
the p-modes.
This reference model is stored for later use, and only updated if the
atmosphere of the simulation has drifted, as will be the case during
relaxation of a new simulation. During production runs, however, the
reference model should not be changed.
This makes the tables of binned opacities specific
for a given simulation, and it makes little sense to use it for another
set of atmospheric parameters.

The 1D radiative transfer calculation on the reference model results in 2D
arrays (as a function of wavelength and depth) 
of monochromatic absorption
coefficients, $a_\lambda^{\rm 1D}$, scattering coefficients,
$\sigma_\lambda^{\rm 1D}$, optical depths, $\tau_\lambda^{\rm 1D}$, and
0$^{\rm th}$ and 2$^{\rm nd}$ angular moments, $J_\lambda^{\rm 1D}$ and
$K_\lambda^{\rm 1D}$, respectively, of the specific
intensities $I_\lambda^{\rm 1D}$. 
This radiative transfer calculation is carried out for $N_\mu=4$
\citet{radau:quadrature}-distributed $\mu$-angles,
and since it is in 1D, no explicit $\phi$-integration is needed.

%++++++++++++++++++++++++++++++++++++++++++++++++++++++++++++++++++++++++++++++
\subsection{Sorting Wavelengths into Bins }
%++++++++++++++++++++++++++++++++++++++++++++++++++++++++++++++++++++++++++++++
\label{sect:lambdaSort}

A particular wavelength is ascribed a bin according to its monochromatic \emph{opacity
strength}. This opacity strength, $x$, is based on results of the 1D reference
calculation of Sect.\ \ref{sect:meanrhoT}, and evaluated as either
\be{eq:OpacStrength1}
    \log_{10}x =-\log_{10}\tau_{\rm Ross}^{\rm 1D}(\tau_\lambda^{\rm 1D}=1)\ ,
\ee
the location on the Rosseland optical depth-scale of the monochromatic
photosphere, \emph{or}
\be{eq:OpacStrength2}
    \log_{10}x = \log_{10}(a_\lambda+\sigma_\lambda)(\tau_\lambda^{\rm 1D}=1)\ ,
\ee
the monochromatic extinction in the monochromatic photosphere
at wavelength $\lambda$.

The Stagger Code gives the option between two methods of opacity binning: Binning with scaled opacities
as described by \citet{aake:numsim1} and in Sect.\ \ref{sect:OpacBinScaled},
using Eq.\ (\ref{eq:OpacStrength2}); and Binning with individual opacities, which is based largely
on the work by \citet{skartlien:bin-rad}, but with additional improvements as
described in Sect.\ \ref{sect:OpacBinIndiv} and using
Eq.\ (\ref{eq:OpacStrength1}) together with bin divisions in wavelength.

%------------------------------------------------------------------------------
\subsubsection{Opacity binning --- scaled opacities}
%------------------------------------------------------------------------------
\label{sect:OpacBinScaled}

\citet{aake:numsim1} introduced the method of opacity binning 
as a first attempt at a realistic treatment of line-blanketing in 3D
convection simulations. The complete spectrum is binned according to the
monochromatic opacity strength, $\log_{10}x_\lambda$ of
Eq.\ (\ref{eq:OpacStrength2}), based on an optical depth of the Rosseland mean
across all wavelengths, but excluding line-opacity. The bins are assumed to be
equidistant in $\log_{10}x$
\be{eq:x_i}
    x_i = 10^{i\Delta x}\ ,\quad{\rm with}\quad
            \Delta x=1\quad{\rm and}\quad i=0,1,2,3\ .
\ee
Furthermore, the bin-wise opacities are approximated as simply proportional to
the $i=0$ (continuum bin) opacity
\be{eq:kapi_sc}
    \kappa_i = x_i \kappa_0\ .
\ee
This was done to optimize the convection code for speed, memory and disk-space,
in order to make the original convection simulations \citep{dravins:conv-spec0,
aake:numsim1,stell-gran3} feasible at all. The subsequent increase in
computational power has made it possible to abandon these simplifications,
as described in Sect.\ \ref{sect:OpacBinIndiv}, below.

The continuum bin opacity, $\kappa_0$, is computed as a smooth transition
between a Rosseland mean of the continuum, $\widetilde{\kappa}_0$, below and
an intensity mean, $\langle\kappa\rangle$, above the photosphere,
\be{eq:kap0}
    \kappa_0 = e^{-c_\tau\widetilde{\tau}_0}\langle\kappa\rangle
         + \left(1-e^{-c_\tau\widetilde{\tau}_0}\right)\,\widetilde{\kappa}_0\ ,
\ee
where $\widetilde{\tau}_0$ is the optical depth integrated from
$\widetilde{\kappa}_0$ of Eq.\ (\ref{eq:kapRi}).
This ensures correct asymptotic behavior in the free-streaming regime in the
optically thin case, as well as in the diffusion regime in the optically thick
case.
The bridging constant, $c_\tau$, has a limited effect on the results and
defaults to a value of $c_\tau=2$, which has been tested against
1D reference models for a range of stellar atmosphere parameters.

The bin-wise Rosseland opacity is
\be{eq:kapRi}
    \frac{1}{\widetilde{\kappa}_i} =
        \frac{\sum_\bin\displaystyle\frac{w_\lambda}{a_\lambda+\sigma_\lambda}
                                 \dd{B_\lambda}{T}}
             {\sum_\bin\displaystyle w_\lambda\dd{B_\lambda}{T}}\ .
\ee
This definition is not confined to the domain of the 1D reference calculation,
since we can evaluate $B_i$ for any temperature and therefore for every point
in a 2D table..

For the scaled opacity binning method, only the continuum mean $\widetilde{\kappa}_0$ is used,
resulting in the effective bin-wise opacity, Eq.\ (\ref{eq:kapi_sc}) via
Eq.\ (\ref{eq:kap0}).

The intensity mean, $\langle\kappa\rangle$, is an average over all wavelengths
\be{eq:kapJ}
    \langle\kappa\rangle^{\rm 1D}
    = \frac{\sum_\lambda a_\lambda J_\lambda^{\rm 1D} e^{-\tau_\lambda/2}w_\lambda}
           {\sum_\lambda J_\lambda^{\rm 1D} e^{-\tau_\lambda/2}w_\lambda}\ ,
\ee
where $J_\lambda^{\rm 1D}$ is the angular average of the monochromatic specific
intensity, $I_\lambda$. The exponential factor ensures that the contribution
of $\langle\kappa\rangle$ to $\kappa_0$ is suppressed in the photosphere and
below. $J_\lambda^{\rm 1D}$ is a result of the 1D reference calculation of
Sect.\ \ref{sect:meanrhoT} and hence necessitates an extrapolation away from that
1D stratification to span the 2D table, as detailed in
Sect.\ \ref{sect:xcorr-extrapol}. It is not, however,
$\langle\kappa\rangle^{\rm 1D}$ itself that we extrapolate, but rather a
correction factor 
\be{eq:xcorr1D}
    x_{\rm corr}^{\rm 1D}
        =  \frac{\langle\kappa\rangle^{\rm 1D}}{\widetilde{\kappa}_0}
     e^{-c_\tau\tau_i^{\rm 1D}} + \left(1-e^{-c_\tau\tau_i^{\rm 1D}}\right)\ ,
\ee
from which the total, bin-wise opacity can be computed as
\be{eq:kapi}
    \kappa_i = x_i x_{\rm corr}\widetilde{\kappa}_0\ ,
\ee
and $x_{\rm corr}$ is the extrapolation of $x_{\rm corr}^{\rm 1D}$ away from the 1D
reference model.

The harmonic mean of the bin-wise, scaled, $\widetilde{\kappa}_0$ Rosseland
opacity does not in general converge to the total
$\kappa_{\rm Ross}$ in the limit of many bins, but rather overestimates
$\kappa_{\rm Ross}$ from the photosphere and below. This over estimation can
range from factors 2--5 in the photosphere, depending on stellar atmospheric
parameters.

This version of opacity binning was employed in the work of, e.g.,
\citet{AGS05,bob:SuperGran,trampedach:3Datmgrid,fabbian:FeAbundsWithBfield}.

%------------------------------------------------------------------------------
\subsubsection{Opacity binning --- individual opacities}
%------------------------------------------------------------------------------
\label{sect:OpacBinIndiv}

\begin{figure}
%\centerline{\includegraphics[width=0.5\textwidth]{figs/cbin_plot_sune250.4bin.150.bak5.v02.jpg}\\
%            \includegraphics[width=0.5\textwidth]{figs/cbin_plot_sune250.12bin.150.bak5.v02.jpg}}
\centerline{\includegraphics[width=0.5\textwidth]{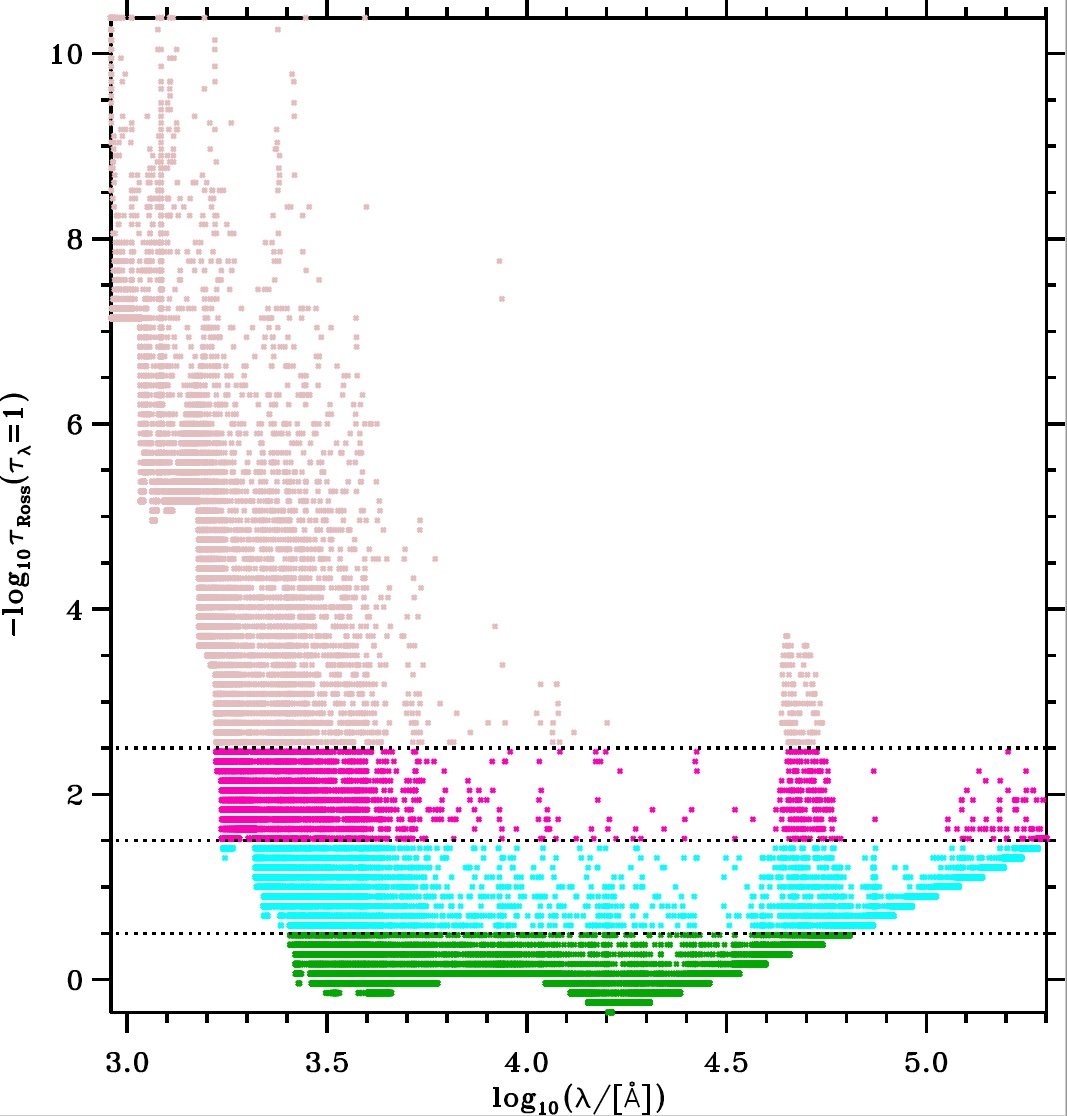}\\
            \includegraphics[width=0.5\textwidth]{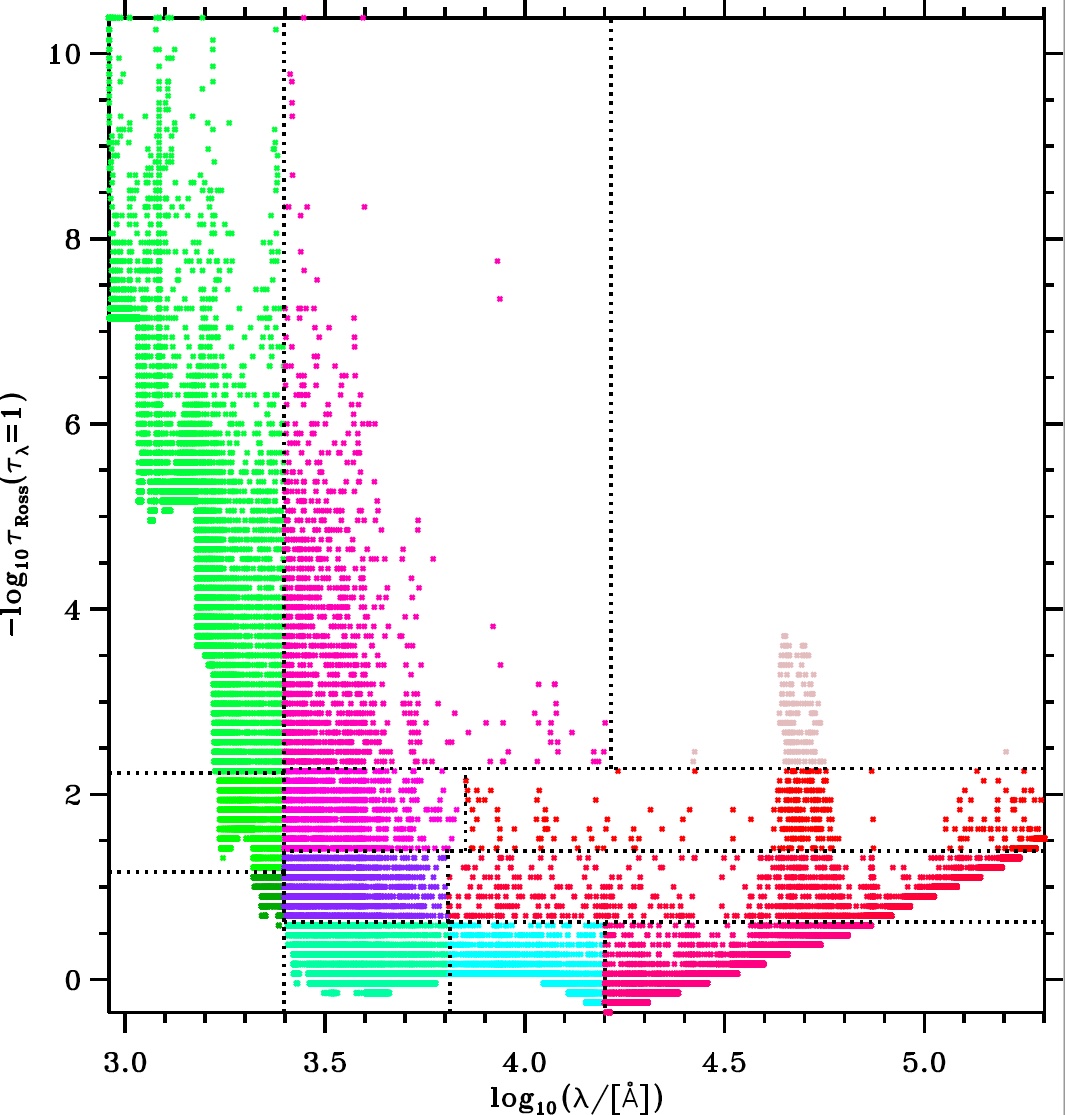}}
\caption{Illustration of the opacity binning of a solar simulation, showing
     opacity strength, Eq.\ (\ref{eq:OpacStrength1}), as function of wavelength
     for the 107\,855 wavelengths included in this monochromatic calculation.
     Bin membership is shown with color, and the borders between the bins are
     also shown with dashed black lines. Various continuum opacity contributions
     are recognized in the lower envelope of these points, including the
     H$^{-}$-bump centered at $\log\lambda=3.9$, the free-free absorption
     red-ward of that, the molecular bands around $\log\lambda=4.2$, the
     hydrogen Lyman edge just short of $\log\lambda=3$, and the absorption
     edges of Al\,I and Ca\,I at $\log\lambda=3.3$, etc.
     Left hand panel: The four equidistant bins of
     Sect.\ \ref{sect:OpacBinScaled}. Right hand panel: A 12 bin
     implementation of the optimized binning of Sect.\ \ref{sect:OpacBinIndiv}.
     \label{fig:cbin_plot_sune250}}
\end{figure}

This method generalizes the bin assignments to also include wavelength.
This was done in recognition of various prominent bound-free absorption edges
in the ultra violet (UV),
which can give rather different behavior of opacities between the blue and red
sides. The molecular bands in the infra red (IR), also exhibit different
behavior than atomic lines in the optical. An example of this generalized
binning scheme is shown in Fig.\ \ref{fig:cbin_plot_sune250} for a solar
simulation.
The case of Eq.\ (\ref{eq:x_i}) can be trivially reproduced in this more
general version.

\begin{figure}
%\centerline{\includegraphics[width=0.5\textwidth]{figs/qrad_cmp_sune250.4bin.150.bak5.v02.pdf}\\
%            \includegraphics[width=0.5\textwidth]{figs/qrad_cmp_sune250.12bin.150.bak5.v02.pdf}}
\centerline{\includegraphics[width=0.5\textwidth]{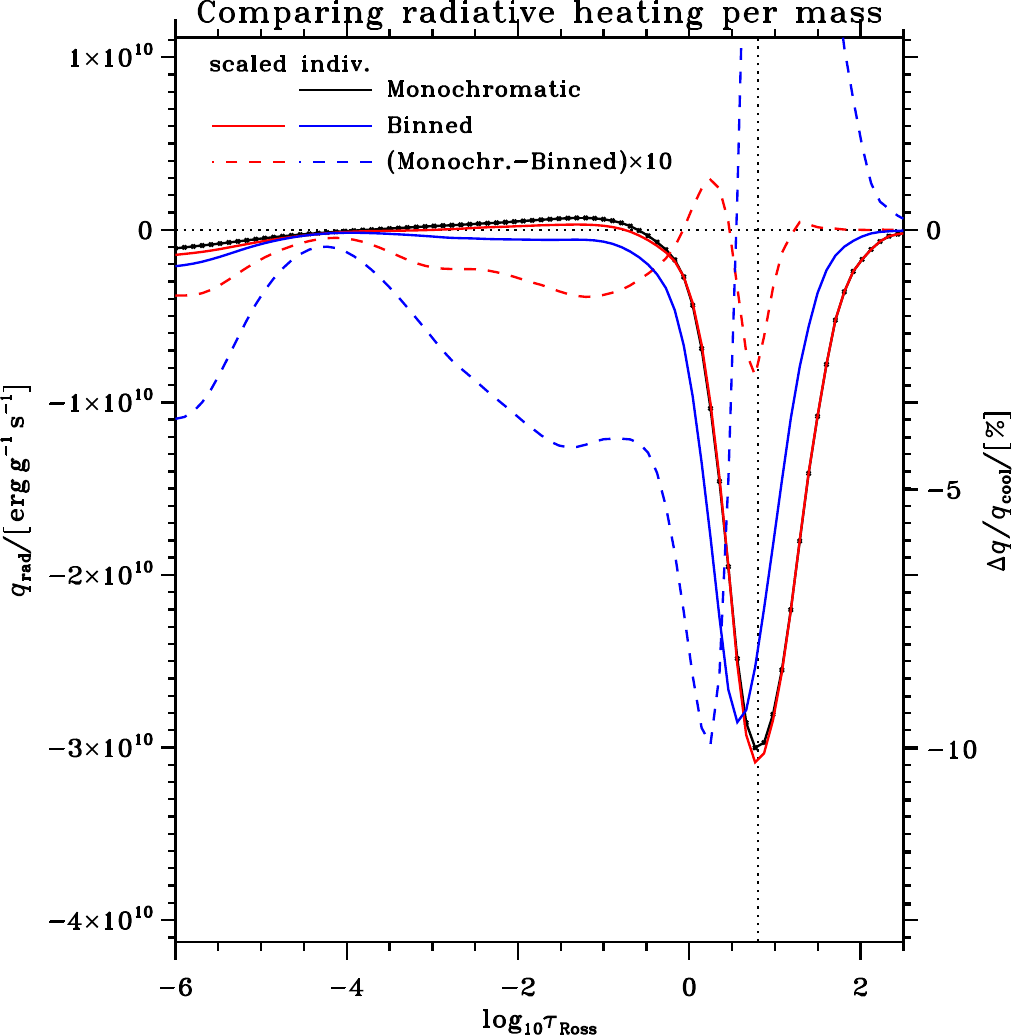}\\
            \includegraphics[width=0.5\textwidth]{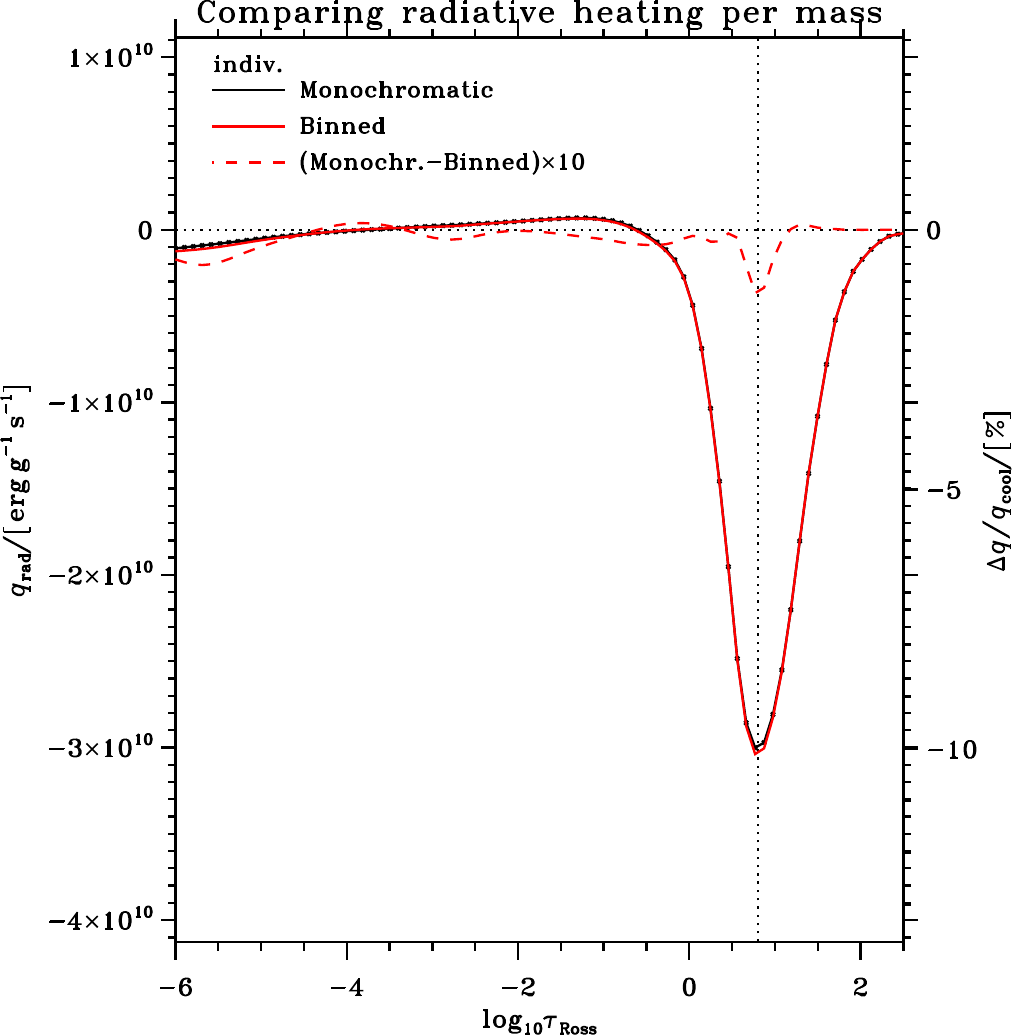}}
\caption{A comparison of the 1D radiative cooling of the full monochromatic
     solution for a solar simulation, and the cooling for
     the binned solutions shown in Fig.\ \ref{fig:cbin_plot_sune250}.
     The difference between the monochromatic and the binned solutions are
     shown, exaggerated by a factor of 10, with the red and blue dashed lines
     on the right-hand scale.
     Left hand panel: The four equidistant bins of
     Sect.\ \ref{sect:OpacBinScaled}, also shown in left hand panel of
     Fig.\ \ref{fig:cbin_plot_sune250}. Scaled opacity calculation shown in
     blue, and individual bin-wise opacities, as in
     Sect.\ \ref{sect:OpacBinIndiv}, in red. With scaled opacities the maximal
     difference w.r.t. the monochromatic calculation, relative to the magnitude
     of the cooling peak (shown with vertical dotted line), is 21.75\%, the
     overall RMS deviation is 6.58\%, and the RMS deviation just in the
     cooling peak is 10.32\%. For individual opacties the deviations
     are 2.81\%, 0.63\% and 1.13\%.
     Right hand panel: A 12 bin
     implementation of the optimized binning of Sect.\ \ref{sect:OpacBinIndiv},
     also shown in right-hand panel of Fig.\ \ref{fig:cbin_plot_sune250},
     giving deviations of just 1.22\%, 0.26\% and 0.45\%, respectively.
     \label{fig:qrad_cmp_sune250}}
\end{figure}

The binning can be optimized, i.e., moving the borders between the bins in
order for the radiative cooling of the binned solution, to reproduce that of
the full monochromatic solution.
In Fig.\ \ref{fig:qrad_cmp_sune250} we show an optimized solution, corresponding
to the binning shown in Fig.\ \ref{fig:cbin_plot_sune250}.
The cooling peak, seen in Fig.\ \ref{fig:qrad_cmp_sune250}, shapes the
photospheric transition. It is crucial to reproduce this feature accurately in
order for the simulation to equilibrate to the correct $T_{\rm eff}$ and in
order to produce realistic synthetic spectra from the simulation.

For the intensity mean this scheme uses the bin-wise version of Eq.\ (\ref{eq:kapJ})
\be{eq:kapJi}
    \langle\kappa\rangle_i
    = \frac{\sum_\bin a_\lambda J_\lambda e^{-\tau_\lambda/2}w_\lambda}
           {\sum_\bin J_\lambda e^{-\tau_\lambda/2}w_\lambda}\ .
\ee
This opacity does not include scattering, unless the table is being made for
non-LTE source functions. For that case scattering is simply added to the
absorption, above.

As for $\langle\kappa\rangle$, the bin-wise opacity is also confined to the
domain of the 1D reference calculation, and an extrapolation is needed in order
to span the extent of the table.

Analogous to the case of scaled opacity binning, the quantity we extrapolate
is the correction factor, $x_{{\rm corr,}i}^{\rm 1D}$, that turns the bin-wise
Rosseland mean, $\widetilde{\kappa}_i$ into the effective, bin-wise opacity,
$\kappa_i$,
\be{eq:xcorr1Di}
    x_{{\rm corr,}i}^{1D}
        =  \frac{\langle\kappa\rangle_i^{\rm 1D}}{\widetilde{\kappa}_i}
     e^{-c_\tau\tau_i^{\rm 1D}} + \left(1-e^{-c_\tau\tau_i^{\rm 1D}}\right)\ .
\ee
The extrapolation of this to span the 2D plane of the table is described
in Sect.\ \ref{sect:xcorr-extrapol}.
The final bin-wise opacity is
\be{eq:kapi_indiv}
    \kappa_i = x_{{\rm corr,}i}\kappa_{{\rm Ross},i}\ .
\ee

In contrast to the binning with scaled opacities, this version of $\kappa_i$
converges properly to the Rosseland mean in the limit of
$N_{\rm bin}\rightarrow\infty$. This ensures the photosphere emerges in the
correct location, whereas with scaled opacities it is too high up in the
atmosphere, i.e., at lower pressure/density.
This version with individual opacities was employed by, e.g.,
\citet{AGSS2009,remo:3Dscatter,magic:stagger-grid},
and is a major cause for the differences
between the solar abundances of \citet{AGS05} and \citet{AGSS2009}.

\citet{collet:3DabundsHaloGiant2} performed a thorough analysis of
different lay-outs and numbers of bins, concluding that even 48 bins
does not result in radiative heating that is significantly closer to
the monochromatic calculation, compared to the standard 12 bins.

%If the source functions are chosen to be non-LTE, then a $J$-averaged, bin-wise,
%photon destruction probability is also added to the table
%\be{eq:epsJi}
%    \epsilon_{J,i} =
%        \frac{\sum_\bin w_\lambda a_\lambda                 J_\lambda}
%             {\sum_\bin w_\lambda(a_\lambda+\sigma_\lambda) J_\lambda}\ ,
%\ee
%and the tabulated Planckian source function is adjusted to
%\be{eq:epsBiBi}
%    \epsilon_{B,i} B_i =
%        \frac{\sum_\bin w_\lambda a_\lambda                 B_\lambda}
%             {\sum_\bin w_\lambda(a_\lambda+\sigma_\lambda) J_\lambda}\ ,
%        \frac{\sum_\bin w_\lambda J_\lambda}{\sum_\bin w_\lambda B_\lambda}
%            \times B_i\ .
%\ee
%With this, the source function for bin $i$ is
%\be{eq:Si}
%    S_i = (1-\epsilon_{J,i}) J_i + \epsilon_{B,i} B_i\ ,
%\ee
%where $J_i$ is the result of the radiative transfer calculation, clearly
%turning this into an iterative process.
%
%------------------------------------------------------------------------------
\subsubsection{Extrapolation of the $x_{\rm corr}^{\rm 1D}$-factor}
%------------------------------------------------------------------------------
\label{sect:xcorr-extrapol}

We need to extrapolate the factor $x_{\rm corr}^{\rm 1D}$ from its 1D stratification
and out to the 2D, ($\rho, e$)-plane of a table, so that it can affect the whole simulation
in a plausible way. In Fig.\ \ref{fig:isotau} we show the distribution of
$\rho$ and $T$ for 17 horizontal planes in a convection simulation. It is
clear that the distribution of points is far from uniform, and we can fit them
fairly well with
\bea{eq:rhoextr}
    \ln\rho = \ln\rho_j - a_1(\log T - \log T_j)
      &+& a_2e^{\displaystyle{-(\log T_j-a_3-a_4\log\tau_j)^2/a_5}}\mideqn
      &-& a_2e^{\displaystyle{-(\log T  -a_3-a_4\log\tau_j)^2/a_5}}\ ,\nonumber
\eea
which connects a point ($T$, $\rho$) in the table, to a point $j$ in the
reference stratification, via the fitted black lines of Fig.\ \ref{fig:isotau}
and the coefficients $a_1$--$a_5$.
\mfigur{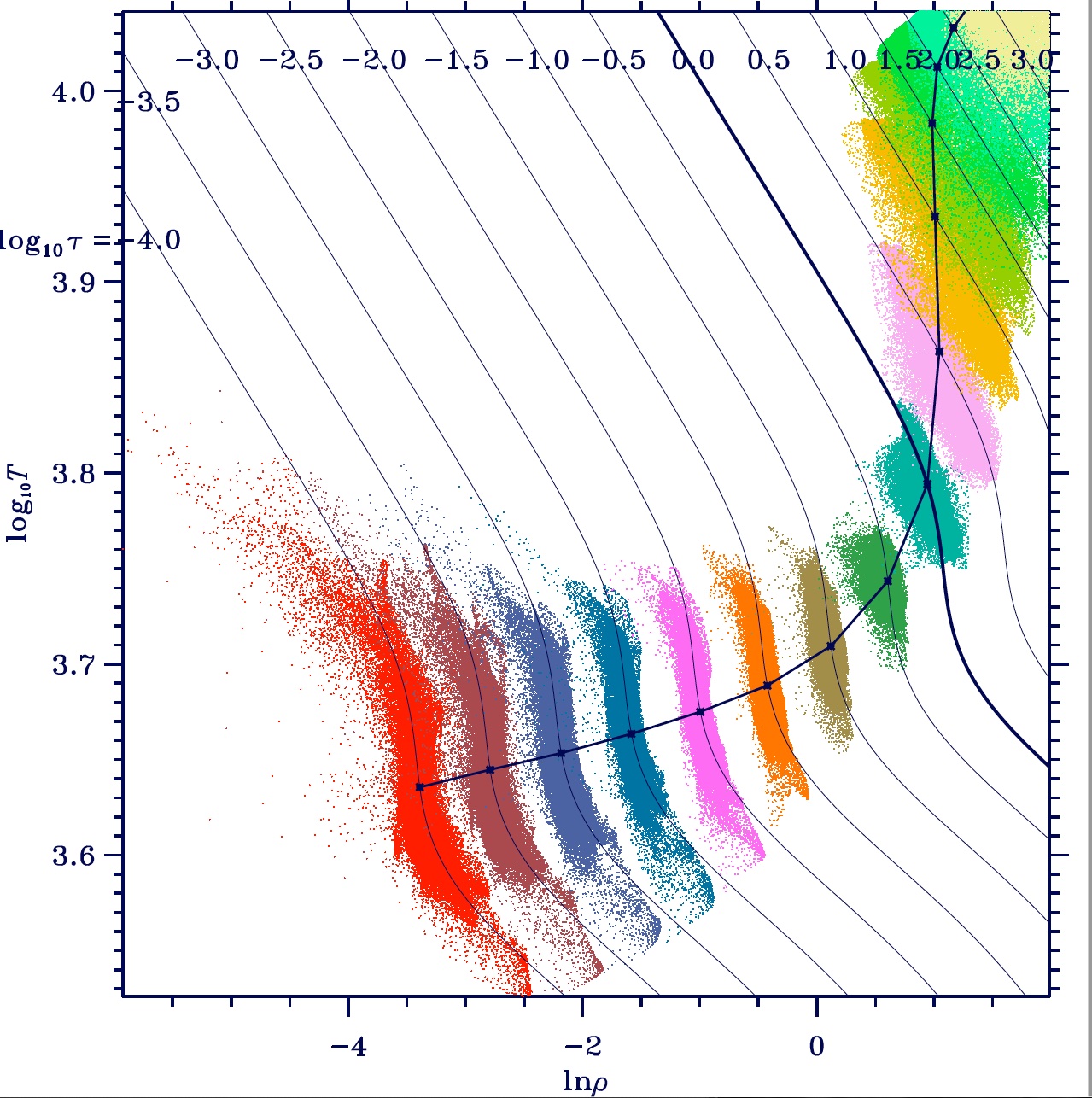}{12cm}
    {Extrapolation of the 1D mean stratification to the 2D table. The points
     are a random subset of the simulations cubes upper 17 layers, and the
     lines are the fits described in the text, with coefficients $a_1$--$a_5$.
     This example is for the simulation {\tt t47g22m+00} at $t=10^6$\,s.}

%------------------------------------------------------------------------------
\subsubsection{Vertical scale for radiative transfer}
%------------------------------------------------------------------------------
\label{sect:yrad}

The radiative transfer calculation can often benefit greatly from a
vertical scale that is explicitly optimized for that purpose. There is
therefore an option for stellar atmosphere calculations, to compute such
a vertical scale for the transfer calculations that better resolves the
radiation field
\citep{collet:3DabundsHaloGiant2}. The radiative heating/cooling is evaluated
on this dynamically optimized radiative grid, and then interpolated back to the
hydrodynamic grid of the simulation for addition to the energy conservation,
Eq.\ (\ref{eq:ien}).
% ##### CHECK #####

%%%%%%%%%%%%%%%%%%%%%%%%%%%%%%%%%%%%%%%%%%%%%%%%%%%%%%%%%%%%%%%%%%%%%%%%%%%%%%%
\section{Equation of State and Opacity}
%%%%%%%%%%%%%%%%%%%%%%%%%%%%%%%%%%%%%%%%%%%%%%%%%%%%%%%%%%%%%%%%%%%%%%%%%%%%%%%
\label{sect:Opac+EOS}

Simple opacities, such as $H^-$, can be calculated analytically as a function of 
$\rho$ and $T$ as needed. Complete stellar atmosphere opacities, however, are
too complex for that and instead auxiliary codes computes tables that are
used by the Stagger Code for looking up opacities and EOS variables.

The Stagger Code also has some simplified options for equations of state (EOS)
with only few elements, that are computationally cheap to solve.

A couple of more realistic sets of atomic physics, relevant for stellar
atmospheres in particular, are elaborated on below.

%------------------------------------------------------------------------------
\subsection{Original Tables for Stellar Atmospheres}
%------------------------------------------------------------------------------
\label{sect:OrigAtmTabs}

The original version of atomic physics tables for stellar atmosphere simulations
consists of a table in $\ln\rho$ and $\theta=5039.78/T$, that gets inverted
into tables in $\ln\rho$ and $\rho e$ as needed by the Stagger Code. Both tables
contain gas pressure, $\ln p$, continuum bin opacity, $\ln\kappa_0$, the
$N_{\rm bin}$ source functions, $\ln B_i$ and the complement to the independent
temperature variable, $\rho e$ and $T$, respectively. Each variable in the
final table is also accompanied by numerical $\d\ln\rho$- and
$\d(\rho e)$-derivatives, to enable bicubic interpolation in the tables by the
Stagger Code. Only the final table is used by the Stagger Code, but the
initial table can be useful in post-processing.

These tables are specific to a particular simulation, and have irregular shapes
that just covers the extent of the simulation, with a small buffer. If the
simulations strays outside this table it is automatically expanded where
needed. The $\theta$-resolution of each iso-chore in the table is determined
from a requirement that $\theta$-interpolations in any of the table variables
do not deviate more than $\epsilon$ from a direct calculation. This parameter
is typically set to $\epsilon=10^{-5}$.

This version assumes the scaled opacity binning of
Sect.\ \ref{sect:OpacBinScaled}, with LTE source functions, $S_i=B_i$.
Both the opacities and thermodynamics are based on the atomic physics of the
original MARCS stellar atmosphere package \cite{b.gus,Gustafsson1975}.

%..............................................................................
\paragraph{EOS}

This EOS is based on an interpretation of the \citet{Deb:Huck} theory of
electrolytes, which lowers the ionization potential of all species according
to the formulation by \citet{griem:PlasmaSpec}. The partition functions of
atoms and ions include an asymptotic part that includes some accounting for
the above mentioned potential-lowering \citep{dejager:part-func}.
This alters the populations of the various species, but there are no direct
effects on the internal energy and the pressure, making this EOS
thermodynamically inconsistent (e.g., values of the adiabatic gradient will
depend on which valid expression is used).

These tables are computed for a 16 element mixture of H, He, C, N, O, Ne, Na Mg,
Al, Si, S, K, Ca, Cr, Fe and Ni. Molecules
of the most abundant elements: H$_2$, H$_2^+$, H$_2$O, OH, CH, CO, CN, C$_2$,
O$_2$, N$_2$, NH and NO are included by means of the equilibrium constants by
\citet{tsuji:molecules}.

%..............................................................................
\paragraph{Opacities}

The bound-free (bf) opacities are listed in input tables as function of wavelength,
and also as function of $T$ if the source has excited states. The included
sources are H$_2^+$ bf and free-free (ff), H$_2^-$ ff, C\,I, Mg\,I, Al\,I and
Si\,I bf, H$^-$ bf and ff as well as ff and bf of the first 15 states of
hydrogen. Scattering by free electrons and Rayleigh scattering by H and H$_2$
is included via analytical expressions.

Spectral lines are included in the statistical approach of opacity distribution
functions (ODF), with 42 distribution functions spanning 3\,000--7\,200{\AA},
each 100{\AA} wide and sampled at four points. These ODFs are tabulated for nine
temperatures, $T=$3--9\,kK and 15 electron pressures, $\log p_{\rm e}=-3$--4.
They involve about 50\,000 atomic lines
on more than that of molecular lines of MgH, CH, OH, NH and CN. For
$\lambda = $7\,200--125\,000{\AA}, a separate set of 23 ODFs
are used for lines of CN and CO molecules, only, provided for six temperatures
between 2 and 7\,kK.

The opacity in these tables, is the continuum bin opacity, $\kappa_0$,
of Eq.\ \ref{eq:kapi}.

%------------------------------------------------------------------------------
\subsection{New Tables for Stellar Atmospheres}
%------------------------------------------------------------------------------
\label{sect:NewAtmTabs}

This version consists of a rectangular table in $\ln\rho$ and $\ln T$ containing quantities
needed for radiative transfer (source functions and opacities for each
bin or wavelength), and a rectangular table in $\ln\rho$ and $\ln(\rho e)$ 
of the EOS, as well as some opacities for post-processing.

Tables have been computed for a variety of scaled solar abundances of the 17 elements:
H, He, C, N, O, Ne, Na, Mg, Al, Si, S, Ar, K, Ca, Cr, Fe and Ni, by
\citet{trampedach:T-tau}. The first simulation to be computed with these tables
is the solar atmosphere that the \citet{AGSS2009} abundances are based on.
Tables have also been computed for the \citet{AG89}, \citet{GS98} and \citet{AGS05} mixtures.

%..............................................................................
\paragraph{EOS}

The EOS table contains {\tt Ntbvar}$\ge 6$ rectangular arrays of
gas pressure $\ln P_{\rm gas}$, Rosseland mean opacity, $\ln\kappa_{\rm Ross}$,
temperature, $\ln T$, electron number density, $n_e$,
Planck mean opacity, $\ln\kappa_{\rm Planck}$ and the
monochromatic opacity at 5\,000\,{\AA}, $\ln\kappa_{5000}$. Other variables
can be added at the end.

These variables are functions of a regular grid of density, $\ln\rho$, and
internal energy per mass, $\ln e$. The extent and resolution of the grid is given
in the header of the table, together with the number of elements and their
abundances used in the calculation, the option flags that were set for the
calculation and the revisions of the codes used in the calculation, as well as
date- and time-stamps, uniquely identifying the table.

The thermodynamics is supplied by the Mihalas-Hummer-D{\"a}ppen (MHD) EOS
\citep{mhd1,mhd2}, which is a model for the Helmholtz free energy, from which
all thermodynamic quantities can be evaluated consistently via the Maxwell
relations \citep[e.g.,][]{mhd3}. Non-ideal effects (a.k.a. pressure ionization)
is physically and smoothly accomplished through the so-called occupation
probabilities assigned for each excited state, and based on a formulation
of perturbations of bound levels from the fluctuating electric field of
passing ions. This gives a both physical and differentiable truncation of the
otherwise diverging partition functions.
All ionization stages and all excited levels are explicitly included, but only two molecules, H$_2$ and H$^+_2$, are treated in the MHD EOS.
The independent variables are $\log_{10}T$ and
$\log_{10}\rho$, which needs to be inverted to $\ln\rho$ and $\ln e$ for the
Stagger simulations. This is done by exploiting the
analytical partial derivatives of the MHD EOS, that allows for an accurate
and physical third-order interpolation to a new regular $\ln e$-grid, and with
analytical and consistent $\ln e$-derivatives.

Convective stability, particularly in deep layers where the velocities and
entropy fluctuations are very small, are sensitive to errors in the EOS, so a
bicubic interpolation in the tables is used.
Each of the {\tt Ntbvar} table entries, $f$, are accompanied by their partial
derivatives, $\dxdy{f}{\ln e}\Delta\ln e$ and $\dxdy{f}{\ln\rho}\Delta\ln\rho$,
in part to facilitate this higher order interpolation, but also to provide high
quality, analytical thermodynamic derivatives for a consistent computation of any needed
thermodynamic quantity.

The opacity calculations do not provide derivatives so they have to be evaluated
numerically. We do so by computing opacities at two guide points at
$\pm\Delta\ln e/3$ from the table points (and inverted to $\log_{10}T$ with the
same function as above), giving second-order derivatives, and similarly for
$\pm\Delta\ln\rho /3$, providing smooth and accurate interpolations in the tables.

The Rosseland opacity in this table is used for optical depth calculations
in the Stagger Code, but the other opacities are merely there for convenience.

The EOS tables are \emph{universal} in the sense that they are
applicable to any simulation with the given composition. Therefore we choose a
large extent of these table, to encompass most stellar atmospheres, with a
default of $\log_{10}\rho=[-14;0]$, $\log_{10}e=[11;14]$, and a resolution of
$N_e=300$ and $N_\rho=57$. The latter is dictated by the original MHD EOS
calculation, which typically use $\Delta\rho=0.25$.

%..............................................................................
\paragraph{Opacities}

The tables are rectangular and regular in the independent variables $\ln\rho$ and $\ln T$.
They contain the relative opacity, $x_i=\kappa_i/\kappa_{\rm Ross}$ of bin $i$,
the Planck part of the source function, $\ln(\epsilon_{B,i} B_i)$, with
$B_i$ from Eq.\ (\ref{eq:Bi}), and the photon
destruction probability, $\ln(\epsilon_{J,i})$ Eq.\ (\ref{eq:eps-lmbd}), of each of
the $N_{\rm bin}$ bins. The tabulation of $\epsilon_{B,i}$ allows for running simulations including scattering, although this is not implemented in the Stagger Code, yet.
The LTE case is recovered when all
$\epsilon_{B,i}=1$ ($\sigma_i = 0$). The three last entries are occupied by the opacities
$\rho\kappa_{\rm Ross}$, Eq.\ (\ref{eq:kapR}), $\rho\kappa_B$, Eq.\ (\ref{eq:kapB}), and
the 5\,000\,{\AA} monochromatic opacity, $\rho\kappa_{5000}$.

Current tables have a resolution of $N_T=221$ and $N_\rho=157$, with an extent
of $\log_{10}\rho=[-14;-1.25]$ and $\log_{10}T=[3;5]$.
Since the evaluation of radiative transfer in an atmosphere simulation takes up
a significant part of the computational cost, and interpolation in the opacity
table is a significant part of that, it is most efficient to increase the
resolution (compared to that of the EOS table) so only linear
interpolations in the table are needed (and no derivatives are stored in these
tables). The simulations are not critically sensitive to small interpolation
errors in opacity. 

The extent and resolution of a table is stored in its header, as are the
elemental abundances, the option flags for the
calculation, the revisions of the codes used in the calculation, as well as
date- and time-stamps, uniquely identifying the table.

Tables are calculated prior to the simulation runs. Line-opacities are
currently supplied by the MARCS stellar atmosphere modeling group
\citet{MARCS-2} in the form of opacity sampling (OS) tables on a 108\,405-point
wavelength
grid covering from 910\,{\AA} to 20\,$\mu$ equidistantly in $\log\lambda$. The
58 included sources of continuum opacities are added in as detailed in
\citet{hayek:Parallel3Dscatter}, and includes molecular sources and collision
induced absorption (CiA). The photoionization cross sections of metals have
been sourced from the NORAD database by \citet{nahar:HiFiOpacs}, based on
close-coupling (CC) calculations, abandoning the usual $LS$-coupling approximation.

Since we want to include molecules in the opacities, and only H$_2$ and H$_2^+$ are included in the MHD EOS, we revert to the
original EOS, from Sect.\ \ref{sect:OrigAtmTabs}, for computing
the population of ionization and dissociation stages, based on the electron pressure from the MHD EOS. This introduces a slight inconsistency in the atomic physics.

%------------------------------------------------------------------------------
\subsection{Free EOS and Blue opacities}
%------------------------------------------------------------------------------
\label{sect:Free+BlueAtmTabs}

Recently, \citet{zhou:FreeEOSStaggerCode} implemented the
Free EOS by \citet{irwin:FreeEOS}, which is constructed to
emulate various commonly used EOS, including the MHD EOS,
described in Sect.\ \ref{sect:NewAtmTabs}. The Free EOS is
simpler and faster than the MHD EOS, and the code is publicly
available, enabling calculation of tables for custom abundances.
It also includes five more elements than the custom computed
MHD EOS tables of Sect.\ \ref{sect:NewAtmTabs}, but still only
the two molecules H$_2$ and H$_2^+$, and no H$^-$ anion. This necessitates
a transition to an atmospheric EOS for $T<10\,000$\,K that \emph{does}
include a comprehensive set of molecules, as well as H$^-$.

Opacities and atmospheric EOS are provided by Blue, 
\citet{amarsi:NLTE-Fe-ForGALAH,amarsi:NLTE-Balmerlines}, which is a
major update to the atomic physics package employed by the MARCS
1D stellar atmosphere code \citep{MARCS-2}. Blue includes opacities from
21 diatomic molecules and H$_2$O, and mostly the same continuum 
opacities as described in Sect.\ \ref{sect:NewAtmTabs}. Atomic and
ionic lines are an improved implementation of the same line data also
used for both MARCS and the Stagger Code physics of
Sect.\ \ref{sect:NewAtmTabs}.

This package is a stand-alone code that produces tables in the
Stagger table format, and is not part of the Stagger repository.
Please address inquiries about this package to its authors.

%%%%%%%%%%%%%%%%%%%%%%%%%%%%%%%%%%%%%%%%%%%%%%%%%%%%%%%%%%%%%%%%%%%%%%%%%%%%%%%
\section{Forcing}
%%%%%%%%%%%%%%%%%%%%%%%%%%%%%%%%%%%%%%%%%%%%%%%%%%%%%%%%%%%%%%%%%%%%%%%%%%%%%%%
\label{sect:Forcing}

%{
Forcing physics is specified by a separate subroutine. Many
routines already exist in the directory FORCING:
\begin{center}
Table 1: Forcing Routines\\
\begin{tabular}{|l|l|}
\hline
\hline
{\bf physics} & {\bf possible use} \\
\hline
no forcing & testing \\
\hline
constant gravity & stellar atmospheres \\
\hline
self gravity & disks \\
\hline
random forcing & turbulence \\
\hline
pattern forcing & \\
\hline
explosion & \\
\hline
piston & waves \\
\hline
\end{tabular}
\end{center}
Additional routines can be written for other specific situations.
%}

%%%%%%%%%%%%%%%%%%%%%%%%%%%%%%%%%%%%%%%%%%%%%%%%%%%%%%%%%%%%%%%%%%%%%%%%%%%%%%%
\section{Boundary Conditions}
%%%%%%%%%%%%%%%%%%%%%%%%%%%%%%%%%%%%%%%%%%%%%%%%%%%%%%%%%%%%%%%%%%%%%%%%%%%%%%%
\label{sect:Bdry}

%{
The Stagger Code has been used for many different types of simulations each with its own set of
boundary conditions.  Variable values extending beyond the boundary are needed by the derivative
and interpolation routines.  The x and z boundaries are always periodic.  The y-boundary can be 
periodic, closed, open in various ways, or forced.  Boundary conditions, when needed, are 
imposed in the y-direction by specifying variable values in ghost zones at each boundary.  
The number of ghost zones is determined by the number of derivative and interpolation operations 
on a variable that
are needed for each time step, as each operation destroys data in the outermost valid ghost zone.
If the domain is stratified, the stratification is taken to be in the y-direction.  
The MHD equations with the sixth order derivatives and fifth order interpolations require five ghost zones.
In this case, the actual boundary point is 5 zones in from the edge of the numerical box.
The so-called \emph{lower} boundary is located at small indices in the $y$-direction, $i_y=${\tt lb}, with ghost zones $i_y=$[1; {\tt lb}], and is normally at the top of the simulation domain, as determines by the direction of gravity, if applied.

%++++++++++++++++++++++++++++++++++++++++++++++++++++++++++++++++++++++++++++++
\subsection{Periodic}
%++++++++++++++++++++++++++++++++++++++++++++++++++++++++++++++++++++++++++++++

If no boundary conditions are imposed, then all three directions are periodic
and there are no ghost zones.
The needed variable values for derivatives and interpolations that extend beyond
a boundary are taken from the corresponding locations on the opposite boundary.
This is the case, for example, for simulations of turbulence, star formation, 
and disks.

%++++++++++++++++++++++++++++++++++++++++++++++++++++++++++++++++++++++++++++++
\subsection{Closed}
%++++++++++++++++++++++++++++++++++++++++++++++++++++++++++++++++++++++++++++++

To impose a closed boundary:

The density or its log is extrapolated with a fifth order extrapolation.  The velocity components
parallel to the boundary are symmetric about the boundary point and the normal component of the
velocity is anti-symmetric about the boundary.  The energy per unit mass is extrapolated and the energy
per unit volume is obtained by multiplying by the extrapolated density.  The parallel components of
the magnetic field are anti-symmetric about the boundary point.

%++++++++++++++++++++++++++++++++++++++++++++++++++++++++++++++++++++++++++++++
\subsection{Rigid}
%++++++++++++++++++++++++++++++++++++++++++++++++++++++++++++++++++++++++++++++

For a rigid boundary: 

The velocity in the ghost zones is taken to be the velocity at the boundary point
and the time derivatives of all variables in the ghost zones are set to zero.

%++++++++++++++++++++++++++++++++++++++++++++++++++++++++++++++++++++++++++++++
\subsection{Stellar Atmosphere}
%++++++++++++++++++++++++++++++++++++++++++++++++++++++++++++++++++++++++++++++

Stellar atmospheres are highly stratified.  The direction of gravity is taken to be downward in
the y-direction of increasing index.  In this case boundary conditions, via ghost zones, must be
imposed in the y-direction.  Time derivatives of all the variables are set to
zero in the ghost zones.

%++++++++++++++++++++++++++++++++++++++++++++++++++++++++++++++++++++++++++++++
\subsubsection{Interior Boundary Inside the Convection Zone}
%++++++++++++++++++++++++++++++++++++++++++++++++++++++++++++++++++++++++++++++

Density and energy are needed to relate the momenta and velocities and determine the pressure in 
the ghost zones.  Their ghost zone values are determined by linear extrapolation of the log density 
and the energy per unit mass.
The momenta are calculated to be symmetric about the boundary point and velocities are calculated from the
momenta and density.  

In the inflows, the density and energy are specified ($e_{\rm bot}$
and $\rho_{\rm bot}$) 
in order to fix the pressure and entropy,
\eab{rein}
    \ddt{\rho} &=& (\rho_{\rm bot}-\rho)/t_{\rm bdry} \\ \nonumber
    \ddt{e}    &=& (e_{\rm bot}-e)/t_{\rm bdry}\ ,  \nonumber
\ean
where $t_{\rm bdry}$ is a relaxation time scale for the boundary.
Fluctuations in the horizontal momenta are similarly damped. 

Outflows should be as unconstrained as possible. However, horizontal pressure fluctuations need
to be minimized for stability, without affecting their entropy.  This is done as follows:
Entropy perturbations per unit mass have 
\eab{entropy}
    T \ddt{S} & = & \ddt{E} - \frac{P}{\rho}\ddt{\ln \rho} \\ \nonumber
    \rho T \ddt{S} & = & \ddt{e} - \left(E +\frac{P}{\rho}\right)\ddt{\rho},
\ean
where $E$ is the internal energy per unit mass and $e$ the internal energy per unit volume.
Thus, perturbations at constant entropy must have
\eb{constS}
    \left(\frac{\partial e}{\partial\rho}\right)_S = \left(E +\frac{P}{\rho}\right)\ = H \ ,
\en
the enthalpy per unit mass.  We also want to cancel the horizontal fluctuations in pressure in the
outflows,
\eab{dPdt}
  \rho \ddt{\ln P} & = & \left[\left(\frac{\partial\ln P}{\partial\ln\rho}\right)_E \right.
    -\left.E\left(\frac{\partial\ln P}{\partial E}\right)_{\rho}\right]\ddt{\rho} +
   \left(\frac{\partial\ln P}{\partial E}\right)_{\rho}\ddt{e} \\ \nonumber
    & = & \left[\left(\frac{\partial\ln P}{\partial\ln\rho}\right)_E\right.
    + \left.\frac{P}{\rho}\left(\frac{\partial\ln P}{\partial E}\right)_{\rho}\right]\ddt{\rho}
\ean
(where $\left({\partial\ln P}/{\partial\ln\rho}\right)_E$ and 
$\left({\partial\ln P}/{\partial E}\right)_{\rho}$ can be obtained from the equation of state).
Hence, it is necessary to cancel the pressure derivative at the boundary and replace it with
\eab{constP}
 {- \rho \ddt{\ln P} + \rho \left(\frac{\ln P_{\rm bot}-\ln P}{t_{\rm
bdry}}\right)} \hspace{2in} \\ \nonumber
   =\left[\left(\frac{\partial\ln P}{\partial\ln\rho}\right)_E\right.
    -\left.E\left(\frac{\partial\ln P}{\partial E}\right)_{\rho}\right]
    \left[-\ddt{\rho} + \left(\frac{\rho_{\rm bot}-\rho}{t_{\rm
bdry}}\right)\right] 
    +\left(\frac{\partial\ln P}{\partial E}\right)_{\rho}
    \left[-\ddt{e} + \left(\frac{e_{\rm bot}-e}{t_{\rm bdry}}\right)\right] \ .
\ean
The boundary value of the derivative $\left(\partial\rho/\partial t\right)_{\rm add}$ 
that needs to be added to the existing derivative at the boundary can be obtained 
using Eq.  \ref{eq:constS},
\eab{addderiv}
   \left(\ddt{\rho}\right)_{\rm add} & = & 
      a\left[\left(\frac{\rho_{\rm bot}-\rho}{t_{\rm bdry}}\right) -\ddt{\rho}\right]
     +b\left[\left(\frac{e_{\rm bot}-e}{t_{\rm bdry}}\right) -\ddt{e}\right]  \\
   \left(\ddt{e}\right)_{\rm add} & = &
       \left(E+\frac{P}{\rho}\right)\left(\ddt{\rho}\right)_{\rm add} \ , \\ \nonumber
   a & = & \left[\left(\frac{\partial\ln P}{\partial\ln\rho}\right)_E\right.
    -\left. E\left(\frac{\partial\ln P}{\partial E}\right)_{\rho}\right]/
      \left[\left(\frac{\partial\ln P}{\partial\ln\rho}\right)_E\right.
       \left.+ \frac{P}{\rho}\left(\frac{\partial\ln P}{\partial E}\right)_{\rho}\right]\\ \nonumber
    b & = & {\left(\frac{\partial\ln P}{\partial E}\right)_{\rho}}/
      \left[\left(\frac{\partial\ln P}{\partial\ln\rho}\right)_E\right.
      \left.+ \frac{P}{\rho}\left(\frac{\partial\ln P}{\partial E}\right)_{\rho}\right] \ .  \nonumber
\ean
These changes preserve the boundary entropy of the fluid.

The magnetic field time
derivative is the curl of the electric field, which preserves $\div \BB = 0$. However, we often want to
impose a boundary condition on the magnetic field.  We therefor store the original magnetic field
values in the ghost zones and apply our boundary conditions to the magnetic field.  First extrapolate
all three magnetic field components into the ghost zone.  Then set the horizontal magnetic field
components to their prescribed boundary values $B_{\rm bdry}$ 
(Bx0,By0,Bz0 see table \ref{sect:BDRYparms})
\eb{bin}
  B_{\rm horiz}=(1-f) B_{\rm horiz}+f B_{\rm bdry} \ ,
\en
where $f$ is a factor that transitions smoothly from zero in the outflows and one in the inflows,
\eb{smooth}
  f=1/\left(1+\exp\left(U_{\rm vert}/U_{\rm bdry}\right)\right) \ .
\en
The electric field in the ghost zones is then given by Ohm's law from the velocity and magnetic 
field ghost zone values.  Alternatively, the electric field could be calculated by solving
\eb{Ealt}
    \left(\curl E\right)_{\rm horiz} = \left(B_{\rm horiz}-B_{\rm bdry}\right)/{\Delta t}\ .
\en
The original ghost-zone values of the magnetic field are
restored and $\partial\BB/\partial t$ is calculated from $\nabla\times\EE$.

%++++++++++++++++++++++++++++++++++++++++++++++++++++++++++++++++++++++++++++++
\subsubsection{Exterior Boundary In the Atmosphere}
%++++++++++++++++++++++++++++++++++++++++++++++++++++++++++++++++++++++++++++++

The log density and internal energy per unit mass are extrapolated into the ghost zones.  
This extrapolation may be unstable so the values of the ghost zone density and energy  are limited.
The density is limited to some minimum fraction and a maximum multiple of the average ghost zone 
value at that level.  The energy per unit mass is limited to a range of values about its boundary
value.  The density derivative at the boundary is further modified by replacing the vertical 
derivative of vertical momentum with the momentum divided by the scale height,
\eb{rbdrylo}
\left(\ddt{\rho}\right)_{\rm bdry}=\left(\ddt{\rho}\right)_{\rm bdry}  + \left(\ddy{\rho u_y}\right)_{\rm bdry} 
- (\rho u_y)/H_P \ ,
\en 
where $H_P$ is the pressure scale height and $y$ is the vertical direction pointed inward.

The ghost zone velocities are set to their value at the boundary and are limited to a maximum
absolute value for both in and outflows and which can not be greater than their value at the boundary
for inflows.

The magnetic field is made to tend toward a potential field at the boundary.  This is achieved by
creating a potential field extrapolation, $B_{\rm potential}$, of the magnetic field into the ghost zones 
and using this to set the vertical derivative of the horizontal components of the electric field 
proportional to $\left(B_{\rm potential}-B\right){\Delta y}/{\Delta t}$.
%}

%%%%%%%%%%%%%%%%%%%%%%%%%%%%%%%%%%%%%%%%%%%%%%%%%%%%%%%%%%%%%%%%%%%%%%%%%%%%%%%
\section{Input / Output}
%%%%%%%%%%%%%%%%%%%%%%%%%%%%%%%%%%%%%%%%%%%%%%%%%%%%%%%%%%%%%%%%%%%%%%%%%%%%%%%
\label{sect:io}

A snapshot consists of the nine variables: density, three components of momentum, internal energy per
unit volume, temperature, and three magnetic field components.  If a \verb:from='from.dat': file is
specified in the \verb'&io' namelist it is used to read an initial snapshot specifying all
nine variables that will be calculated evolving in time, otherwise the same \verb:file='run.dat':
is used for both input and output.  In addition, a mesh file must be selected
that gives the coordinate, coordinate spacing and Jacobian between physical and
index space at each grid location.  The first line are the three dimensions.  Then
for each dimension there is a line giving: the cell centered grid spacing, the face
centered grid spacing, the cell centered location, the face centered location, 
1/derivative of the cell centered locations with respect to the cell index,
1/derivative of the face centered locations with respect to the cell index.
If there is a \verb'from.dat' file, these values will be taken from the file \verb'from.msh'
if it exists.  Similarly a \verb'from.tim' file is sought with the time information.

Output consists of eleven files based on the root name \verb'run' of the \verb'run.dat' file.  One is
\verb'run.dat' itself to which snapshots are written every \verb'nstep' timesteps or \verb'tsnap' time
intervals.  Another is a \verb'run.scr' file of snapshots that overwrite each other every \verb'nscr'
time steps or \verb'tscr' time intervals.  These are direct access unformatted files. 
At the same interval as the saved snapshots are
written,  file \verb'run.tim' has the current time information added to it, a file 
of the various energy fluxes is written to \verb'run.flux' and various
statistical quantities are written to  \verb'run.stat'.
A log file is written for each run.  The script that starts the run begins it
by recording the date and time
of the start of the run, text describing the run, the size and date of the
executable, the number of CPUs/node, the number of nodes and a list of the nodes.
As each namelist is read the value of
each parameter, whether default or read in from the namelists is
printed to the log file.  Next if a snapshot of the variables is read the name of the "from" file
is given, then the time, if a time file is found,  and then the first several values of each
variable near the top (lb) is printed as they are read in..
Every time step the log file is appended with the time step number, the time, time
interval and the location of the most restrictive of each of the Courant conditions
due to viscosity, resistivity, density and energy changes.  Each time step a small 
text file  \verb'run.chk' is also written that gives similar time step information.
At the start of each run a copy
of the input file is saved as well as added to a file of all the input files for
that run, a copy of the mesh file is saved, a file giving the
dimensions, grid spacing, number of variables and mpi decomposition is written.

%%%%%%%%%%%%%%%%%%%%%%%%%%%%%%%%%%%%%%%%%%%%%%%%%%%%%%%%%%%%%%%%%%%%%%%%%%%%%%%
\section{Tests, Comparisons with Observations}
%%%%%%%%%%%%%%%%%%%%%%%%%%%%%%%%%%%%%%%%%%%%%%%%%%%%%%%%%%%%%%%%%%%%%%%%%%%%%%%
\label{sect:tests}

%{
%(
We have performed several basic tests of the Stagger Code as well
as comparing its results with observations of solar spectral lines,
the solar velocity spectrum, and solar p-mode properties.

%++++++++++++++++++++++++++++++++++++++++++++++++++++++++++++++++++++++++++++++
\subsection{Advection}
%++++++++++++++++++++++++++++++++++++++++++++++++++++++++++++++++++++++++++++++

\begin{figure}[!htb]
   \centerline{
   \includegraphics[width=0.5\textwidth]{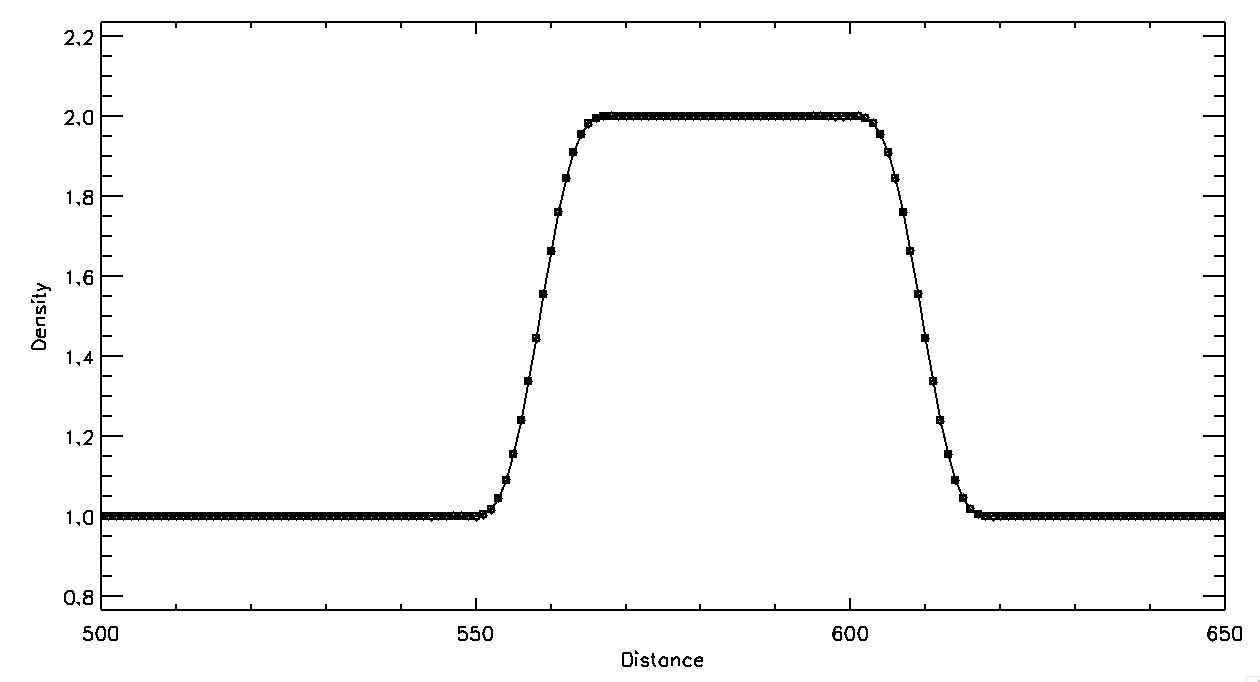}
   \includegraphics[width=0.5\textwidth]{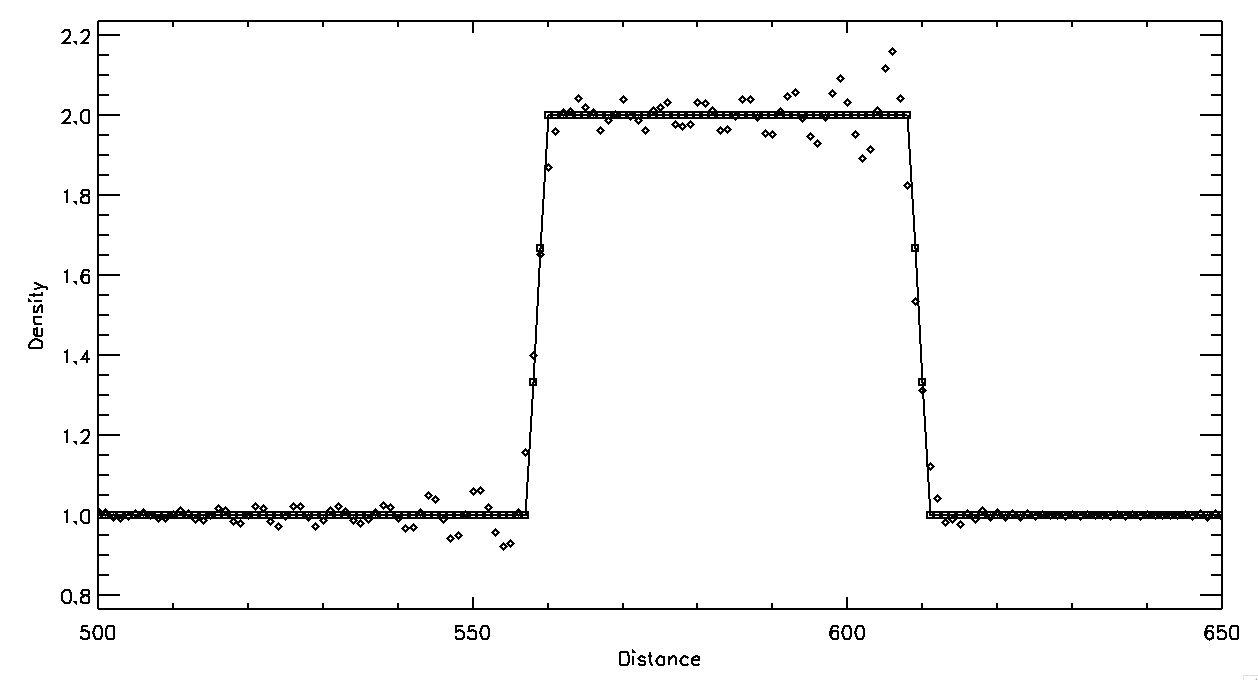}}
   \caption {Density advection of a smooth profile (left) and a sharp profile
    (right): the solid line (and squares) is the initial density distribution 
    displaced 480 grid cells.  The diamonds are the density after being
    advected 480 grid cells to the right.
   \label{fig:advect}}
\end{figure}

To test the advection properties of the Stagger Code, we advected a 
density hat with a constant velocity towards the right, in Fig.\ \ref{fig:advect}
The \emph{smooth} density profile is barely altered after being advected
480 grid cells.  The density profile is advected properly but
develops oscillations behind the sharp fronts.  The Stagger Code
is conservative (the total mass is unchanged) and non-dissipative.
The Stagger Code is, however, dispersive. That is, shorter
wavelength components travel slower than the longer wavelength
components, which leads to short wavelength oscillations
in the density after advection.  The conservative, non-dissipative
behavior is not affected.  In order to minimize the dispersive
effects, it is necessary to use sixth order derivatives and fifth
order interpolations on of the density on the staggered grid.
In the presence of steeper fronts (shocks) these dispersive effects appear 
and extra diffusion in equations \ref{eq:mom}, \ref{eq:itv} and \ref{eq:ien} is 
needed to damp out the short, unphysical wavelengths.

%++++++++++++++++++++++++++++++++++++++++++++++++++++++++++++++++++++++++++++++
\subsection{Brio-Wu MHD Shock Tube}
%++++++++++++++++++++++++++++++++++++++++++++++++++++++++++++++++++++++++++++++

A Brio-Wu shock tube \citep{BrioWu:shtube} was calculated.  Initial state is
a single step in density, pressure = energy ($\gamma=2$) and transverse
magnetic field, near the middle of the domain.  Velocity is zero everywhere to begin with.

\begin{center}
 Table 2: Initial State \\
\begin{tabular}{|l|c|l|}
\hline
Variable & left & right  \\
\hline
Density & 1 & 0.125  \\
\hline
Pressure, $\gamma = 2$ & 1 & 0.1  \\
\hline
Transverse B & 1 &  -1  \\
\hline
\end{tabular}\\
\end{center}

\begin{figure}[!htb]
   \centerline{\includegraphics[width=0.5\textwidth]{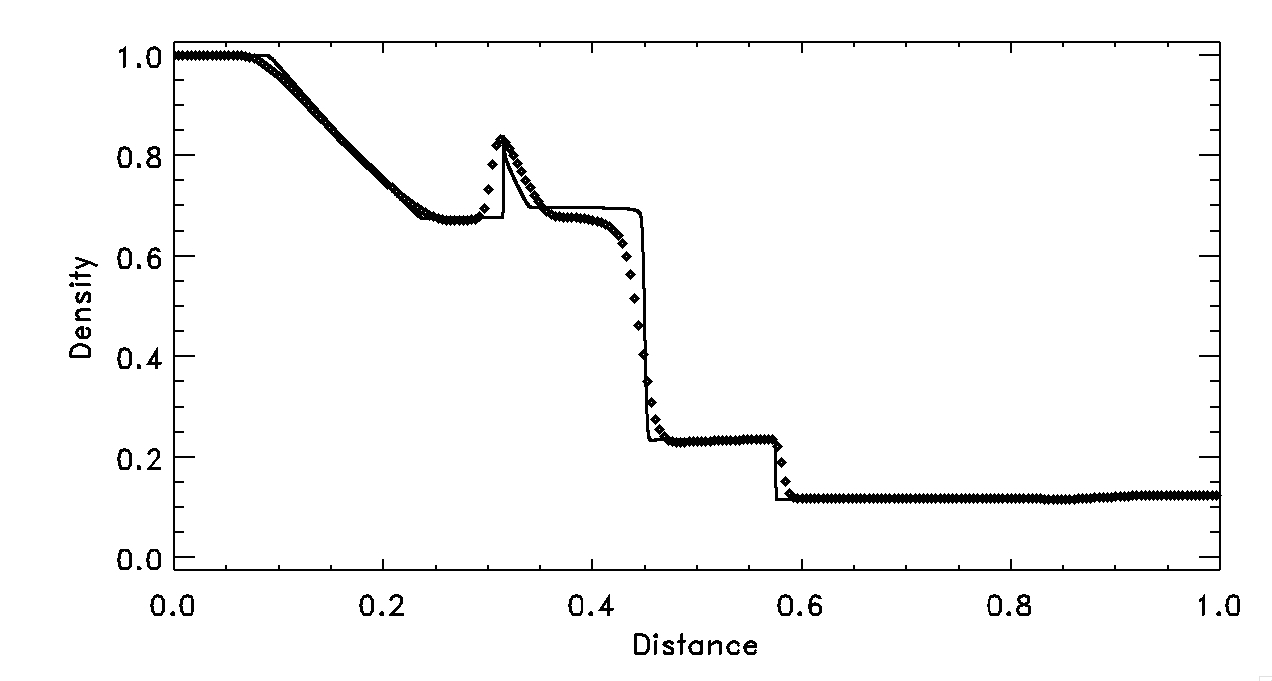} \\
           \includegraphics [width=0.5\textwidth]{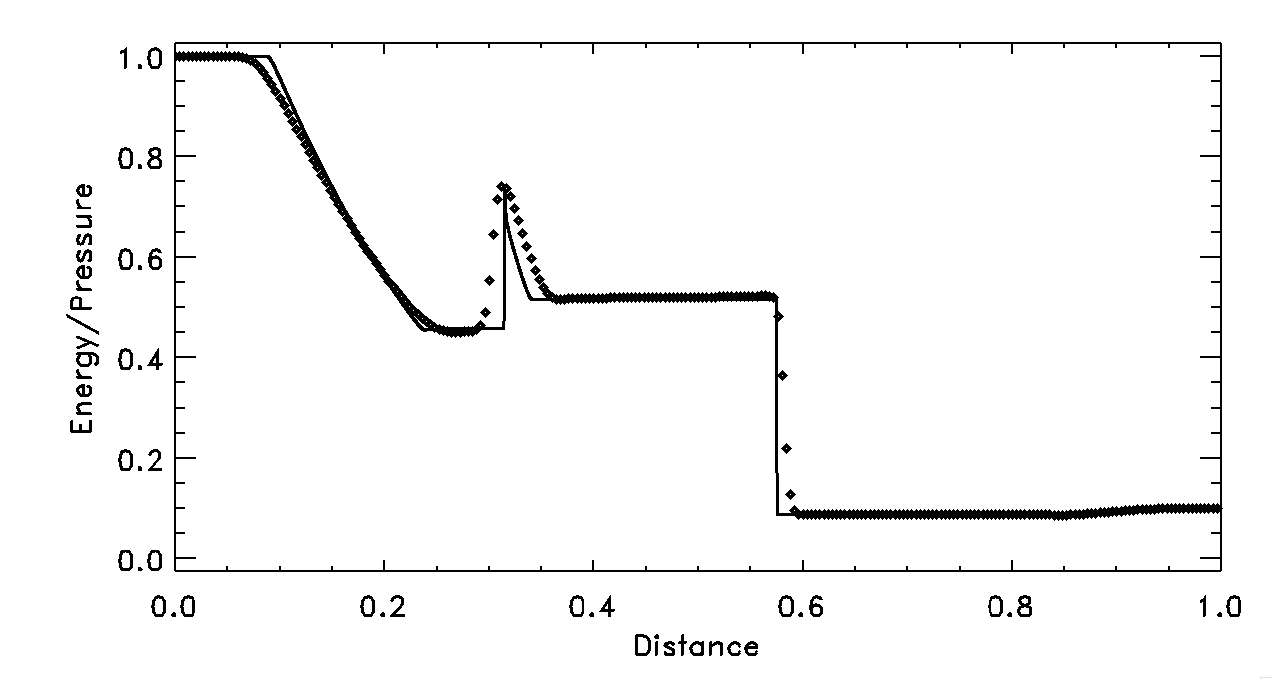}}
   \centerline{\includegraphics[width=0.5\textwidth]{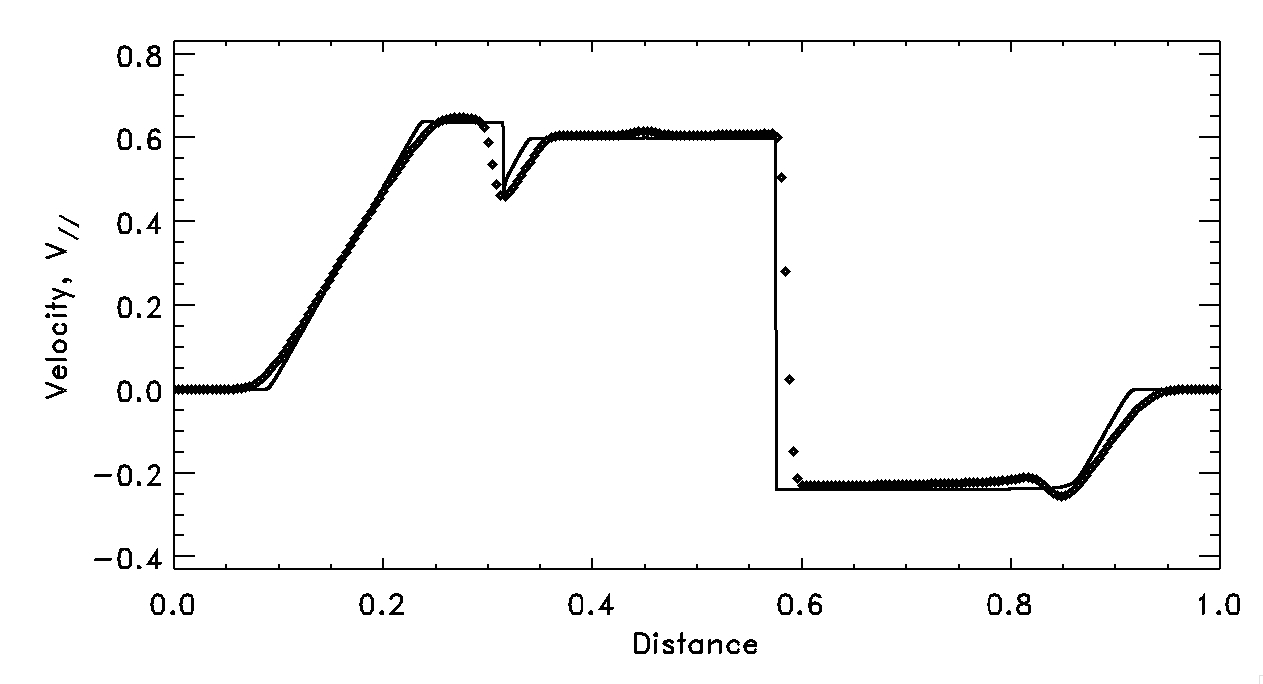} \\
           \includegraphics [width=0.5\textwidth]{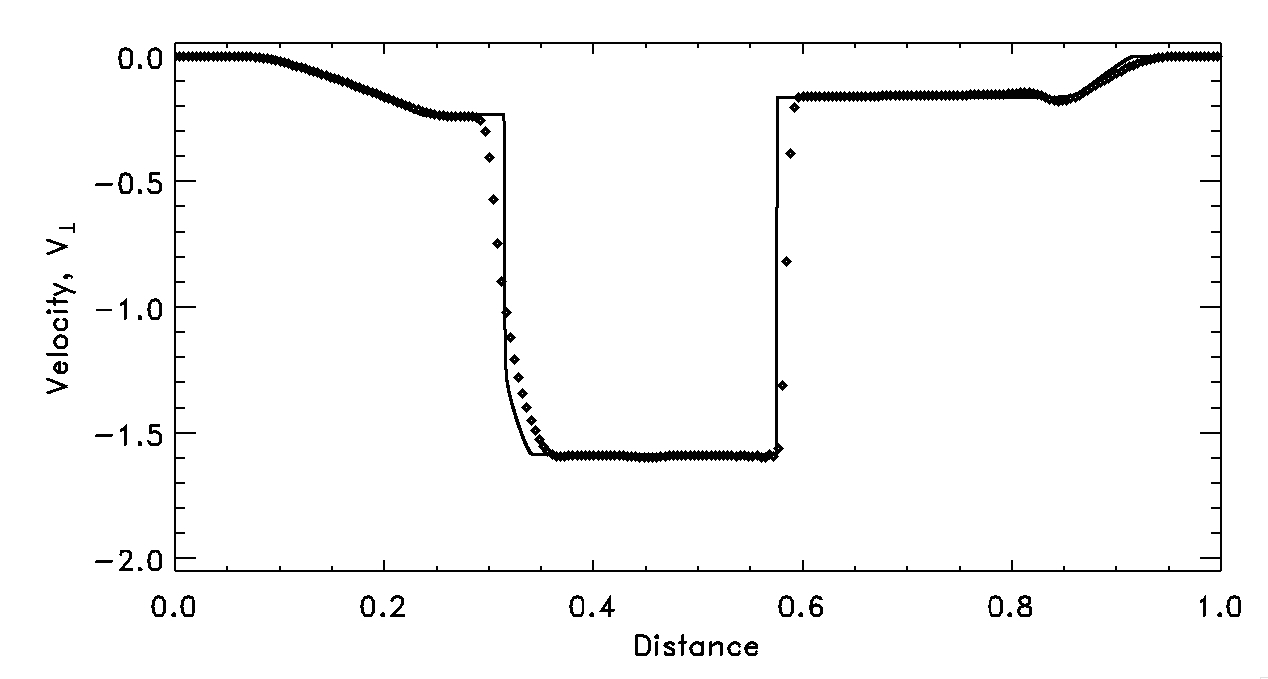}}
   \centerline{\includegraphics[width=0.5\textwidth]{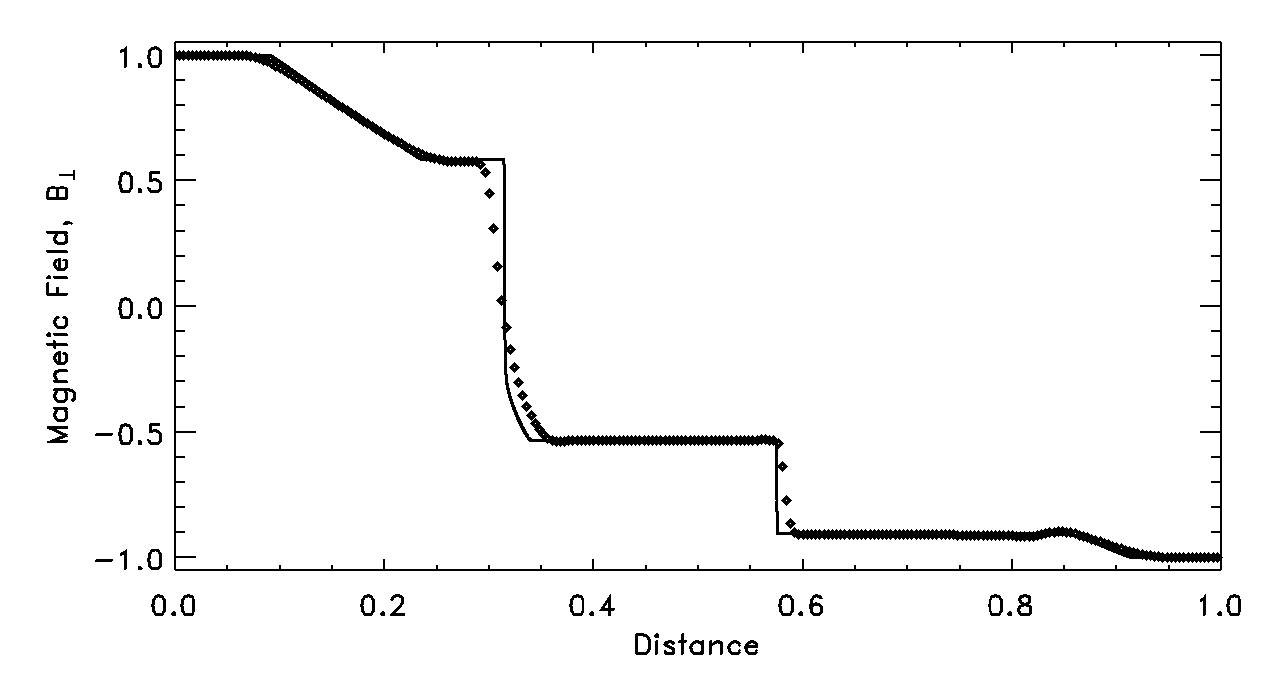}}
   \caption{Brio-Wu shock tube results for density, pressure, parallel
           and transverse velocities and transverse magnetic field at
           time of t=0.15 computed on a grid of 250 cells (diamonds) 
           compared to a calculation with 8000 cells (solid lines).
           The viscosity parameters were $c_v$=nu1=0.04, $c_s$=nu2=2.5 
           and $c_w$=nu3=0.04 (eqn. \ref{eq:RMnu} and appendix).
   \label{fig:shtube}}
\end{figure}

Results were computed on a grid of 250 cells and compared with a
calculation using 8000 cells at time t = 0.15 (Fig. \ref{fig:shtube}).
A fast rarefaction is the first wave to the right. It has a small
overshoot in the velocities at low resolution.  Behind is a shock
that is well handled at both resolutions for all the variables and
which moves slightly faster at low resolution.  Density and momenta
have jumps at the contact discontinuity, which is broader at low
resolution.  A slow compound  wave and rarefaction farther to the
left are slightly broader at low resolution.  There is also some 
rounding at transitions at low resolution.

\begin{figure}[!htb]
   \centerline{ \includegraphics [width=0.5\textwidth]{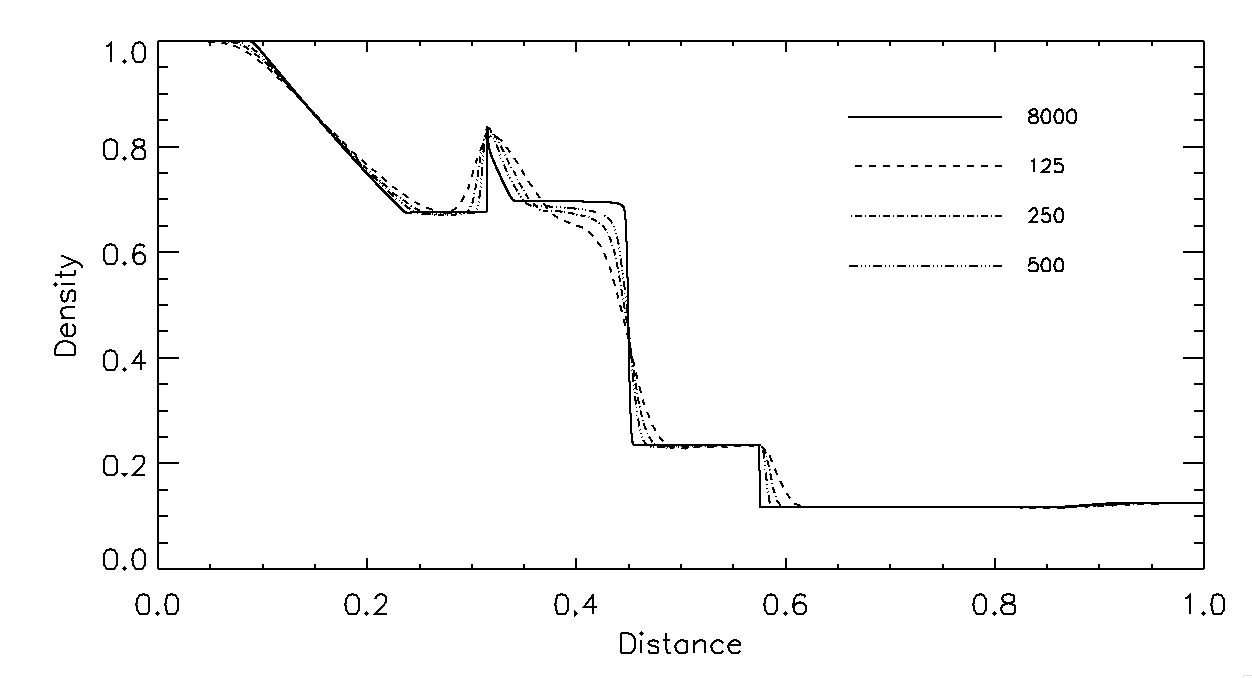} \\
                \includegraphics[width=0.5\textwidth]{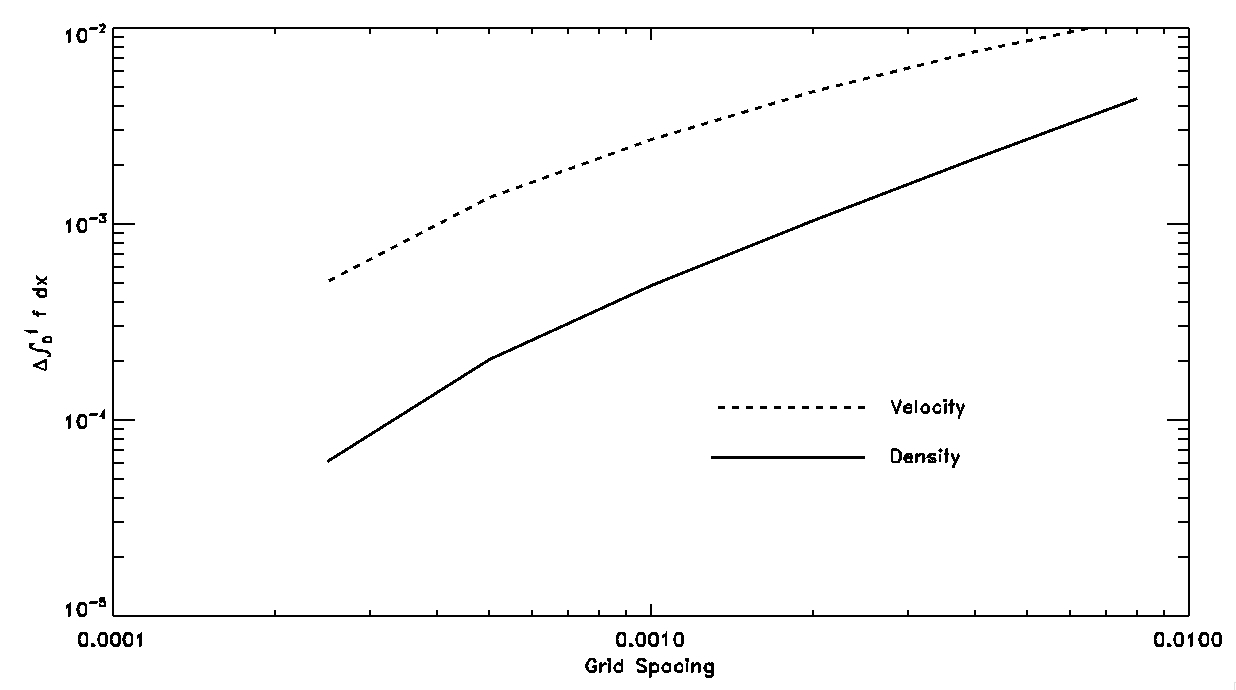}} 
   \caption{Convergence properties.
   \label{fig:BWconvg}}
\end{figure}

Convergence properties are shown in Fig. \ref{fig:BWconvg}. Comparing
density profiles at different resolutions shows that resolution  
primarily affects the various fronts, while smooth portions of the profile are 
nearly resolution independent.  The difference in the integral of the density 
and absolute value of the velocity at low resolutions with that at the
highest resolution shows that the convergence at these fronts is
approximately linear in the grid spacing.

%++++++++++++++++++++++++++++++++++++++++++++++++++++++++++++++++++++++++++++++
\subsection{Orszag-Tang}
%++++++++++++++++++++++++++++++++++++++++++++++++++++++++++++++++++++++++++++++

The Orszag-Tang 2D MHD test was also run \citep{OT79}.
It tests how well a code handles the formation of MHD shocks and 
shock-shock interactions, and also reveals if any directionality is
imparted on the dynamics by the grid.  
The initial state is,
\eab{initOT}
\rho & = & 25/36\pi \\
P_{\rm gas} & = & 5/12\pi/(\gamma-1), \gamma=5/3 \\
u_x & = & - \sin(2\pi y) \\
u_y & = & \sin(2\pi x) \\
B_x & = & -\sin(2\pi y)/4\pi \\
B_y & = & \sin(4\pi x)/4\pi \ .
\ean

\begin{figure}[!htb]
%  \centerline{\includegraphics[width=0.5\textwidth]{figs/Pgas.jpg} \\
%        \includegraphics [width=0.5\textwidth]{figs/BB.jpg}}
   \centerline{\includegraphics[width=0.8\textwidth]{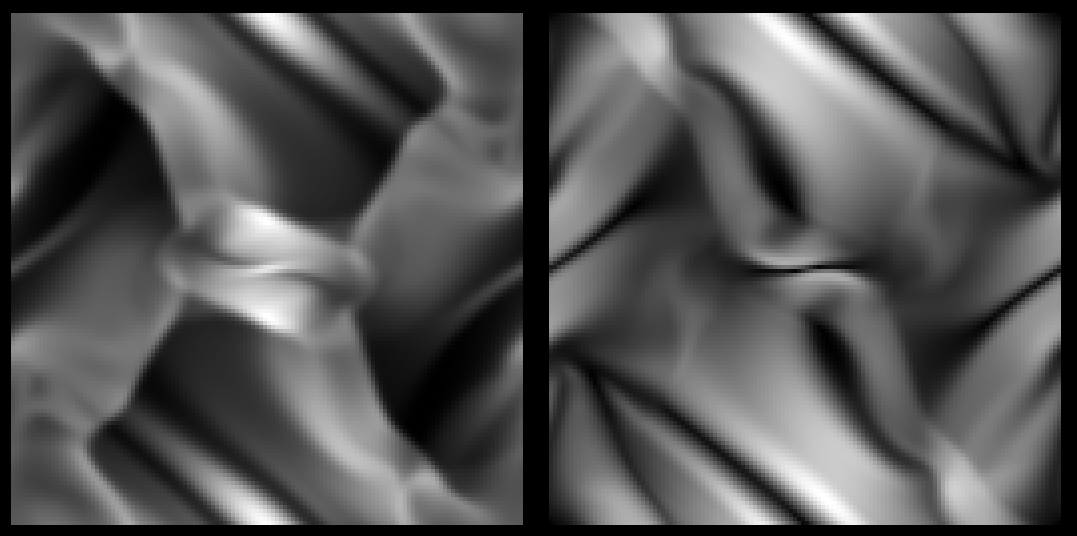}}
   \centerline{\includegraphics[width=0.8\textwidth]{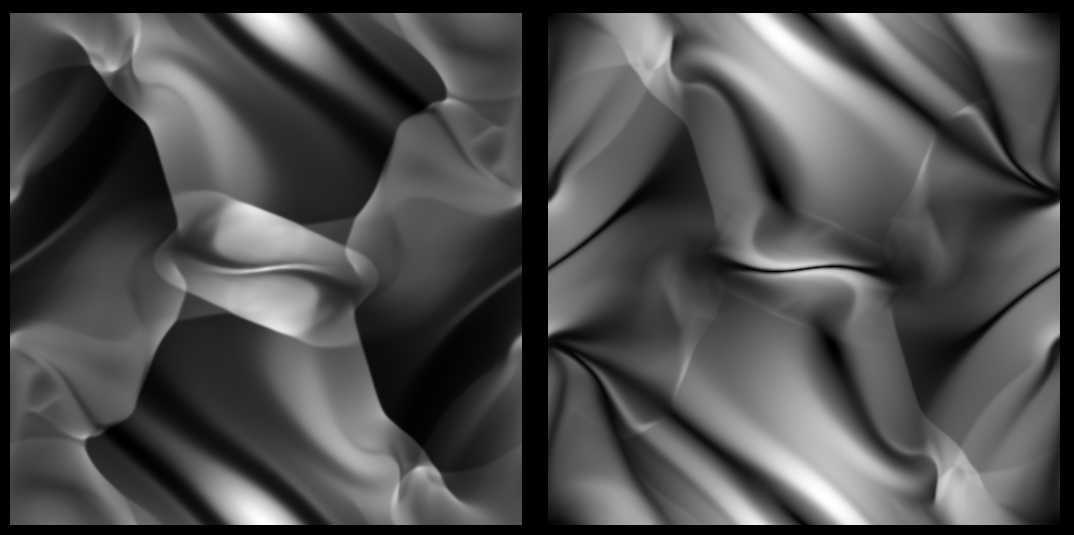}}
   \centerline{\includegraphics[width=0.8\textwidth]{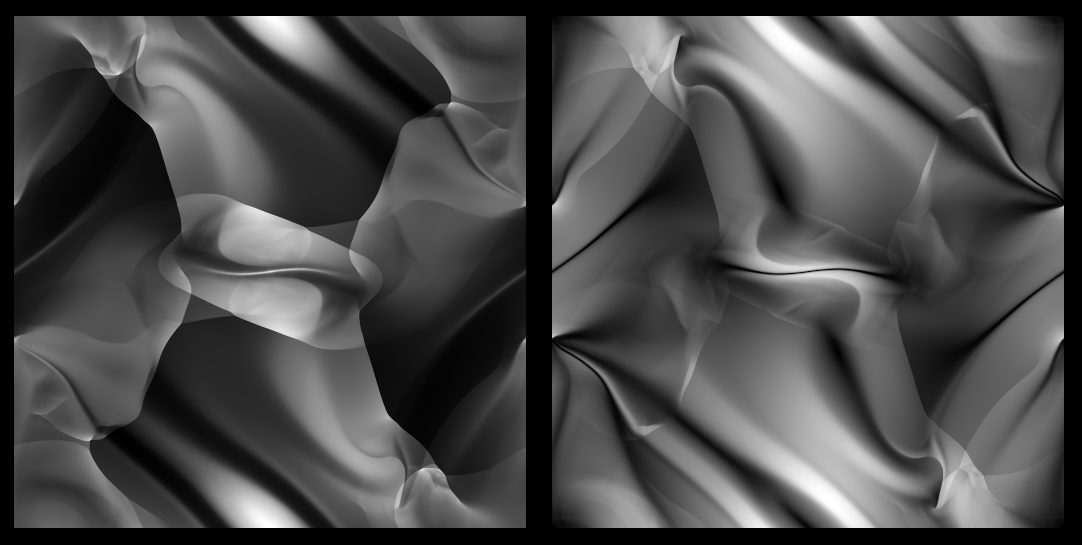}}
%  \centerline{\includegraphics[width=0.3\textwidth]{OT_eb_128.jpg} \\
%  \includegraphics[width=0.3\textwidth]{OT_eb_512.jpg} \\
%  \includegraphics[width=0.3\textwidth]{OT_eb_2048.jpg}}
   \caption{Orszag-Tang 2D MHD test results for gas pressure (left) and magnetic
           field magnitude (right) at time t=0.5.  Results for grids $128^2$ (top), 
           $512^2$ (middle) and $2048^2$ (bottom) are compared. }
   \label{fig:OT}
\end{figure}

The resulting gas and magnetic pressure are shown in Fig. \ref{fig:OT}.
The performance of the Stagger Code is evaluated by comparing these results with
similar calculations using other codes, e.g. \citep{DaiOT98}.  

\begin{figure}[!htb]
   \centerline{ \includegraphics [width=0.5\textwidth]{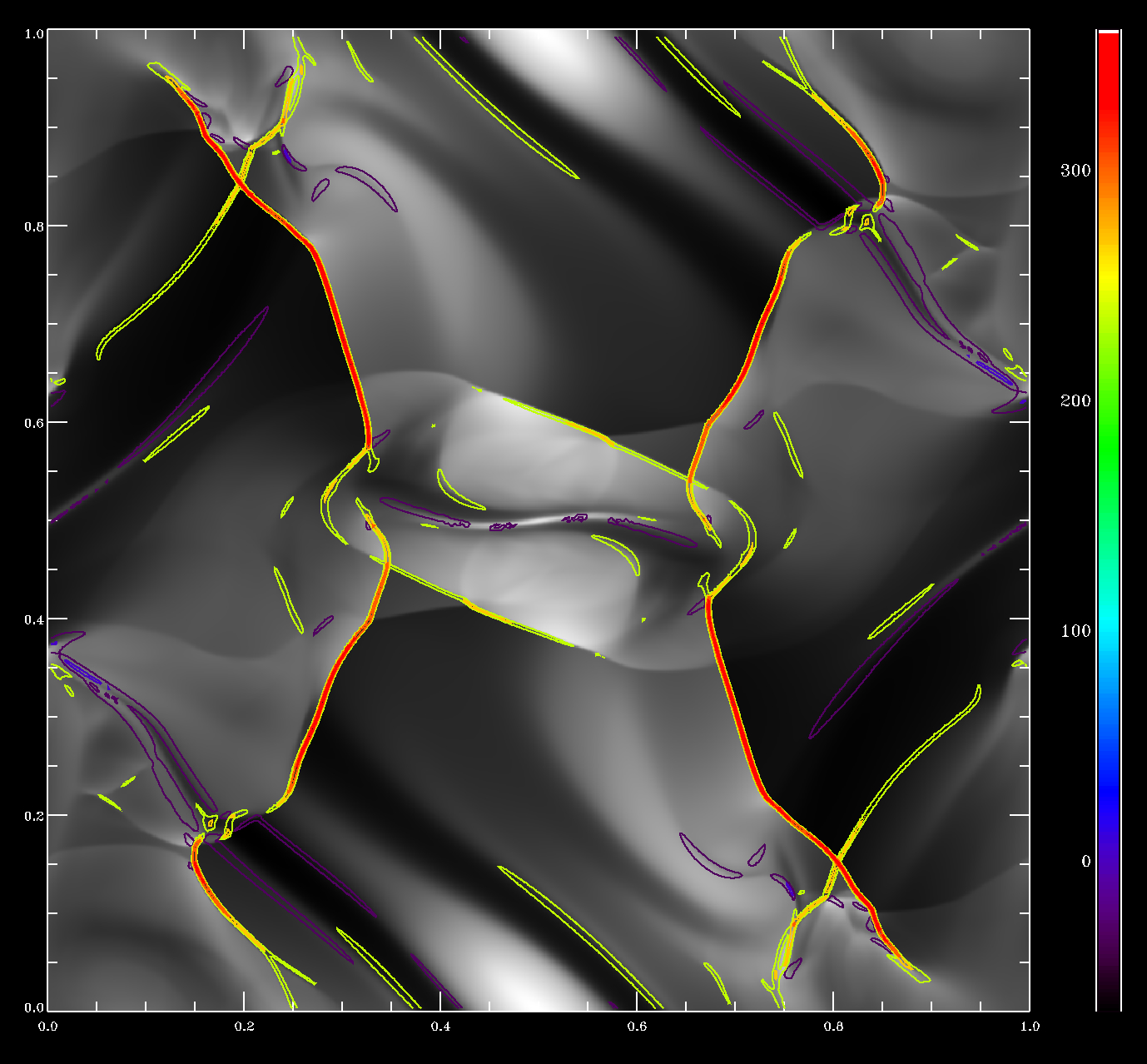} \\
                \includegraphics[width=0.5\textwidth]{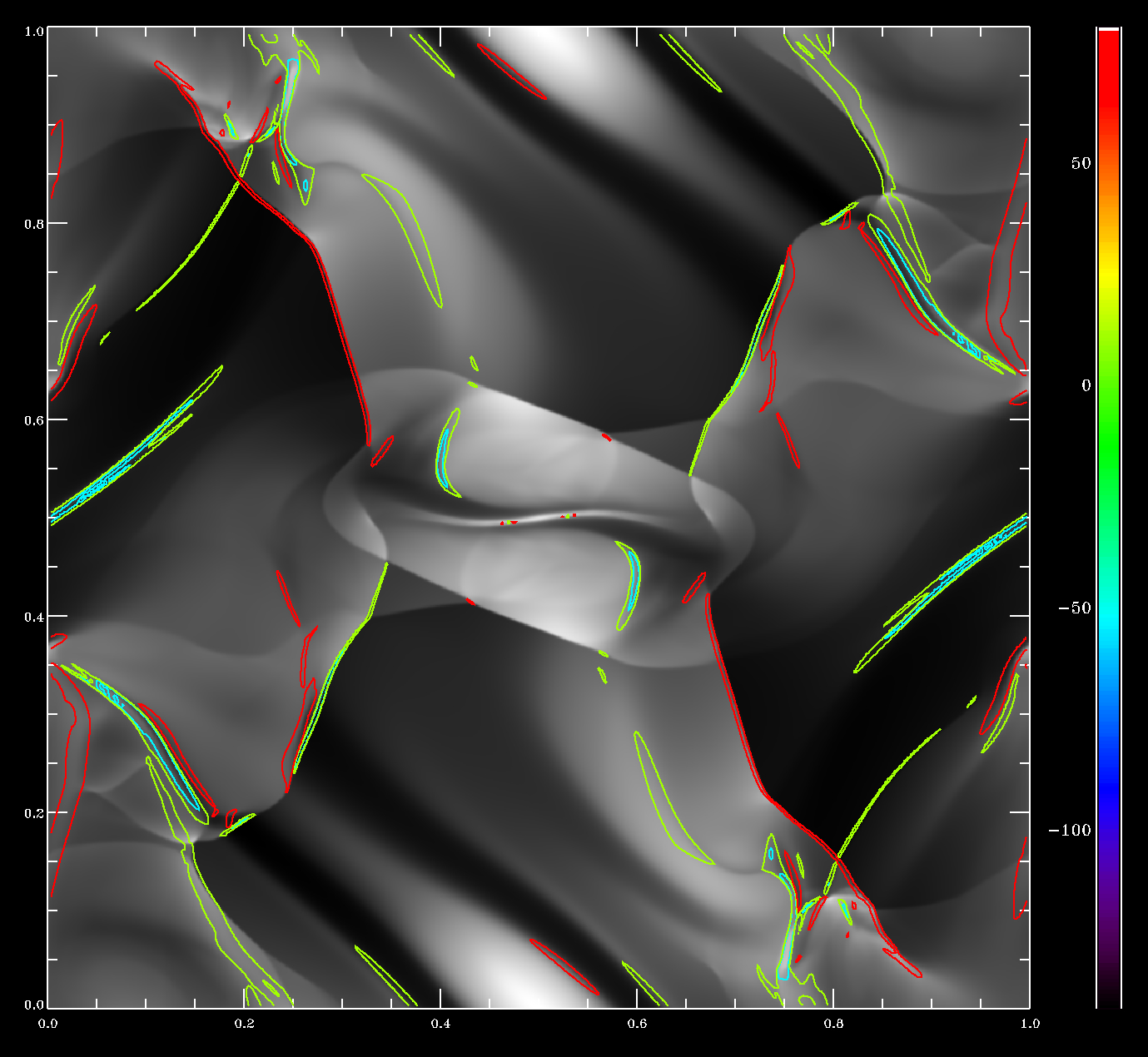}}
   \caption{{Image of the pressure with superimposed contours of the negative
             velocity divergence (left) and current density (right).}
   \label{fig:OT+JV}}
\end{figure}

Shocks are visible as sharp jumps in the pressure (Fig. \ref{fig:OT}).
Interactions of shocks and waves are further illustrated in 
Fig.  \ref{fig:OT+JV} showing contours of the negative of the velocity divergence 
and of the current density superimposed on the gas pressure image.
The shocks around the central oval and the long (red) divergence trails 
from it are fast shocks \citep{Snowetal2021}. 

%++++++++++++++++++++++++++++++++++++++++++++++++++++++++++++++++++++++++++++++
\subsection{Spectral Lines}
%++++++++++++++++++++++++++++++++++++++++++++++++++++++++++++++++++++++++++++++

\begin{figure}[!htb]
  \centerline{
  \includegraphics[width=0.5\textwidth]{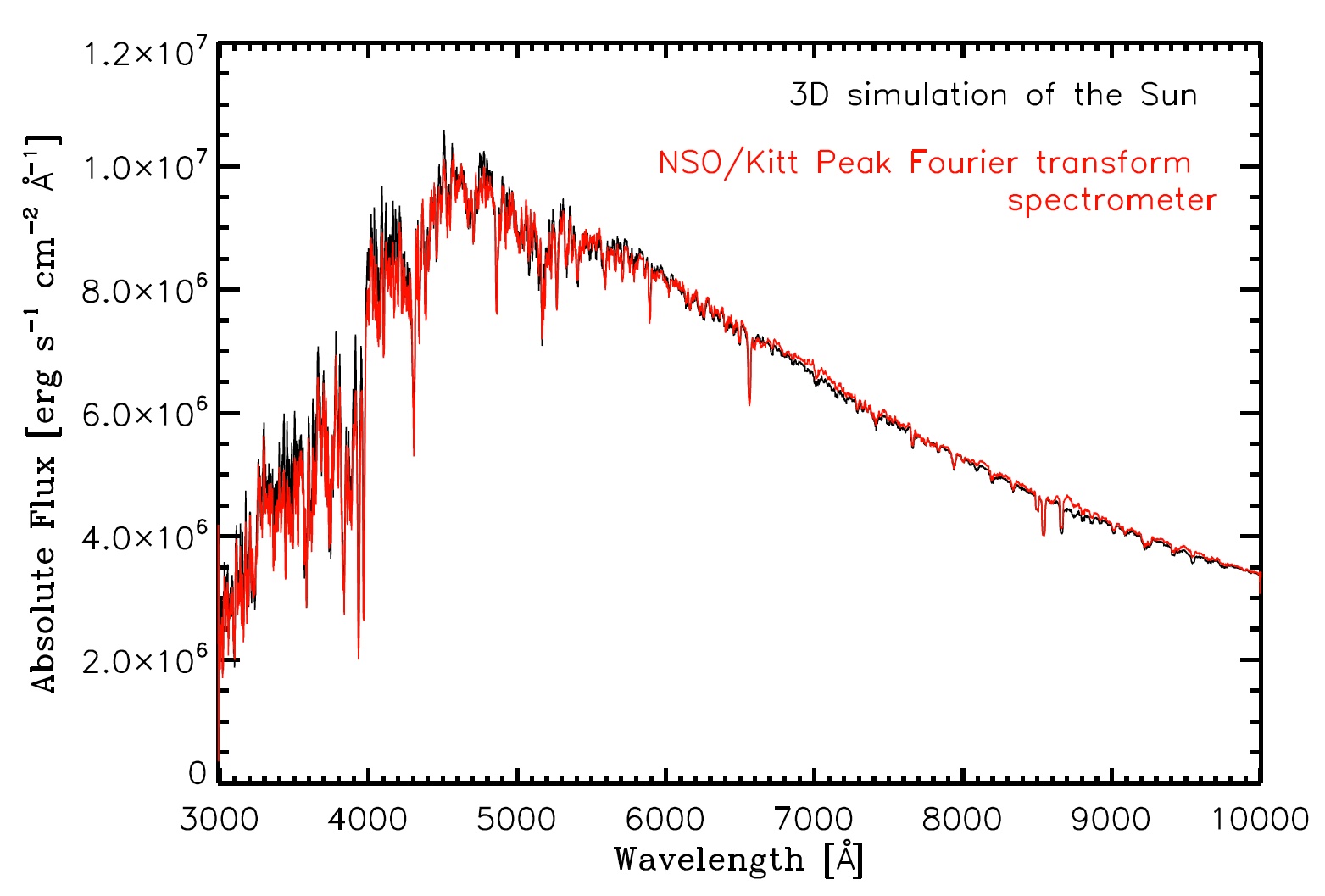}}
    \caption{Comparison of the simulated solar flux (black) from 300 to 1\,000\,nm with
    the observed \citep[red][]{kurucz:NewSolarAtlases} solar surface flux \citep{chiavassa:StaggerSpectra}.
    The good agreement shows that the thermodynamic structure of the simulation
    near the surface is realistic.
    \label{fig:sunflux}}
\end{figure}

A sensitive test of both the dynamic and thermal structure of the simulations are
given by comparing observed solar spectra with spectral synthesis
performed on a realistic solar simulation.  The spectral energy
distribution (i.e., low resolution spectrum) from 300 to 1\,000\,nm is in good agreement with
observations Fig.\ \ref{fig:sunflux}, showing that the mean thermal structure of the
Stagger solar simulations is realistic.

\begin{figure}[!htb]
  \centerline{
  \includegraphics[width=0.9\textwidth]{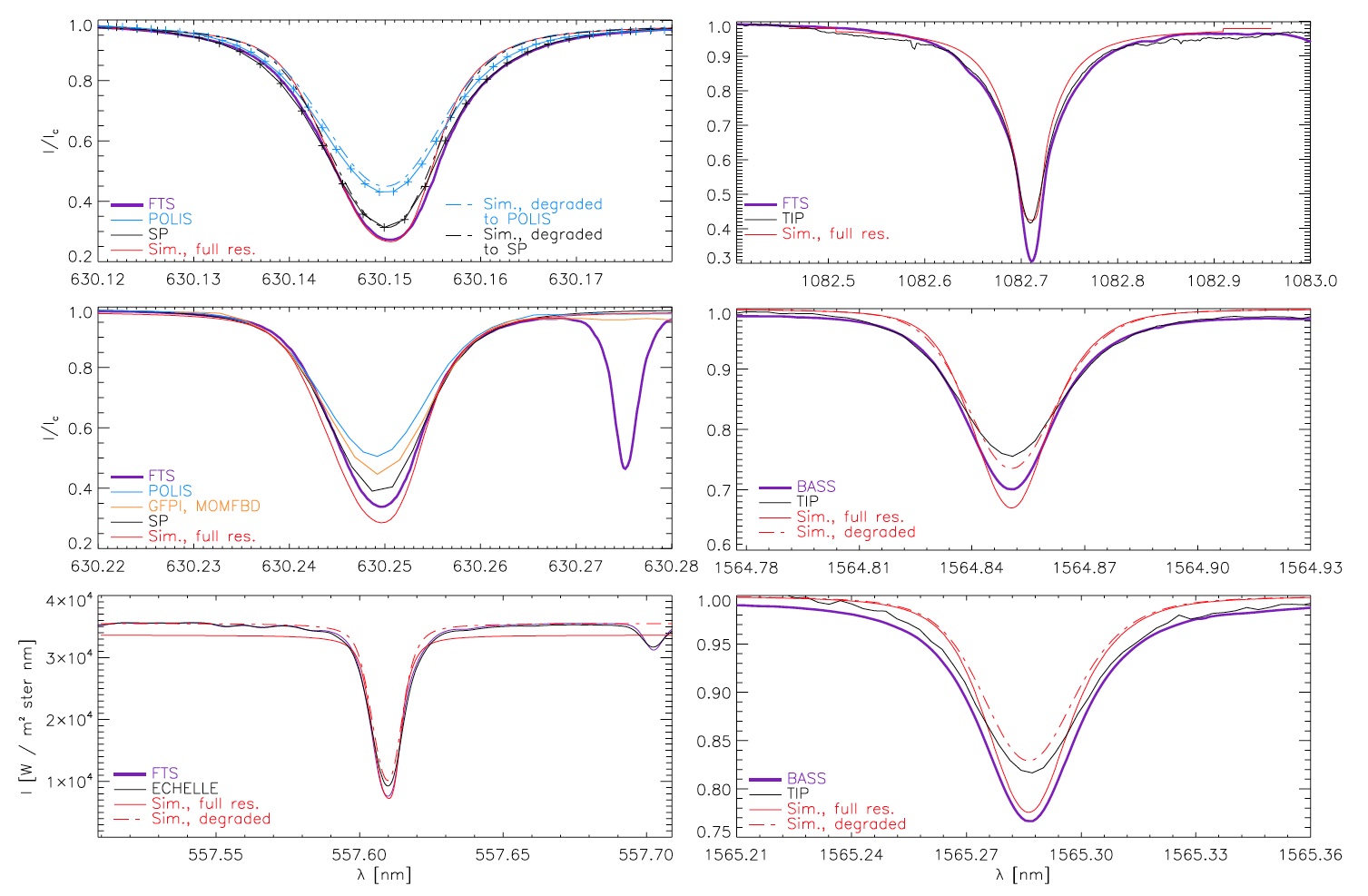}}
    \caption{Comparison of line spectra calculated from a hydrodynamic
    6$\times$6\,Mm Stagger simulation and observations of the Fe\,I
    557.6, 630.15, 630.25, 1564.8 and 1565.2\,nm lines and the Si\,I
    1082.7\,nm line with the German Vacuum Tower Telescope and Hinode %\ref{Beck2013}
    \citep{beck:SolarConvFlucts}.
    The continuum intensities were matched except for the 557\,nm
    line which is in absolute flux units. The Si\,I line is formed in the
    upper photosphere where both non-LTE effects neglected in the
    simulations and as well as boundary effects reduce the agreement
    with the observations.
    \label{fig:linespectra}}
\end{figure}

Spectral line profiles depend sensitively on both the temperature
structure and the convective velocities.  The line profiles at any
give location vary in depth, wavelength and width.  To compare with
observations, the simulated profiles are convolved with a spatial 
Gaussian times a Lorentzian kernel (representing the telescope point
spread function) and convolved with a Gaussian in wavelength (representing
the instrument degradation) to obtain the same 
resolution as the observations and then spatially averaged to produce
the mean profile.  These mean profiles are an accurate check on the
detailed correlation of temperature and velocity in the solar Stagger
simulations.  Figure \ref{fig:linespectra} shows comparisons with
several iron lines in the visible and infrared formed in the low
to mid photosphere \citep{beck:SolarConvFlucts}.  The simulation covered a
region 6 $\times$ 6\,Mm wide and extended from the temperature minimum
down to a depth of 2\,Mm below unit continuum optical depth.  The
horizontal grid spacing was 24\,km and the vertical grid spacing was
15\,km.  The simulations used four (continuum, weak, medium and strong lines)
scaled opacity bins (see Sect.\ \ref{sect:OpacBinScaled}). The continuum and line opacities and the equation of
state, for the simulation, were taken from the Uppsala stellar atmospheres 
package and ODF tables
(see Sect.\ \ref{sect:OrigAtmTabs}). The spectral line synthesis was subsequently performed
with the LTE LILIA code \citep{socas-navarro:StokesInvTechniqs}, based on a
number of simulation snapshots.
It is the convective velocities that provide the non-thermal Doppler
broadening in the simulation with no free parameters, that in 1D
calculations is represented by micro and macro turbulence parameters.

\begin{figure}[!htb]
  \centerline{
  \includegraphics[width=0.8\textwidth]{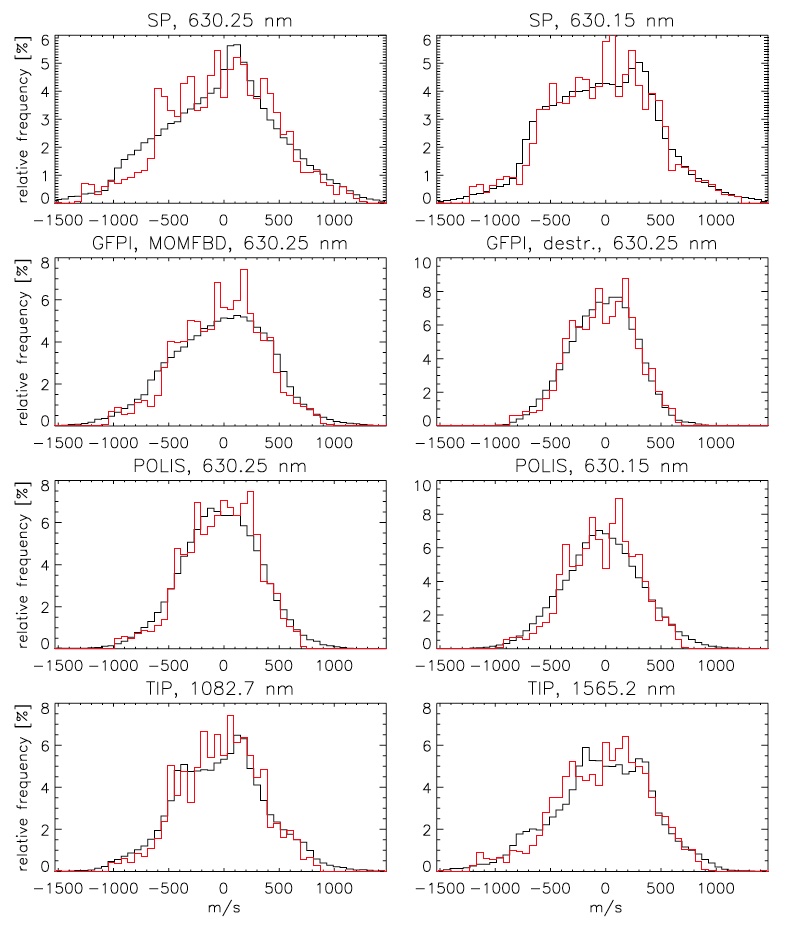}}
   \caption{Histogram of line core velocities from the simulation (red) and
   observations (black).
   \label{fig:corevel}}
\end{figure}

The line core velocities also match well between the degraded simulation and the
observations (Fig.\ \ref{fig:corevel}).

Further comparisons between the simulation generated and observed spectral lines:
intensity histogram, line depth, line asymmetry, equivalent width and full width at
half maximum can be found in \cite{beck:SolarConvFlucts}.
With the current, larger, NST and DKIST solar telescopes more detailed comparisons
will be possible since the telescope resolution reaches that of the simulations (24\,km).
A few simulation snapshots with 12\,km horizontal resolution are available, but may not
be adequately relaxed to the higher resolution.

%++++++++++++++++++++++++++++++++++++++++++++++++++++++++++++++++++++++++++++++
\subsection{Center to Limb Variation}
%++++++++++++++++++++++++++++++++++++++++++++++++++++++++++++++++++++++++++++++

\begin{figure}[!htb]
  \centerline{
  \includegraphics[width=0.6\textwidth]{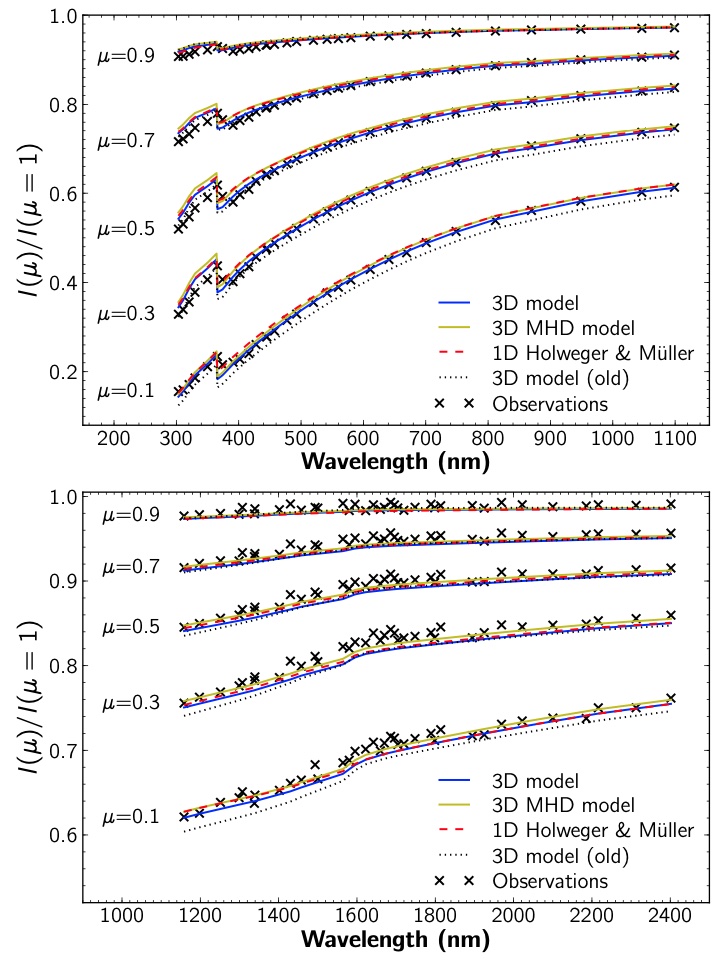}}
   \caption{CLVs in the continuum intensity from \cite{Pereira13}.
    Top panels: comparison with the visible/infrared observations of
    \cite{neckel-labs:solar-LD94}.
    Bottom panels: comparison with the near-infrared observations
    of \cite{pierce:limb-dark-1,pierce:limb-dark-2}, for wavelengths between 1158.35--2401.8 nm.
   \label{fig:CLV}}
\end{figure}

The center to limb variations of continuum intensities provide a
sensitive probe of the solar photosphere. Because the continuum
intensity is proportional to the local source function of continuum
forming regions, its center to limb variation is a measure of the
temperature variation with depth (the closer to the solar limb, the
higher up in the atmosphere).  \cite{Pereira13} have made such a
comparison shown in Fig. \ref{fig:CLV}.  For this work Stagger simulations
based on the equation-of-state and opacities detailed in sections \ref{sect:OrigAtmTabs} and \ref{sect:NewAtmTabs} were compared, both adopting the solar chemical
composition by \cite{AGS05}.  The emergent
intensities were spatially and temporally averaged over 90 snapshots.  There is
a significant improvement between the older 4 bin scaled opacities and the
newer 12 bin optimized opacities that are a divided in both opacity strength and
wavelength (Fig. \ref{fig:cbin_plot_sune250}).

%++++++++++++++++++++++++++++++++++++++++++++++++++++++++++++++++++++++++++++++
\subsection{Convection Velocity Spectrum}
%++++++++++++++++++++++++++++++++++++++++++++++++++++++++++++++++++++++++++++++

\begin{figure}[!htb]
  \centerline{
  \includegraphics[width=0.6\textwidth]{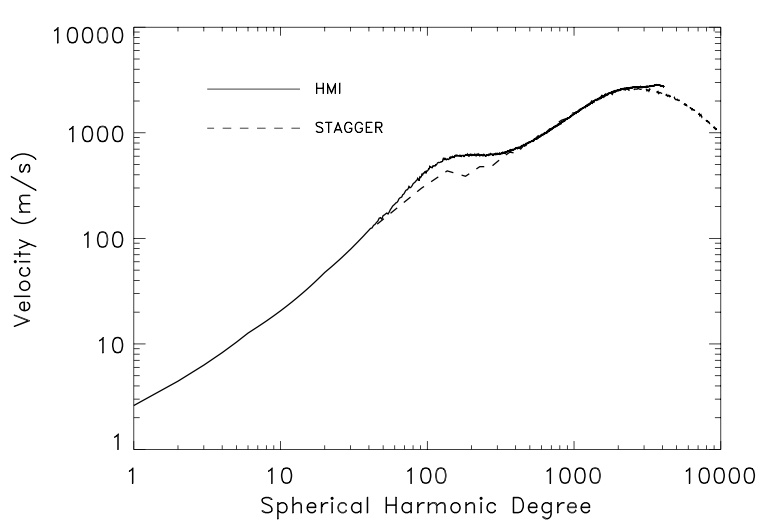}}
  \caption{Simulated convective total velocity spectrum compared to the observed 
  solar spectrum as as calculated by \citet{hathaway:SunConvSpectr2} from HMI data.  Plotted is
  $[kP(k)]^{1/2}$ where $P(k)$ is the velocity power at  wavenumber k.
  Note that the spectrum is featureless except for the peak at granulation scales,
  $\sim 1\,Mm$, and excess power at supergranulation scales.  The amplitude at very
  low degree is uncertain.
  \label{fig:velspectrum}}
\end{figure}

The convective velocity spectrum is a good test of the velocity amplitudes at
different spatial scales.  This in turn gives information on the velocities at
different depths since the scale of the dominant convective structures increases
with depth because of mass conservation in the stratified atmosphere.
Figure \ref{fig:velspectrum} compares the spectrum of simulated velocities at
optical depth unity with those determined by \citet{hathaway:SunConvSpectr2} from HMI data.
The simulation data are from a 96 $\times$ 96\,Mm wide and 20\,Mm deep run
covering 25\,hours of solar time.  The horizontal resolution was 24\,km and the
vertical resolution varied from 12\,km near optical depth unity to 70\,km near 20
Mm depth. A horizontal magnetic field was advected into the computational domain by
inflows at the bottom boundary.  The field strength was slowly increased to 1\,kG and
thereafter held fixed.  So magnetic flux is constantly being advected into the
simulation domain.  There is no net vertical magnetic field at any horizontal plane in
the simulation, but of course in magnetic concentrations there are local strong vertical
fields.
There is excellent agreement in the range of granulation.  However, the
simulation does not show the observed excess power at supergranule scales.  The
reason for this is still an issue for ongoing research.

\citet{yelles_chaouche:SunConvFlowPowSpctr} has compared the simulation vertical and
horizontal velocities with IMaX observations at disk center of the
Fe\,I 525.02\,nm line.  The simulations had a box size of 6 $\times$
6\,Mm horizontally and extended from 0.43\,Mm above the average
$\tau_{500} = 1$ level to 2\,Mm below this level.  Runs were made
with zero magnetic field and with 50, 100, 200\,G average signed
vertical field.  

\begin{figure}[!htb]
  \includegraphics[width=0.8\textwidth]{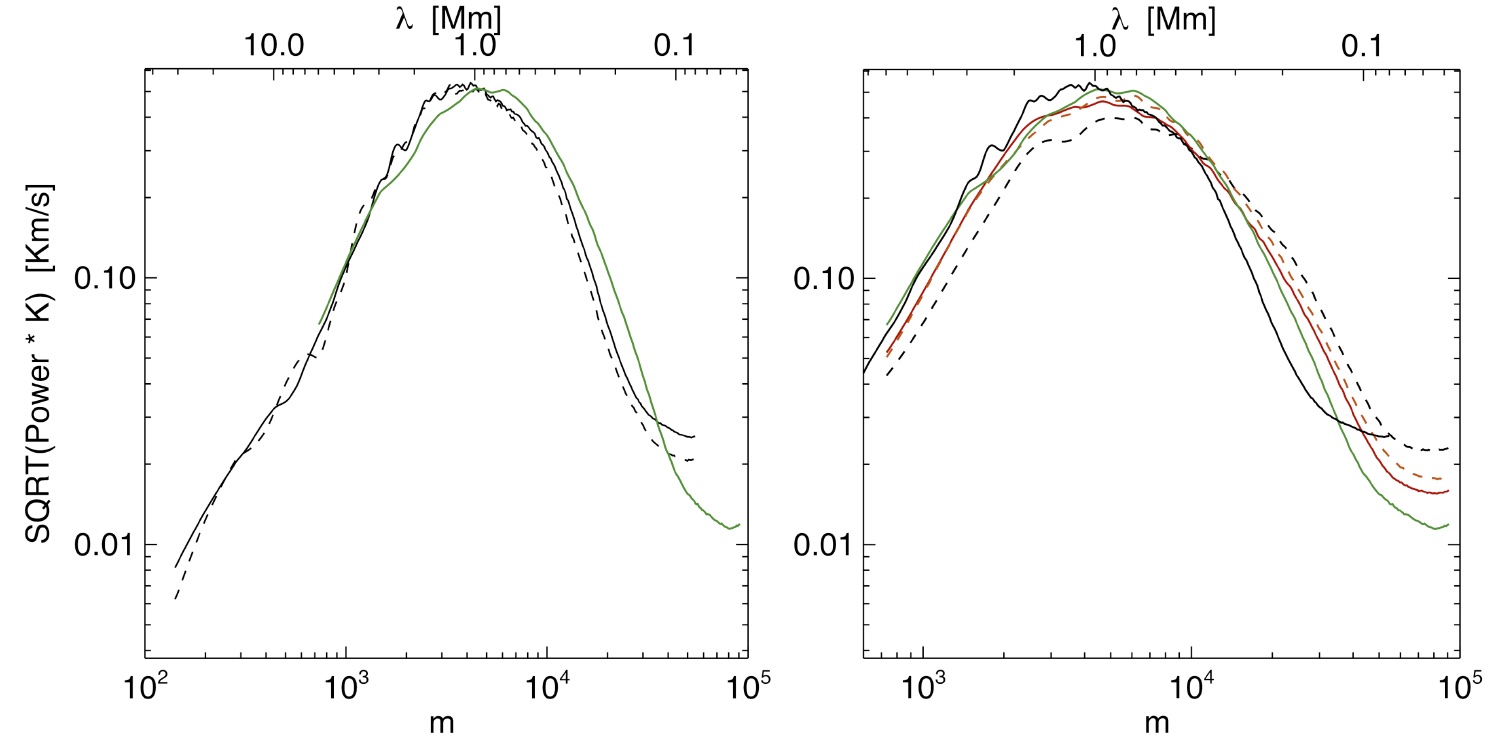}
    \caption{Vertical velocity spectrum $V(k)=[kP(k)]^{1/2}$ of simulation 
    compared to IMaX observed spectrum \citep{yelles_chaouche:SunConvFlowPowSpctr}.  The left figures shows 
    the hydrodynamic simulation (green) and two IMaX runs (black).  The right 
    figure shows the IMaX (solid black) and hydro plus three MHD simulations: 
    0\,G (hydro green), 50\,G (red), 100\,G (orange) and 200\,G (dashed black). 
    \label{fig:vvertspectrum}}
\end{figure}

\begin{figure}[!htb]
  \includegraphics[width=0.8\textwidth]{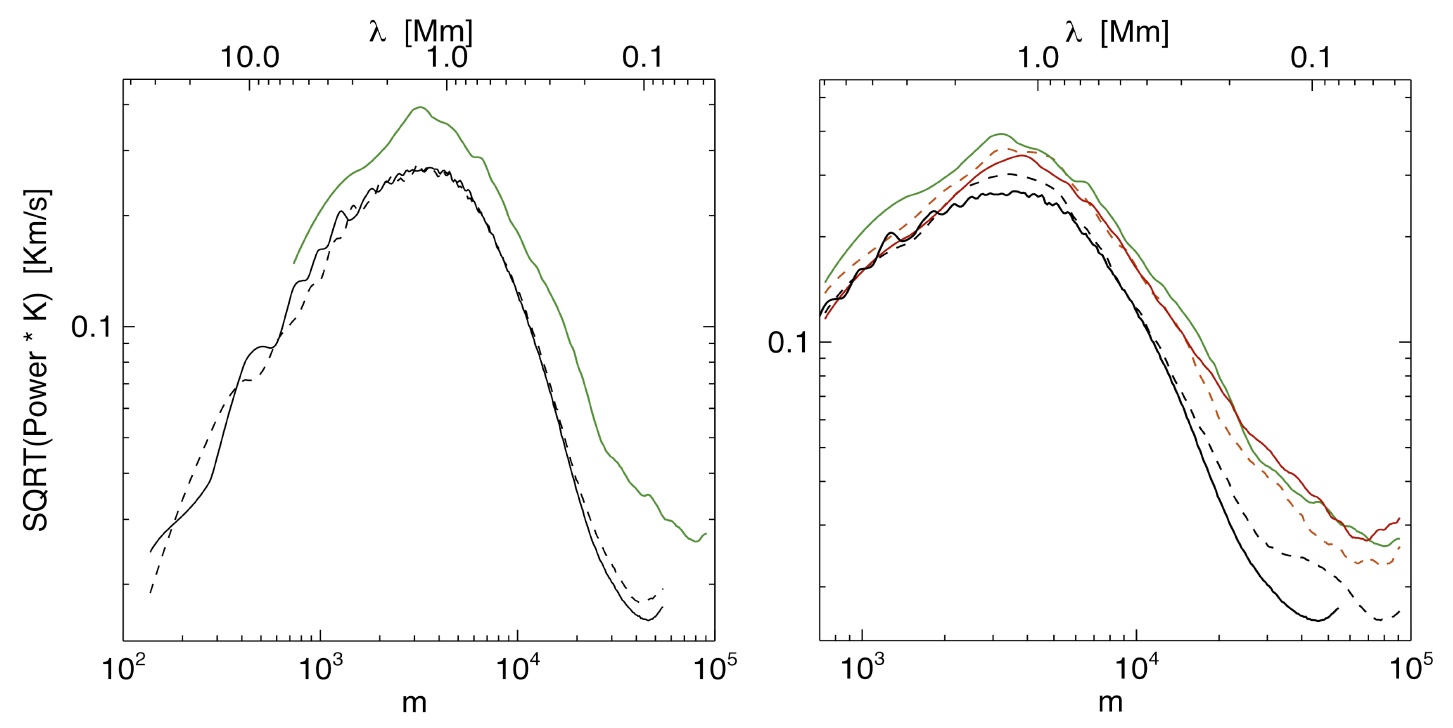}
    \caption{Horizontal velocity spectrum computed by Local Correlation Tracking
    of simulation compared to IMaX observed spectrum \citep{yelles_chaouche:SunConvFlowPowSpctr}.  The left
    figure is for the hydrodynamic run and the right shows results for the three 
    MHD runs as well.  The color scheme is the same as for 
    Fig.\ \ref{fig:vvertspectrum}.
    \label{fig:vhorizspectrum}}
\end{figure}

To compare with the observations, synthetic spectral
lines were calculated for the vertical columns in the simulation
results using the Nicole code \citep{socas-navarro:HinodeEmpiricalSunAtm} assuming LTE.
These are then modified to reproduce the instrumental degradation of
the observations.  The vertical velocity is determined from the Doppler shift of
the spectral lines and 
the horizontal velocities are determined by Local Correlation Tracking applied
to successive continuum images for both the simulation and the observations
Figs.\ \ref{fig:vvertspectrum} and \ref{fig:vhorizspectrum}.

The velocities are then compared with the actual simulation velocities on
surfaces of constant optical depth ($\tau_{500} = $ const.) corresponding to 
the formation region of the spectral line .
The best correlation for the vertical component (0.96) is obtained at $\tau_{500}
= $ 0.1.  The best correlation for the horizontal component is lower (0.5) at
$\tau_{500}=$ 1, which provides an estimate of the accuracy of LCT.

%++++++++++++++++++++++++++++++++++++++++++++++++++++++++++++++++++++++++++++++
\subsection{P-Modes}
%++++++++++++++++++++++++++++++++++++++++++++++++++++++++++++++++++++++++++++++

The p-mode spectrum samples the mean structure throughout the simulation domain.
We restrict comparisons with solar observations to non-radial modes that have 
their lower turning points well within the computational box of 20\,Mm depth.

\begin{figure}[!htb]
  \centerline{
  \includegraphics[width=0.6\textwidth]{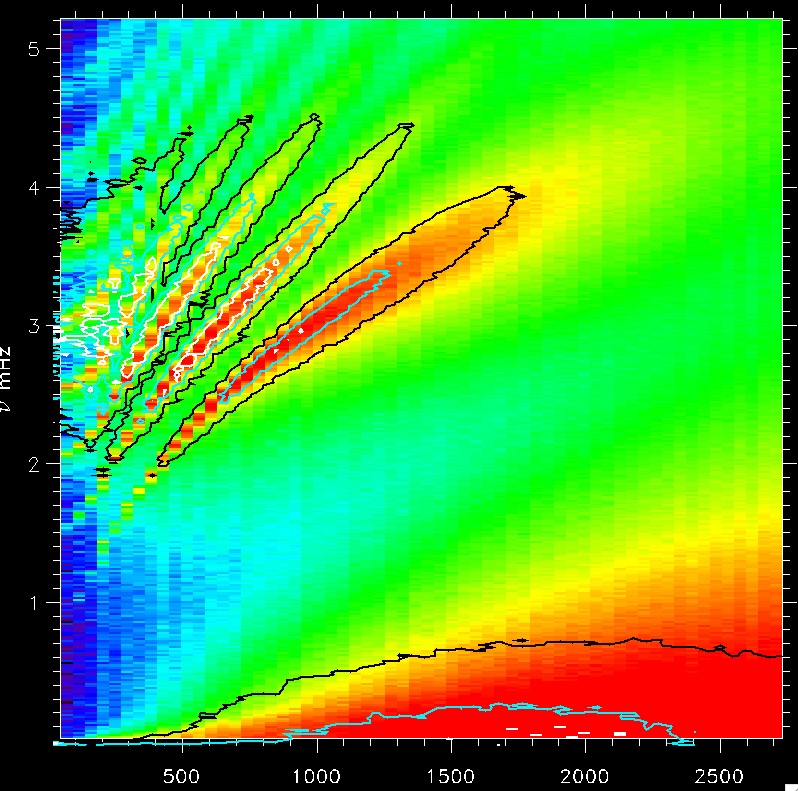}}
    \caption{k-$\Om$ diagram of p-mode power. Images are from the an 18 hour 
    time sequence of a 96\,Mm square by 20.5\,Mm deep simulation and
    contours are from a 36 hour time sequence of 93.3\,Mm square HMI data.
    The abscissa is $\ell$ and the ordinate is the frequncy in mHz.
    \label{fig:kw}}
\end{figure}

The solar MHD convection simulation were run with the {\tt mconv} experiment. 
We make a direct comparison of an 18 hour time sequence of a 96\,Mm wide by 20.5\,Mm 
deep simulation with a  36 hour 93.3\,Mm square 
patch of the HMI Doppler velocity observations at disk center.
For these simulations we used the 4-bin scaled opacities and an equation of
state both from the Uppsala stellar atmospheres package \citep{Gustafsson1975}.
The simulation had 1\,kG horizontal magnetic field advected into the computational
domain by inflows at the bottom.  At continuum optical depth 0.1, the average unsigned vertical field is 2\,G and
its maximum is 2\,kG in either direction.  The average horizontal field is 3.5\,G with
a maximum of 650\,G.
We take the Fourier transform in space and time of both
data sets.  Because of the slight difference in horizontal
dimensions, we linearly interpolate the Fourier transformed HMI data to the same
wavenumber grid as the simulations.  The
$k-\Om$ diagram is shown in Fig. (\ref{fig:kw}).

\begin{figure}[!htb]
  \centerline{
   \includegraphics[width=0.6\textwidth]{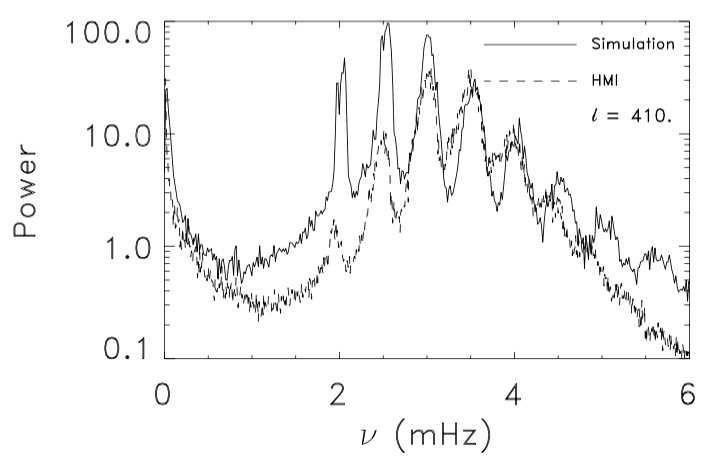}}
  \centerline{
   \includegraphics[width=0.6\textwidth]{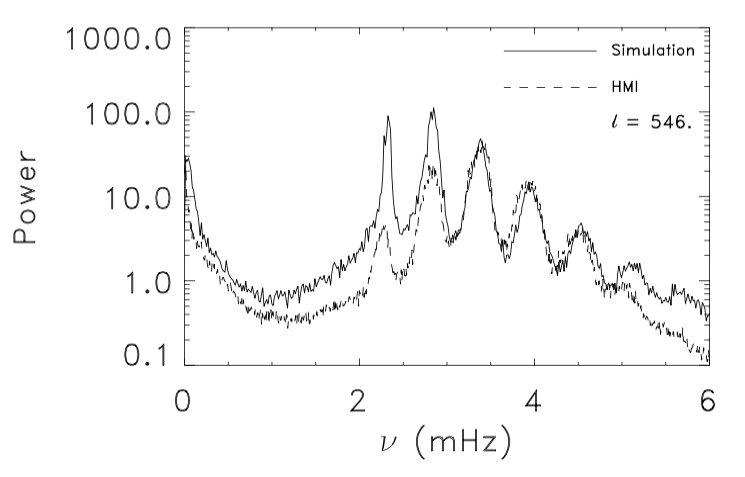}}
  \centerline{
   \includegraphics[width=0.6\textwidth]{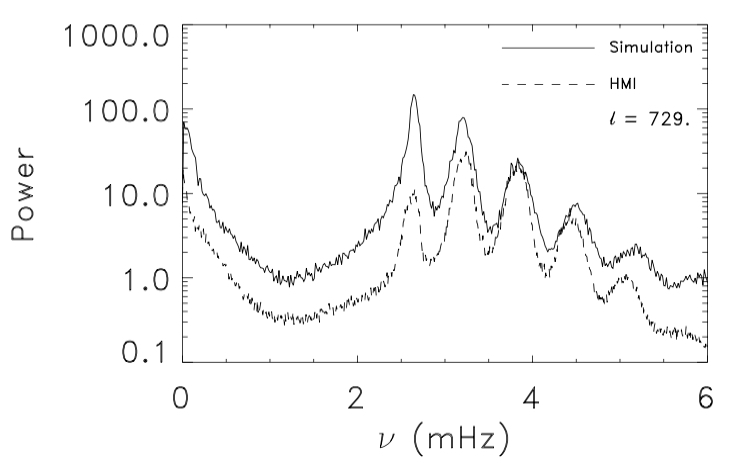}}
    \caption{The p-mode spectra comparing the 96\,Mm wide simulation with the 93.3\,Mm
    wide HMI observations for horizontal wavelengths $\ell = 410,546,729$ corresponding to
    a horizontal wavelengths 11, 8, and 6\,Mm respectively.
    The simulation was an 18 hour time sequence and the HMI observations were
    a 36 hour time sequence.  The power is summed over all direction at each wavenumber.
    The units of power are $m^2 s^{-2}$.
%{\tiny Do I have the correct units for the power?  I %tried to find some
%solar observations with the same type of analysis, %but only found values per
%Hz.}
    \label{fig:pmodespectrum}}
\end{figure}

Individual spectra for $\ell=$ 410, 546 and 729 are shown in Fig.
(\ref{fig:pmodespectrum}).  
The good, but not perfect, agreement in mode frequency, width and amplitude between 
the simulation and the observed modes indicates that
the simulated convective structure is an accurate representation of the solar
structure in the top 20\,Mm of the convection zone and that the driving and
damping processes are fairly accurately represented in the simulations. 
The small deviations provide important data for further analysis of 
p-mode driving and damping.   Theoretically
\citep{stein2001,goldreich1994} we expect the modes to be excited
stochastically by rapid variations in the Reynolds stresses and
surface entropy fluctuations.  The mode widths 
represent the mode lifetimes, which range from 1-4 hours for 
these modes. Lifetimes should depend primarily on the mode damping. 
The significant deviations between the simulated and observed
modes are: 
(1) the lowest frequency simulated mode amplitudes are larger than those
observed for all wavelengths.
(2) Simulated mode widths increase with decreasing wavelength relative to those
observed, from slightly narrower than observed to wider than observed.
(3) The simulated mode amplitudes at each wavelength decrease relative to those
observed with increasing frequency and become equal for the second or third 
harmonic.
(4) The intermode power in the simulations is larger than observed.

%++++++++++++++++++++++++++++++++++++++++++++++++++++++++++++++++++++++++++++++
%\subsection{Turbulence}
%++++++++++++++++++++++++++++++++++++++++++++++++++++++++++++++++++++++++++++++

%++++++++++++++++++++++++++++++++++++++++++++++++++++++++++++++++++++++++++++++
\subsection{Scaling}
%++++++++++++++++++++++++++++++++++++++++++++++++++++++++++++++++++++++++++++++

\begin{figure}[!htb]
  \includegraphics[width=0.6\textwidth]{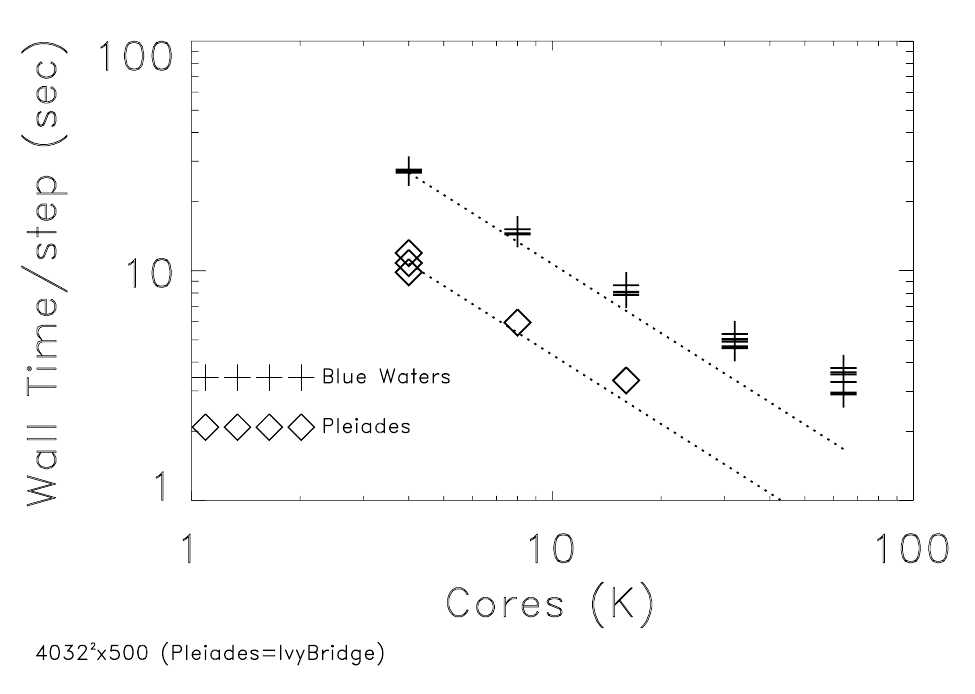}
    \caption{Wall time per time step as a function of the number of processors (in units of 1K)
    for a $4032^2 \times 500$ solar convection simulation.
    \label{fig:scaling}}
\end{figure}

We have tested the scaling of the Stagger Code on the Blue Waters and Pleiades
supercomputers using 48 $\times$ 48\,Mm wide by 20.5\,Mm vertical computational box
of dimensions $4032^2 \times 500$.  It included radiative transfer with a 4 bin 
scaled opacity and one vertical and four slanted rays.  
The runs were 100 time steps, using 4032 MPI tasks.
Figure \ref{fig:scaling} shows the wall time per time step for 3,200 to 64,000 processors.

Where there are multiple data points, they refer to multiple runs with the same setup.
The difference in times is due to different loads on the computers.
The difference between Pleiades and Blue Waters is due to the
different processors on the machines.
Significant departures from linearity occurs at $50^3$ grid points per
processor.
%)
%}

%%%%%%%%%%%%%%%%%%%%%%%%%%%%%%%%%%%%%%%%%%%%%%%%%%%%%%%%%%%%%%%%%%%%%%%%%%%%%%%
\section{Conclusion}
%%%%%%%%%%%%%%%%%%%%%%%%%%%%%%%%%%%%%%%%%%%%%%%%%%%%%%%%%%%%%%%%%%%%%%%%%%%%%%%
\label{sect:conclusion}

%{
%(
The Stagger Code is a realistic, three-dimensional, magneto-hydrodynamic
code.  It is an excellent tool for modeling many astrophysical
situations and has been used to study stellar atmospheres, star
formation, galaxy evolution and supersonic turbulence.  It is modular
and can easily be modified to include the needed physics.  Such
codes can be used to both make predictions that can guide and be
confirmed by observations (edge brightenings of solar granules
were predicted before they were observed) and aid in the interpretation
of observations by producing models with much more information than
is observable.  In this way our understanding of astrophysical
phenomena is increased.

In this paper we have given a detailed description of the methodology
used in the Stagger Code: the sixth order centered derivatives and
fifth order centered interpolations on a staggered grid; the third
order, low memory Runge-Kutta time advance; the various methods of
applying boundary conditions using ghost zones in the non-periodic
direction.  The different modules for treating energy transfer were
presented, including a detailed description of the non-gray radiative
transfer solution.  The Feautrier equations are solved on a rotating
set of long characteristics with one vertical and several slanted
rays.  Opacities are binned to drastically reduce the number of
frequencies solved for.  Local Thermodynamic Equilibrium is assumed.
We described the physics used in the equation of state and opacity
calculations.  Parameters are input via a text file containing a
number if namelists with their values.  We provide several tests
of the Stagger Code, both for idealized problems and in comparison
with solar observations.  These show that the Stagger Code is highly
conservative, non-disapative with low dispersion.  The behavior of
the dynamic surface layers of the solar atmosphere is accurately
modeled.  The Stagger Code has excellent parallel scaling,
including with a realistic non-gray radiative transfer solution,
when there are more than $50^3$ cells per processor, 

Much progress has been made in astrophysics with past and current
computing capacity.  Exabyte computing capability and new codes to
take advantage of them will significantly increase the extent and
complexity of astrophysical phenomena that can be simulated.

%)
%}

\begin{acknowledgments}
RT acknowledge financial support from NASA grant 80NSSC18K0559.
This research has made extensive use of the indispensable NASA
High-End Computing (HEC) Program through the NASA Advanced
Supercomputing (NAS) Division at Ames Research Center and the
Astrophysics Data System.  \end{acknowledgments} %\clearpage

\appendix

%{
\section{Namelists}
Parameters for the Stagger Code are input via an input text file containing a number of namelists with input parameters.  The namelists \verb'&io', \verb'&run', \verb'&var', \verb'&grid', and \verb'&pde'  are read in the routine \verb'params.f90', \verb'&bdry' in \verb'BOUNDARIES/boundary', \verb'&slice' in \verb'slice.f90', \verb'&quench' in \verb'STAGGER/quench.f90', \verb'&eqofst' in \verb'EOS/eos', \verb'&part' in \verb'particles.f90', \verb'&force' in \verb'FORCING/gravit' or another forcing routine, \verb'&init' in an \verb'INITIAL/*.f90' initialization routine, \verb'&expl' in \verb'FORCING/noexplode' or a stars or sn routine, and \verb'&cool' in \verb'COOL/radiation' or a similar cooling routine.  
Default values are specified in the routines that read each namelist.  If the parameters in a given namelist are not needed the namelist can be left blank, or with only selected parameters present. 
In addition there are some initialization routines that may or may not be compiled into the executable that have additional namelist inputs. The most important namelists are shown below, with typical input parameters commented:

\subsection{io -- I/O parameters}
\begin{verbatim}
&io
!--- should always be set ---
 file='snapshot.dat'         ! output snapshot file
 iread=-2                    ! -2: last available snapshot; n: snapshot n; -1: do not read
 iwrite=-2                   ! -2: next available snapshot; n: snapshot n; -1: do not write
 tsnap=0.3                   ! write snapshot every tsnap time units
 nscr=50                     ! overwrite scr file every nscr steps
!--- often used -------------
 do_scp=.true.               ! automatically start an 'scp' transfer
 scp_dir='host:data/runxx'   ! remote directory to scp results to
 do_subdir=.false.           ! use subdirectories for rank-specific files
 do_divb_clean=.true.        ! apply div(B) cleaning to initial snapshot
!--- sometimes used ---------
 from='input.dat'            ! input snapshot file (default = file)
 do_init_smooth=.false.      ! apply smoothing to initial snapshot
 do_init_smooth_hd=.false.   ! apply smoothing to HD part of initial snapshot
!--- defaults normally ok ---
 do_parallel_io=.true.       ! use MPI parallel I/O
 do_master_io=.false.        ! alternative I/O method (obsolete)
 nsnap=0                     ! write snapshot every nsnap steps (for tests)
 tscr=0.0                    ! overwrite scr file every tscr time units
 nstep=9999999               ! maximum time steps in run
 tstop=999999                ! maximum time of run
 divb_iter=15                ! max number of div(B) iterations
 divb_eps=1e-4               ! required accuracy of div(B) cleaning
/
\end{verbatim}

\subsection{run -- Run control parameters}
\label{sect:RUNparms}
\begin{verbatim}
&run
!--- should always be set ---
 Cdt=0.4                     ! Courant condition on velocity 0.3-0.5 
(Eqn. 19)
 Cdtd=0.7                    ! Courant condition on viscosity 0.6-0.8
 Cdtr=0.1                    ! Courant condition on density & energy changes  0.1-0.4
!--- defaults normally ok ---
 do_flags=.true.             ! control execution with flag files
 timeorder=3                 ! Time integration order
 iseed=-77                   ! Random number seed
!--- debug parameters -------
 do_trace=.false.            ! trace calls
 do_dump=.false.             ! enable dumps
 do_compare=.false.          ! compare results of two runs
 do_logcheck=.false.         ! write check info to log file false      
 do_check=.false.            ! write check file
 do_timer=.false.            ! timer calls
 do_validate=.false.         ! bitwise reduction precision etc
 dbg_select=0                ! select topic for debugging
 dbg_barrier=0    ´          ! barrier number to test
 dbg_level=0                 ! debug level
 idbg=0                      ! Debug printout control
 rank_timer=.false.          ! timers for each rank
 spin_barrier=0              ! spins to do at barrier
 t_prop=0                    ! proper time, for field upping
 E_prop=0                    ! proper energy, for field upping
 E_mag=0                     ! actual energy
 check_barrier               ! check that barriers are in sync
 use_isend                   ! default: use MPI_Issend
/
\end{verbatim}

\subsection{vars -- Controls which equations are active}
\begin{verbatim}
&vars
!--- defaults normally ok ---
 do_density=.true.           ! include density evolution?
 do_momenta=.true.           ! include momemtum evolution?
 do_energy=.true.            ! include energy evolution?
 dio_mhd=.true.              ! include the induction equation?
 do_pscalar=.false.          ! include a passive scalar equation?
/
\end{verbatim}

\subsection{grid -- Sets grid dimensions}
\begin{verbatim}
&grid
!--- should always be set ---
 mx=33 my=33 mz=33           ! dimensions of computational box
 mpi_nx=0 mpi_ny=0 mpi_nz=0  ! dimensions of mpi domains
 sx=1 sy=1 sz=1              ! physical dimensions of box
 xmin=1e6 ymin=1e6 zmin=1e6  ! minimum of [xyz]-axis
                             ! if ==1e6 init_mesh determins value
!--- defaults normally ok ---
 lb=3*1                      ! lower index boundary
 ub=3*0                      ! upper index boundary
 omega=3*0                   ! coordinate system rotation rate
´meshfile='mesh.dat'         ! file with mesh information
/
\end{verbatim}

\subsection{pde -- PDE parameters}
\label{sect:PDFparms}
\begin{verbatim}
&pde
!--- defaults often ok -----
 do_stratified=.false.      ! special treatment of some operations
 stratified=0               ! subtract average energy
 stratified=1               ! subtract average energy, with quench
 stratified=2               ! subtract average flux
 stratified=3               ! subtract average flux, with quench      
 do_loginterp=.true.        ! interpolate in the log
 do_2nddiv=.true,           ! use second order derivatives
 do_lnpg=.false.            ! compute pressure gradient in the log
 do_nodrag=.true.           ! subtract net viscous drag
 do_kepler=.false.          ! subtract Kepler velocity
 do_massexact=.true.        ! enforce exact mass conservation
 do_max5=.false.            ! average & smooth over 5 zones
 do_max5_hd=.false.         ! average & smooth over 5 zones in HD
 do_max5_mhd=.false.        ! average & smooth over 5 zones in MHD
 gamma=1.666667             ! gas gamma
 nu1=0.003-0.05             ! normal viscosity (Eqn. 20)
 nu2=0.3-1.5                ! shock viscosity
 nu3=0.003-0.05             ! wave speed viscosity
 nu4=0-1.5                  ! ln(rho) gradient coefficient
 nuS=1                      ! solenoidal fraction
 nur=0-0.1                  ! relative mass diffusion
 nuE=1                      ! relative energy diffusion
 nuB=1                      ! relative magnetic diffusion
 nup=1                      ! relative passive scalar diffusion
 do_aniso_nu=.false.        ! anisotropic nu, for skew aspect ratios
 do_aniso_eta=.false.       ! anisotropic eta, for skew aspect ratios
 d2lnr_lim=0.3              ! 2nd order divergence when $d^2\ln\rho$ larger
 nrho=0                     ! decrease compressiv viscority for depths  > nrho
 Csmag=0.1-0.15             ! Smagorinsky constant
 Ssmag=0.05                 ! Smagorinsky constant
 csound=1                   ! for polytropic EOS
 cmax=20                    ! maximum Alfen speed
 do_massflux=.false.        ! include viscous mass diffusion
 t_Bgrowth=0                ! growth rate of B
 CdivB=0.06-0.11            ! limit divB diffusion
 ambi=0                     ! ambipolar diffusion
 nu_atmos                   ! atmospheric diffusion term
 nu_inner=1.4               ! inner boundary for atmospheric diffusion
 nu_outer=1.6               ! outer boundary for atmospheric diffusion
 do_upping=.false.          ! field upping to keep magnetic energy constant
 do_cylatmos=.false.        ! outer edge for atmospheric diffusion
/
\end{verbatim}

\subsection{bdry -- Boundary parameters}
\label{sect:BDRYparms}
\begin{verbatim}
&bdry
 lb=1                       ! lower boundary index
 ub=my                      ! upper boundary index
 t_bdry=0.01                ! boundary decay time   (Eqn. 57)
 t_Bbdry=0.01               ! boundary decay time for B
 eetop_fraction=0.01-1      ! fraction of old average energy
                            ! at top boundary used in new
 htop=0                     ! density scale height at top boundary
 rmin=1e-3 -- 1e-2          ! minimum $\rho$ at top bounary = rmin*$<\rho>$
 rbot=-1                    ! fiducial density at bottom boundary (Eqn. 57, 61)
 ebot=-1                    ! fiducial energy at bottom boundary
 pbot=-1                    ! fiducial pressure at bottom boundary (Eqn. 61)
 Bx0,By0,Bz0=0              ! B at bottom boundary  (Eqn. 64, 66)
 Binflow=.true.             ! impose B at bottom boundary
 t_bdry=0.05--0.6           ! boundary decay time
 t_Bbdry=0.01               ! boundry decay time for B
 t_vbdry=1.2                ! boundary decay time for V inflow
 Uy_bdry=0.01               ! smoothing transition from in to out flows  (Eqn.
65)
 uy_max=20                  ! limit velocity at top boundary
 pb_fact=1                  ! limit $\rho$ and $e$ perturbations due to B
 lb_eampl=0                 ! ??
/
\end{verbatim}

\subsection{slice -- slice extraction parameters}
\begin{verbatim}
&slice
 do_slice=.false.           ! write slices
 do_vrms=.false.            ! write vrms instead of velocity
 tslice=1                   ! time interval between slices
 ntslice=10                 !
 nslice=1                   ! timestep interval between slices
 jdir=3                     ! normal to direction of slice
 jx,y,z=(m{x,y,z}+1)/2$     ! mid point
 nvar=1                     ! number of slice variables
 var=O                      ! variable number to make slice of
/
\end{verbatim}

\subsection{quench -- quenching parameters}
\label{sect:QUENCHparms}
\begin{verbatim}
&quench
 do_quench=.true.           ! use quenching?
 qmax=8                     ! maximum quenching     (Eqn. 24)
 qlim=2                     ! limit second differences
/
\end{verbatim}

\subsection{eqofst -- Combined EOS, Opacity table}
There are also other, more specialized versions of this namelist
\begin{verbatim}
&eqofst
 do_eos=.true.           ! on or off?
 do_ionization=.true.    ! calculate ionization
 do_table=.false.        ! use table
 tablefile='table.dat'   ! table file of EOS and Opacity
/
\end{verbatim}

\subsection{part -- Trace particles}
\begin{verbatim}
&part
do_particles=.false      ! use trace particles
fraction                 !
/
\end{verbatim}

The namelist \verb'/force/' may take different forms depending on the system being
modeled.  

Namenlist \verb'/init/'
Namelist /\verb'expl/' is used when there are explosions.

Namelist \verb'/cool/' specifies whether there is cooling, and if so the variables needed for a given type of cooling. 

\subsection{cool -- Cooling and Radiative Transfer}
\begin{verbatim}
&cool
 do_cool             ! t or f 
 intfile             ! name 
 nmu=2               ! number of rays in mu-direction
 nphi=4              ! number of azimuthal angles
 dphi=10             ! change in azimuthal angle each time step
 form_factor=1       ! ??
 dtaumin=0.1         ! minimum $\Delta\tau$
 dtaumax=1e2--1e3    ! maximum $\Delta\tau$
 do_newton=.false.   ! do newtonian cooling
 y_newton=-0.3       ! do Newtonian cooling above
 dy_newton=0.05      ! thickness  of tranzition zone
 t_newton=0.01--0.3  ! timescale
 ee_newton=5.265     ! fiducial energy & 5.265
 ee_min=3.6--4.4
 t_ee_min=0.005      ! timescale to raise to minimum
 do_limb=.false.     ! compute limb darkening
 mu0 & & 0
 ny0 & & 0
 mblocks & & 1
 do_fld=.false.      ! do flux-limited diffusion
 do_conduction=.false. ! thermal conduction?
 do_corona=.false.   ! coronal thin cooling?
 do_chromos=.false.  ! simple chromospheric cooling?
 t_chromos=1         ! time scale for ditto
 ee_chromos=7        ! energy per unit mass to cool fo
 do_Casiana=.false.  ! use Casiana expressions
 do_top=.true.       ! use upper boundary
 r_top=1.e-8         ! ??
 t_top=1             ! ??
 ee_top=?            ! damp fluctuations to ee_top 
 h_top               ! transition for Newtonian cooling
 d_top               ! transition width
 do_damp=.false.     ! damping?
 t_damp=1            ! time scale
 iphot               ! photospheric index
 ichrom              ! chromospheric index
 c_scale             ! scaling constant
 do_limit            ! apply limiter
 tt_limit            ! limiting temperature
 cdt_cond=0.05       ! Courant condition
/
\end{verbatim}
%}

%%%%%%%%%%%%%%%%%%%%%%%%%%%%%%%%%%%%%%%%%%%%%%%%%%%%%%%%%%%%%%%%%%%%%%%%%%%%%%%
\bibliography{StaggerCode}
%
% \bibliography{bibs/eos,bibs/opac,bibs/conv,bibs/starmod,bibs/seism,bibs/staratm,bibs/obs,bibs/math}
%%%%%%%%%%%%%%%%%%%%%%%%%%%%%%%%%%%%%%%%%%%%%%%%%%%%%%%%%%%%%%%%%%%%%%%%%%%%%%%

\end{document}